\documentclass[%
 reprint,
superscriptaddress,
nofootinbib,
 amsmath,amssymb,
 aps,
]{revtex4-1}

\usepackage{graphicx}
\usepackage{dcolumn}
\usepackage{bm}
\usepackage{here}

\begin{document}

\preprint{APS/123-QED}

\title{Black hole-neutron star binary merger: Dependence on black hole spin orientation and equations of state}

\author{Kyohei Kawaguchi}
\affiliation{Yukawa Institute for Theoretical Physics, Kyoto University, Kyoto 606-8502, Japan}
\author{Koutarou Kyutoku}

\affiliation{Interdisciplinary Theoretical Science (iTHES) Research Group,
RIKEN, Wako, Saitama 351-0198, Japan\\
}
\author{Hiroyuki Nakano}
\affiliation{Department of Physics, Kyoto University, Kyoto 606-8502, Japan}
\affiliation{Center for Computational Relativity and Gravitation, and School of Mathematical Sciences, Rochester Institute of Technology, 85 Lomb Memorial Drive, Rochester, New York 14623}
\author{Hirotada Okawa}
\affiliation{Yukawa Institute for Theoretical Physics, Kyoto University, Kyoto 606-8502, Japan}
\affiliation{Advanced Research Institute for Science and Engineering, Waseda University, 3-4-1 Okubo, Shinjuku, Tokyo 169-8555, Japan}
\author{Masaru Shibata}
\affiliation{Yukawa Institute for Theoretical Physics, Kyoto University, Kyoto 606-8502, Japan}
\author{Keisuke Taniguchi}
\affiliation{Department of Physics, University of the Ryukyus, 1 Senbaru, Nishihara, Okinawa 903-0213, Japan}
\affiliation{Graduate School of Arts and Sciences, University of Tokyo, Komaba, Meguro, Tokyo 153-8902, Japan}

\date{\today}

\begin{abstract}
	We systematically performed numerical-relativity simulations for black hole (BH) - neutron star (NS) binary mergers with a variety of the BH spin orientation and nuclear-theory-based equations of state (EOS) of the NS. The initial misalignment angles of the BH spin measured from the direction of the orbital angular momentum are chosen in the range of $i_{\rm tilt,0}\approx30^\circ$--$90^\circ$. We employed four models of nuclear-theory-based zero-temperature EOS for the NS with which the compactness of the NS is in the range of ${\cal C}=M_{\rm NS}/R_{\rm NS}=0.138$--$0.180$, where $M_{\rm NS}$ and $R_{\rm NS}$ are the mass and the radius of the NS, respectively. The mass ratio of the BH to the NS, $Q=M_{\rm BH}/M_{\rm NS}$, and the dimensionless spin parameter of the BH, $\chi$, are chosen to be $Q=5$ and $\chi=0.75$, together with $M_{\rm NS}=1.35M_\odot$ so that the BH spin misalignment has a significant effect on tidal disruption of the NS. We obtain the following results: (i) The inclination angle of $i_{\rm tilt,0}< 70^\circ$ and $i_{\rm tilt,0}< 50^\circ$ are required for the formation of a remnant disk with its mass larger than $0.1M_\odot$ for the case ${\cal C}= 0.140$ and ${\cal C}=0.160$, respectively, while the disk mass is always smaller than $0.1M_\odot$ for ${\cal C}\agt 0.175$. The ejecta with its mass larger than $0.01M_\odot$ is obtained for $i_{\rm tilt,0}<85^\circ$ with ${\cal C}=0.140$, for $i_{\rm tilt,0}<65^\circ$ with ${\cal C}=0.160$, and for $i_{\rm tilt,0}<30^\circ$ with ${\cal C}=0.175$. (ii) The rotational axis of the dense part of the remnant disk with its rest-mass density larger than $10^9~{\rm g/cm^3}$ is approximately aligned with the remnant BH spin for $i_{\rm tilt,0}\approx 30^\circ$. On the other hand, the disk axis is misaligned initially with $\sim 30^\circ$ for $i_{\rm tilt,0}\approx 60^\circ$, and the alignment with the remnant BH spin is achieved at $\sim 50$--$60~{\rm ms}$ after the onset of merger. The accretion time scale of the remnant disk is typically $\sim 100~{\rm ms}$ and depends only weakly on the misalignment angle and the EOS. (iii) The ejecta velocity is typically $\sim0.2$--$0.3c$ and depends only weakly on the misalignment angle and the EOS of the NS, while the morphology of the ejecta depends on its mass. (iv) The gravitational-wave spectra contains the information of the NS compactness in the cutoff frequency for $i_{\rm tilt,0}\alt 60^\circ$.

\begin{description}
\item[PACS numbers]
04.25.D-, 04.30.-w, 04.40.Dg
\end{description}
\end{abstract}

\maketitle

\section{Introduction}
 The ground-based gravitational-wave detectors, such as Advanced LIGO~\cite{bib:ligo}, Advanced VIRGO~\cite{bib:virgo}, and KAGRA~\cite{bib:kagra}, will detect signals in the next decade. Black hole (BH) - neutron star (NS) binary mergers are one of the most promising gravitational-wave sources for these detectors~\cite{bib:ps}. Since NSs and BHs are compact objects with strong gravity, detection of gravitational waves from compact binary mergers, including BH-NS mergers, can be a touchstone of the theory of gravity~\cite{bib:grtest}. The gravitational waveforms from BH-NS mergers depend on the BH mass, the BH spin,  the NS mass, and the NS radius, and thus they will carry information of these binary parameters~\cite{bib:bhns9,bib:bhns10}. In particular, the information of the NS radius could be used for  constraining the equation of state (EOS) and makes a big contribution not only to astrophysics but also to nuclear and fundamental physics~\cite{bib:gwi1,bib:gwi2,bib:gwi3,bib:gwi4}. Since the expected signal-to-noise ratio in the gravitational-wave detection is not very high, taking a cross correlation between the observational data and theoretical template is an important method to extract the information from the detected signal at least for the next generation gravitational-wave detectors~\cite{bib:gwlr}, and hence waveforms for a variety of binary parameters have to be derived.
 
 BH-NS mergers are also proposed as a potential progenitor of short-hard gamma-ray-bursts (sGRBs)~\cite{bib:grb1}. If the NS is tidally disrupted during the merger, a hot and massive disk with mass $\agt 0.01M_\odot$ could be formed around the remnant spinning BH. This BH-disk system could launch a relativistic jet by releasing its gravitational energy through the neutrino emission or an electromagnetic energy flow in a short time scale $\alt 2~$s, and hence, could be the central engine of the sGRBs. This scenario for sGRBs, so-called merger scenario, is suited for explaining its duration and the estimated event rates (see Refs.~\cite{bib:grb2,bib:grb3} for reviews and references therein). For extracting physical information of the central engine of sGRB from electromagnetic observations, we have to clarify the formation process of the BH-disk system and the relation between this system and sGRB.

 A fraction of the NS material would be ejected during tidal disruption~\cite{bib:bhns12,bib:bhns13}. Since the NS consists of highly neutron-rich matter, r-process nucleosynthesis is expected to take place in the ejecta~\cite{bib:chemi1}, and the emission powered by decay of the radioactive nuclei would occur (kilonova/macronova)~\cite{bib:eje1,bib:eje2,bib:eje3}. This electromagnetic counterpart of the binary merger is useful for determining the position of the source when gravitational waves are detected. Also, their light curves will reflect the binary parameters, and could be useful for extracting the physical information of the binary. Furthermore, the r-process nuclei produced in the ejecta are considered to contribute to the chemical evolution history of the universe ~\cite{bib:chemi1,bib:chemi2,bib:chemi3}.

 The property of disks and ejecta, particularly their mass, is essential for predicting the electromagnetic counterparts, such as sGRB and kilonova. The mass of the matter that remains outside the BH after the merger depends strongly on whether and at which orbital separation the NS is disrupted by the tidal force of the companion BH~\cite{bib:bhns_liv}. If the tidal disruption occurs at a sufficiently distant orbit from the innermost stable circular orbit (ISCO) of the BH, an appreciable amount of mass would remain outside the remnant BH. On the other hand,  if the tidal disruption does not occur or occurs near the ISCO of the BH, entire NS material would be swallowed by the BH. The orbital separation at which the tidal disruption occurs depends on binary parameters, particularly on the mass ratio of the binary, the NS compactness, and the BH spin. Thus, we have to clarify the dependence of tidal disruption processes on these parameters in order to predict the mass of the remnant disk and the ejecta.
 
 The orbital angular momentum and BH spin can be misaligned. A population synthesis study suggests that about a half of BH-NS binaries can have an initial configuration in which the misalignment angle between the orbital angular momentum and the BH spin is larger than $45^\circ$~\cite{bib:ps2}.  A post-Newtonian (PN) study  shows that the orbit precesses due to spin-orbit coupling (or dragging of inertial frames) in the presence of misalignment~\cite{bib:pn}. This could  affect dynamics of BH-NS binaries not only in the inspiral phase but also in the merger phase. 
  
   Recently, a variety of numerical-relativity (NR) simulations have been performed for BH-NS binaries~\cite{bib:bhns1,bib:bhns2,bib:bhns3,bib:bhns4,bib:bhns5,bib:bhns6,bib:bhns7,bib:bhns8,bib:bhns9,bib:bhns10,bib:bhns11,bib:bhns12,bib:bhns13,bib:bhns_m1,bib:bhns_m2,bib:bhns_m3,bib:bhns_r1,bib:bhns_r2,bib:bhns_t1,bib:bhns_t2}, and quantitative dependence of the merger process on binary parameters have been revealed. Some works were done taking into account magnetic fields~\cite{bib:bhns_m1,bib:bhns_m2,bib:bhns_m3}, nuclear-theory-based EOS~\cite{bib:bhns8,bib:bhns12,bib:bhns13,bib:bhns_r1, bib:bhns_r2}, and neutrino cooling~\cite{bib:bhns_r1,bib:bhns_r2}. However, most of them were done for the case that the BH spin is aligned with the orbital angular momentum and there are only two studies for the misaligned case~\cite{bib:bhns_t1,bib:bhns_t2}. Moreover, mass ejection has not yet been studied for the misaligned-spin case based on NR simulations (there is a study with an approximate treatment of general relativity~\cite{bib:bhns_fm}).

 The primary purpose of this paper is to clarify the quantitative dependence of the disk formation and the mass ejection from BH-NS mergers on BH spin misalignment and nuclear-theory-based EOS by performing NR simulations systematically. We employed four nuclear-theory-based EOSs described by piecewise polytrope with four pieces~\cite{bib:eos}, and varied the misalignment angle between the orbital angular momentum and the BH spin, while fixing the NS mass, the magnitude of the BH spin, and the mass ratio of the binary as $1.35M_\odot$, $0.75$, and $5$, respectively.

 This paper is organized as follows. In Sec.~\ref{sec:sec2}, we describe methods for computing initial conditions, piecewise polytropic EOS, and models of BH-NS binaries we employed in this paper. In Sec.~\ref{sec:sec3}, the formulation and the methods of numerical simulations are summarized. In Sec.~\ref{sec:sec4}, we present the numerical results for misaligned-spin BH-NS mergers. Finally, summaries and discussions of this work are presented in Sec.~\ref{sec:sec5}. Throughout this paper, we adopt the geometrical units in which $G=c=1$, where $G$ and $c$ are the gravitational constant and the speed of light, respectively. Exceptionally, $c$ is sometimes inserted for clarity when we discuss the velocity of ejecta. Our convention of notation for physically important quantities is summarized in Table~\ref{tb:symb}. The dimensionless spin parameter of the BH, total mass of the system at infinite separation, mass ratio, and compactness of the NS are defined as $\chi=S_{\rm BH}/M_{\rm BH}^2$,  $Q=M_{\rm BH}/M_{\rm NS}$, $m_0=M_{\rm BH}+M_{\rm NS}$, and ${\cal C}=M_{\rm NS}/R_{\rm NS}$, respectively. ${\bf J}=\left(J^x, J^y, J^z\right)$ is the total angular momentum of the system (see Sec.~\ref{ssec:diag} for its definition).
  The angle between the BH spin and the orbital angular momentum, $i_{\rm tilt}$, is defined by
\begin{equation}
	i_{\rm tilt}:={\rm cos}^{-1}\left(\frac{{\bf S_{\rm BH}}\cdot {\bf L}}{S_{\rm BH} L}\right),
\end{equation}
where ${\bf L}$ is the orbital angular momentum of the binary, which is defined using the total angular momentum, ${\bf J}$, and the BH spin, ${\bf S}_{\rm BH}$, as ${\bf L}:={\bf J}-{\bf S}_{\rm BH}$ with $S_{\rm BH}=|{\bf S}_{\rm BH}|$ and $L=|{\bf L}|$. We use $i_{\rm tilt,0}$ to describe $i_{\rm tilt}$ at the initial condition. Latin and Greek indices denote spatial and spacetime components, respectively.

\begin{table*}
\caption{Our convention of notation for physically important quantities.}
\begin{center}
 \begin{tabular}{lc} \hline
 Symbol & \\\hline\hline
 $M_{\rm irr}$ & The irreducible mass of the BH\\
 $M_{\rm BH}$ & The gravitational mass of the BH in isolation\\
 $S_{\rm BH}$ & The magnitude of the BH spin angular momentum\\
 $M_{\rm NS}$ & The gravitational mass of the NS in isolation\\
 $R_{\rm NS}$ & The circumferential radius of the NS in isolation\\
 $m_0$ & The total mass of the system at the infinite separation\\
 $M_0$ & The Arnowitt-Deser-Misner mass of the system\\
 $J$ & The total angular momentum of the system\\
 $Q$  &  The mass ratio $M_{\rm BH}/M_{\rm NS}$\\
 ${\cal C}$  &  The compactness parameter of the NS $M_{\rm NS}/R_{\rm NS}$\\
 $\chi$ & The dimensionless spin parameter of the BH $S_{\rm BH}/M_{\rm BH}^2$\\
 $i_{\rm tilt}$ & The misalignment angle of the BH spin\\
 $i_{\rm tilt,0}$ & $i_{\rm tilt}$ at the initial condition
 \\\hline
 \end{tabular}
\end{center}
\label{tb:symb}
\end{table*}

\section{Initial Condition}\label{sec:sec2}
\subsection{Formulation and methods}
We prepare quasiequilibrium states of BH-NS binaries
as initial conditions of numerical simulations in a similar manner to
our previous works for aligned-spin binaries
\cite{kyutoku_st2009,bib:bhns9,bib:bhns10}. Numerical computations are performed by using multidomain spectral method library {\tt LORENE}~\cite{bib:lor}. As gravitational
radiation reaction reduces the orbital eccentricity \cite{peters1964}, typical BH-NS binaries settle to a quasicircular
orbit. For the case that the orbital separation is large, the
time scale of radiation reaction, $\tau_\mathrm{GW}$, is much longer
than the orbital period, $P_\mathrm{orb}$, because their ratio is given
by~\cite{bib:bhns_liv}
\begin{equation}
 \frac{\tau_\mathrm{GW}}{P_\mathrm{orb}} \approx 1.1 \;
  \frac{(1+Q)^2}{4Q} \left( \frac{r}{6m_0} \right)^{5/2} .
\end{equation}
 For the case that
the BH has a spin angular momentum inclined to the orbital
angular momentum, the orbital plane precesses due to the spin-orbit
coupling effect of general relativity, and a closed orbit is not
obtained even in the absence of radiation reaction \cite{bib:pn}. This implies that even in one orbital cycle, the gravitational interaction between two objects varies depending on the angle between the BH spin and the line connecting two centers of mass, and hence, the definition of the 
quasicircular orbit is not trivial. At
a large orbital separation, however, the precession time scale,
$P_\mathrm{prec}$, is also longer than $P_\mathrm{orb}$ as~\cite{bib:pn}
\begin{equation}
 \frac{P_\mathrm{prec}}{P_\mathrm{orb}} \approx 1.7 \;
  \frac{7(1+Q)^2}{4(4Q+3)} \frac{r}{6 m_0} ,
\end{equation}
where we take the small-spin limit. Thus, we can neglect the orbital precession as well as 
the gravitational radiation reaction for computing initial data at
a large orbital separation, and the binary can be regarded approximately
as an equilibrium configuration in the comoving frame.

We then compute quasiequilibrium states assuming the presence of an
instantaneous helical Killing vector field with the orbital angular
velocity $\Omega$,
\begin{equation}
 \xi^\mu := \left( \partial_t \right)^\mu + \Omega \left(
	     \partial_\varphi \right)^\mu .
\end{equation}
For nonspinning or aligned-spin BHs, this
reduces to a genuine helical Killing vector as far as the radiation
reaction is neglected. Accordingly, the orbital plane can be taken to be
a plane perpendicular to the rotational axis. This does not hold for
misaligned-spin BHs, and we do not restrict the orbital plane to
be a plane perpendicular to the rotational axis. Instead, we compute
initial data by requiring that neither the BH nor NS
has the velocity component along the rotational axis
\cite{bib:bhns_t1}. This implies that the binary is located at the
extrema of coordinate separation along the rotational axis, where the
helical symmetry should hold instantaneously.

The line element in the $3+1$ form is written as
\begin{align}
 ds^2 & = g_{\mu \nu} dx^\mu dx^\nu \notag \\
 & = - \alpha^2 dt^2 + \gamma_{ij} \left( dx^i + \beta^i dt \right)
  \left( dx^j + \beta^j dt \right) ,
\end{align}
where $\alpha$, $\beta^i$, and $\gamma_{ij}$ are the lapse function,
shift vector, and three-diemensional spatial metric, respectively. Initial data of the
gravitational field consist of $\gamma_{ij}$ and the extrinsic curvature
defined by
\begin{equation}
 K_{ij} := - \frac{1}{2} \mathcal{L}_n \gamma_{ij} ,
\end{equation}
where $n^\mu$ is the future-oriented timelike unit normal vector field
to the initial hypersurface.

We employ the extended conformal thin-sandwich formalism
\cite{bib03,bib04} with the conformal
transverse-traceless decomposition of Einstein's equation
\cite{bib05} to compute $\alpha$, $\beta^i$, $\gamma_{ij}$, and
$K_{ij}$. We assume the conformal flatness of the spatial metric,
$\gamma_{ij} = \psi^4 \tilde{\gamma}_{ij} = \psi^4 f_{ij}$, the
stationarity of the conformal metric, $\partial_t \tilde{\gamma}_{ij} =
0$, and the maximal slicing condition, $K = 0 = \partial_t K$. Here,
$f_{ij}$ and $K$ are the flat spatial metric and the trace part of the
extrinsic curvature, $K := \gamma^{ij} K_{ij}$, respectively. To handle
a coordinate singularity associated with the BH, we decompose the
conformal factor $\psi$ and a weighted lapse function $\Phi := \alpha
\psi$ into singular and regular parts using constants $M_\mathrm{P}$ and
$M_\Phi$ as \cite{bib:bhns1,kyutoku_st2009}
\begin{align}
 \psi & = 1 + \frac{M_\mathrm{P}}{2 r_\mathrm{BH}} + \phi , \\
 \Phi & = 1 - \frac{M_\Phi}{r_\mathrm{BH}} + \eta ,
\end{align}
where $r_\mathrm{BH} := | \mathbf{x} - \mathbf{x}_\mathrm{P} |$ is the
coordinate distance from the puncture located at
$\mathbf{x}_\mathrm{P}$. We also decompose $\hat{A}_{ij} := \psi^{-2}
K_{ij}$ into regular and singular parts as
\begin{equation}
 \hat{A}_{ij} = \tilde{D}_i W_j + \tilde{D}_j W_i - \frac{2}{3} f_{ij}
  \tilde{D}_k W^k + K_{ij}^\mathrm{P} ,
\end{equation}
where $\tilde{D}_i$ denotes a covariant derivative associated with $f_{ij}$ and the index of $W^i$ is raised/lowered by
$f_{ij}$. The singular part is given by \cite{bowen_york1980}
\begin{align}
 K_{ij}^\mathrm{P} & := \frac{3}{2 r_\mathrm{BH}^2} \left[ \hat{x}_i
 P^\mathrm{BH}_j + \hat{x}_j P^\mathrm{BH}_i - ( f_{ij} - \hat{x}_i
 \hat{x}_j ) \hat{x}^k P^\mathrm{BH}_k \right] \notag \\
 & + \frac{3}{r_\mathrm{BH}^3} \left[ \epsilon_{ikl} S_\mathrm{P}^l
 \hat{x}^k \hat{x}_j + \epsilon_{kjl} S_\mathrm{P}^l \hat{x}^k \hat{x}_k
 \right] ,
\end{align}
where $\hat{x}^i := ( x^i - x_\mathrm{P}^i ) / r_\mathrm{BH}$. The index
of $\hat{x}^i$ is also raised/lowered by $f_{ij}$. The parameters
$P^\mathrm{BH}_i$ and $S_\mathrm{P}^i$ are constants associated with the
linear momentum and spin angular momentum of the puncture, respectively
(see below).

The equations to determine $\phi$, $\beta^i$, $\eta$, and $W_i$ are
derived by combining the Hamiltonian constatint, momentum constraint,
$\partial_t K = 0$, and $\partial_t \tilde{\gamma}_{ij} = 0$ as
\begin{align}
 \Delta \phi & = - 2 \pi \psi^5 \rho_\mathrm{H} - \frac{1}{8} \psi^{-7}
 \hat{A}_{ij} \hat{A}^{ij} , \\
 \Delta \beta^i & + \frac{1}{3} \tilde{D}^i \tilde{D}_j \beta^j = 16 \pi
 \Phi \psi^3 j^i + 2 \hat{A}^{ij} \tilde{D}_j \left( \Phi \psi^{-7}
 \right) , \\
 \Delta \eta & = 2 \pi \Phi \psi^4 \left( \rho_\mathrm{H} + 2S \right) +
 \frac{7}{8} \Phi \psi^{-8} \hat{A}_{ij} \hat{A}^{ij} , \\
 \Delta W_i & + \frac{1}{3} \tilde{D}_i \tilde{D}_j W^j = 8 \pi \psi^6
 j_i .
\end{align}
where $\Delta := f^{ij} \tilde{D}_i \tilde{D}_j$. The matter source
terms are defined by
\begin{align}
 \rho_\mathrm{H} & := T_{\mu \nu} n^\mu n^\nu , \\
 j^i & := - T_{\mu \nu} n^\mu \gamma^{\nu i} , \\
 S^{ij} & := T_{\mu \nu} \gamma^{\mu i} \gamma^{\nu j} ,
\end{align}
with $S := \gamma^{ij} S_{ij}$. The asymptotic flatness gives the outer boundary conditions as
\begin{equation}
 \phi |_\infty \; = \; \beta^i |_\infty \; = \; \eta |_\infty \; = \;
  W_i |_\infty = 0.
\end{equation}
 In contrast to initial data of BH-NS binaries computed in the excision framework
\cite{taniguchi_bfs2007,taniguchi_bfs2008,grandclement2006,grandclement2007,foucart2008}, we do not have to give inner
boundary conditions at the horizon. We also do not have to give nonzero
boost contributions to the shift vector~\cite{bib:bhns_t1}, because the
Arnowitt-Deser-Misner linear momentum of the system can be set to zero
by choosing $P^\mathrm{BH}_i$ appropriately. Hence, BH-NS binaries with misaligned spins do not exhibit the center-of-mass
motion in the puncture framework.

Free parameters associated with the puncture are determined as
follows. The so-called puncture mass, $M_\mathrm{P}$, is adjusted for obtaining a desired BH mass, $M_\mathrm{BH}$. The other mass
parameter, $M_\Phi$, is determined by the condition that the
Arnowitt-Deser-Misner mass and the Komar mass agree  with each other for stationary and asymptotically flat spacetime~\cite{beig1978,ashtekar_magnonashtekar1979}, that is,
\begin{equation}
 \oint_{r \to \infty} \partial_i \Phi dS^i = - \oint_{r \to \infty}
  \partial_i \psi dS^i = 2 \pi M_0.
\end{equation}
 The linear momentum of the
puncture, $P_\mathrm{BH}^i$, is determined by the condition that the
Arnowitt-Deser-Misner linear momentum of the system vanishes as
\begin{equation}
 P^\mathrm{BH}_i = - \int j_i \psi^6 d^3 x .
\end{equation}
The spin parameter of the puncture, $S_\mathrm{P}^i$, is given in
Cartesian coordinates:
\begin{equation}
 S_\mathrm{P}^i = S_\mathrm{P} ( \sin i'_\mathrm{tilt} , 0 , \cos
  i'_\mathrm{tilt} ) .
\end{equation}
The magnitude $S_\mathrm{P}$ is adjusted for obtaining a desired value of
BH spin, $S_\mathrm{BH}$, measured on the
horizon.\footnote{Although we distinguish $S_\mathrm{P}$ from $S_\mathrm{BH}$, the difference between two is
at most $O(10^{-4})$.} The inclination angle $i'_\mathrm{tilt}$ is
chosen to be $30^\circ$, $60^\circ$, and $90^\circ$ in this
study. Because the direction of orbital angular momentum does not agree
with the axis of helical symmetry, $i'_\mathrm{tilt}$ does not alway agree with
$i_\mathrm{tilt}$ , which is defined as the angle between the orbital and BH spin angular momenta. The typical
difference is $3^\circ$, and we do not adjust values of
$i_\mathrm{tilt}'$ to control $i_\mathrm{tilt}$ in this study.

The NS matter is assumed to be composed of an ideal fluid. The
energy-momentum tensor is given by
\begin{equation}
 T_{\mu \nu} := \rho h u_\mu u_\nu + P g_{\mu \nu} ,
\end{equation}
where $\rho$ is the rest-mass density, $P$ is the pressure, $h = 1 +
\varepsilon + P/\rho$ is the specific enthalpy with $\varepsilon$ the
specific internal energy, and $u^\mu$ is the four-velocity of the
fluid. The velocity field of the fluid is expected to be irrotational,
because the viscosity of the NS matter is low
\cite{kochanek1992,bildsten_cutler1992} and the rotational periods of
observed NS in compact binaries are not very short (see,
e.g., \cite{lorimer2008}). The zero relativistic vorticity condition, or
irrotationaliry condition, is written as
\begin{align}
 \omega_{\mu \nu} & := ( \delta^\alpha_\mu + u^\alpha u_\mu ) (
 \delta^\beta_\nu + u^\beta u_\nu ) \left( \nabla_\alpha u_\beta -
 \nabla_\beta u_\alpha \right) \notag \\
 & = h^{-1} \left[ \nabla_\mu ( h u_\nu ) - \nabla_\nu ( h
 u_\mu ) \right] \notag \\
 & = 0,
\end{align}
where the energy-momentum conservation and adiabacity condition are used for deriving the second-line expression~\cite{bib:irr1,bib:irr2}.
This implies the presence of a velocity potential $\Psi$ such that $h
u_\mu = \nabla_\mu \Psi$, and the elliptic equation to determine $\Psi$
is derived from the continuity equation $\nabla_\mu ( \rho u^\mu ) =
0$ together with the helical symmetry. The irrotational conditions and helically symmetric conditions of
the specific momentum, $\mathcal{L}_\xi ( h u^\mu ) = 0$, are combined
to give
\begin{equation}
 h \xi_\mu u^\mu = - C ( = \mathrm{const.} ) ,
\end{equation}
which is used to determine $h$. The integration constant $C$ is determined by the
condition that the baryon rest mass of the NS takes a desired
value. As we explain in the next section, the specific enthalpy
determines all the other thermodynamical quantities in the computation
of initial data.

The relative location of each component of the binary is determined as
follows. We fix the binary separation in the direction perpendicular to
the rotational axis, which is chosen to be the $z$-axis. The centers of
BH and NS are put on the $xz$-plane, but now they are not limited to the
$x$-axis.\footnote{We can also fix the locations of both components by
forcing them to be on the $x$-axis. This method respects the
instantaneous helical symmetry as well as the method adopted in this
study. Taking the fact that no criteria are available to determine which
condition gives superior initial data, we simply follow the method
adopted in \cite{bib:bhns_t1}.} The orbital angular velocity of the
binary is determined by the force-balance condition that the NS does not
move perpendicular to the rotational axis, and this amounts to requiring
$dh/dx=0$ in our coordinates. The location of the rotational axis with
respect to the binary components is determined by the condition that the
magnitude of the orbital angular momentum agrees with the value derived
by the third-and-a-half post-Newtonian formulas for a given value of the
orbital angular velocity (see Appendix~D of
\cite{bohe_mfb2013}). Finally, the binary separation along the
rotational axis is determined by the condition that the NS has no
velocity component along the rotational axis, and this amounts to
requiring $dh/dz=0$ in our coordinates \cite{bib:bhns_t1}.

\subsection{Piecewise polytropic equations of state}
 Since the cooling time scale of NS is much shorter than the lifetime of typical compact binaries, we can employ a zero-temperature EOS for the NS just before binary mergers~\cite{bib:nscool}. Employing a zero-temperature EOS, the thermodynamical quantities, such as $P$, $\varepsilon$, and $h$, can be described  as functions of $\rho$ as
\begin{equation}
	P=P\left(\rho\right),\varepsilon=\varepsilon\left(\rho\right),h=h\left(\rho\right).
\end{equation}
From the first-law of thermodynamics, these quantities satisfy relations,
\begin{eqnarray}
	d\varepsilon&=&\frac{P}{\rho^2}d\rho,\label{eq:fsteps}\\
	dh&=&\frac{1}{\rho}dP,
\end{eqnarray}
which determine $\varepsilon$ and $h$ from given $P\left(\rho\right)$, respectively. 

In this work, we employ a piecewise polytropic EOS~\cite{bib:eos} to describe a zero-temperature EOS of the NS. This is a phenomenologically parametrized EOS, which reproduces a zero-temperature nuclear-theory-based EOS at high density only with a small number of polytropic constants and indices as 
\begin{equation}
	P\left(\rho\right)=\kappa_i\rho^{\Gamma_i}\ {\rm for}\,\rho_{i-1}\le\rho<\rho_i\,\left(1\le i\le n\right),
\end{equation}
where $n$ is the number of the pieces used to parametrize an EOS. $\rho_i$ is the rest-mass density at a boundary of two neighboring $i$-th and $(i+1)$-th pieces, $\kappa_i$ is the $i$-th polytropic constant, and $\Gamma_i$ is the $i$-th adiabatic index. Note, here, $\rho_0=0$ and $\rho_n\rightarrow\infty$. Requiring  the continuity of the pressure,  $\kappa_i\rho_i^{\Gamma_i}=\kappa_{i+1}\rho_i^{\Gamma_{i+1}}\left(1\le i\le n-1\right)$, the EOS is determined completely by giving $\kappa_1$, $\rho_i$ and $\Gamma_i$. $\varepsilon$ is determined by integrating Eq.~(\ref{eq:fsteps}) with the integration constant, $\varepsilon\left(0\right)=0$.
 It was shown that the piecewise polytropic EOS with four pieces reproduces the nuclear-theory-based EOSs within $\sim 5\% $ errors in pressure for the nuclear density range~\cite{bib:eos}.
 
\begin{table*}
\caption{The key quantities for piecewise polytropic EOSs~\cite{bib:eos} which we employ in this paper. $P_2$ is the pressure at $\rho=\rho_2$ shown in the unit of ${\rm dyne/cm^2}$, $\Gamma_i$ is the adiabatic index for each piecewise polytrope,  and $ $$M_{\rm max}$ is the maximum mass of the spherical NS for a given EOS. $R_{1.35}$, $\rho_{1.35}$, $M_{*,1.35}$, and ${\cal C}_{1.35}$ are the radius, the central rest-mass density, the baryon rest mass, and the compactness parameter for the NS with $M_{\rm NS}=1.35M_\odot$, respectively.}
\begin{center}
 \begin{tabular}{l|ccccc|cccc} \hline
 Model & ${\rm log}_{10}P_2$ & $\Gamma_2$ & $\Gamma_3$ & $\Gamma_4$ & $M_{\rm max}[M_\odot]$&$R_{1.35}[{\rm km}]$ & $\rho_{1.35} [{\rm g/cm^{3}}]$ & $M_{*,1.35}[M_\odot]$ & ${\cal C}_{1.35}$  \\ \hline\hline
 APR4 & 34.269 & 2.830 & 3.445 & 3.348 & 2.20  	&  11.1 & 8.9$\times10^{14}$ & 1.50 & 0.180\\
 ALF2 & 34.616 & 4.070 & 2.411 & 1.890 & 1.99   	&12.4 & 6.4$\times10^{14}$ & 1.49 & 0.161\\
 H4 & 34.669 & 2.909 & 2.246 & 2.144 & 2.03     	&13.6 & 5.5$\times10^{14}$ & 1.47 & 0.147\\
 MS1 & 34.858 & 3.224 & 3.033 & 1.325 & 2.77  	&14.4 & 4.2$\times10^{14}$ & 1.46 & 0.138\\\hline
 \end{tabular}
\end{center}
\label{tb:eoslist}
\end{table*}
Table~\ref{tb:eoslist} lists the EOSs which we employ in our study. We employ the models of the NS EOS which can realize the NS with $M_{\rm NS}\approx 2M_\odot$ which satisfies the recent observational constraint~\cite{bib:NSMmax1,bib:NSMmax2}. For these models, the NS radius is in the range $\sim11-15~{\rm km}$ for $M_{\rm NS}=1.35M_\odot$, which is largely consistent with the recent theoretical and observational suggestion~\cite{bib:NSR1,bib:NSR2}. Following~\cite{bib:eos,bib:bhns9,bib:bhns10} , we always fix the parameters of EOS in the subnuclear-density region as
\begin{eqnarray}
	\Gamma_1&=&1.35692395,\\
	\kappa_1/c^2&=&3.998 736 92\times10^{-8}\left({\rm g/cm^3}\right)^{1-\Gamma_1},
\end{eqnarray}
 and we set $\rho_2= 10^{14.7}~{\rm g/cm^3}$ and $\rho_3= 10^{15}~{\rm g/cm^3}$. Here, we insert $c$ for clarity. Instead of giving $\rho_1$, we give $P_2$ in Table~\ref{tb:eoslist} for each EOS, which is the pressure at $\rho=\rho_2$.

\subsection{Models}

\begin{table}
\caption{Key parameters and quantities for the initial conditions adopted in our numerical simulation. The adopted EOS, the initial angle between orbital angular momentum and the BH spin $\left(i_{\rm tilt,0}\right)$, the ADM mass $\left(M_0\right)$, and the total angular momentum $\left(J_0\right)$ , respectively. Note that $M_{\rm NS}=1.35M_\odot$ and $m_0=8.10M_\odot$}
\begin{center}
 \begin{tabular}{l|cccc} \hline
 Model & EOS & $i_{\rm tilt,0}[^\circ]$ & $M_0[M_\odot]$& $J_0[GM_\odot/c]$  \\ \hline\hline
 APR4i30 & APR4  &33&8.04&63\\
 APR4i60 & APR4  &63&8.05&57\\
 APR4i90 & APR4  &94&8.05&47\\ \hline
 ALF2i30 & ALF2 &33&8.04&63\\
 ALF2i60 & ALF2 &63&8.05&57\\
 ALF2i90 & ALF2 &94&8.05&47\\ \hline
 H4i30   & H4   &33&8.04&63\\
 H4i60   & H4   &63&8.05&57\\
 H4i90   & H4   &94&8.05&47\\ \hline
 MS1i30  & MS1  &32&8.04&63\\
 MS1i60  & MS1  &63&8.05&57\\
 MS1i90  & MS1  &93&8.05&48\\ \hline
 \end{tabular}
\end{center}
\label{tb:model}
\end{table}

 As we already mentioned, we choose $i_{\rm tilt,0}\approx30^\circ, 60^\circ$, and $90^\circ$.  We  employ four different piecewise polytropic EOSs, APR4, ALF2, H4, and MS1, for each value of $i_{\rm tilt,0}$. On the other hand,  we set the NS mass, $M_{\rm NS}$, the mass ratio, $Q$, and dimensionless spin paramter, $\chi$, to be fixed values $\left(M_{\rm NS}, Q, \chi\right)=\left(1.35, 5, 0.75\right)$, for which the misalignment of the BH spin has a  significant effect on tidal disruption. For all the models, the initial angular velocity $\Omega_0$ normalized by the total mass is set to be $m_0\Omega_0=0.036$. We summarize several key quantities for the initial condition in Table~\ref{tb:model}. The label for the model denotes the EOS name and the value of $i_{\rm tilt,0}$. Specifically, ``i30", ``i60", and ``i90" denote the models with $i_{\rm tilt,0}\approx30^\circ, 60^\circ$, and $90^\circ$, respectively. For all the models, we rotate the initial data before we start the simulation so that the initial direction of the total angular momentum agrees with the direction of the $z$-axis.

\section{Methods of Simulations}\label{sec:sec3}
Numerical simulations are performed using an adaptive mesh refinement (AMR) code {\tt SACRA}~\cite{bib15}. Here, we employ a Baumgarte-Shapiro-Shibata-Nakamura (BSSN) formulation partially incorporating Z4c prescription~\cite{bib14}. The gauge conditions, the numerical scheme, and the diagnostics are essentially the same as those described in~\cite{bib:bhns9,bib:bhns10}.
\subsection{Formulation and Numerical Methods}
We solve Einstein's evolution equation partially incorporating the Z4c formulation (see~\cite{bib:eccred} for our prescription) with a moving puncture gauge. The Z4c formulation is a modified version of the BSSN-puncture formulation, introducing a new variable. In the BSSN formulation, the long-term simulation causes a slow accumulation of the numerical error that leads to gradual violation of the constraints, which should be zero if the evolution equations are solved exactly. This accumulation comes from the fact that in the BSSN formulation, the evolution equations of the constraints have a non-propagating mode. In the Z4c formulation, by introducing a new variable, $\Theta$, the evolution equation of the constraints changes entirely to a wave equation. Therefore, the violation of the constraints propagates away, and the local accumulation of the numerical error can be suppressed.

In the Z4c formulation, we evolve the conformal factor, $W:=\gamma^{-1/6}$, the conformal three-metric, ${\tilde \gamma}_{ij}=\gamma^{-1/3}\gamma_{ij}$, a variable slightly modified from the trace of the extrinsic curvature, ${\hat K}:=K-2\Theta$, the conformal trace-free part of the extrinsic curvature, ${\tilde A}_{ij}=\gamma^{-1/3} \left(K_{ij}-K\gamma_{ij}/3\right)$, an auxiliary variable, ${\tilde \Gamma}^i$, and the new variable, $\Theta$. Here, $\gamma:={\rm det} \gamma_{ij}$. The evolution equations are written as,
\begin{eqnarray}
	\left(\partial_t-\beta^i\partial_i\right)W&=&\frac{1}{3}W\left[\alpha \left({\hat K}+2\Theta\right)-\partial_i\beta^i\right],\\
	\left(\partial_t-\beta^k\partial_k\right){\tilde \gamma}_{ij}&=&-2\alpha{\tilde A}_{ij}+{\tilde \gamma}_{ik}\partial_j\beta^k+{\tilde \gamma}_{jk}\partial_i\beta^k\nonumber\\
	&-&\frac{2}{3}{\tilde \gamma}_{ij}\partial_k\beta^k,
\end{eqnarray}
\begin{eqnarray}
	\left(\partial_t-\beta^i\partial_i\right){\hat K}&=&-D^iD_i\alpha\nonumber\\
	&+&\alpha\left[{\tilde A}_{ij}{\tilde A}^{ij}+\frac{1}{3}\left({\hat K}+2\Theta\right)^2\right.\nonumber\\
	&+& \left. 4\pi\left(\rho_{\rm H}+S\right)\right]\nonumber\\
	&+&\alpha\kappa_1\left(1-\kappa_2\right)\Theta,
\end{eqnarray}
\begin{eqnarray}
	\left(\partial_t-\beta^j\partial_j\right)\Theta&=&\bigg[\frac{1}{2}\alpha\left\{R-{\tilde A}_{ij}{\tilde A}^{ij}+\frac{2}{3}\left({\hat K}+2\Theta\right)^2\right\}\nonumber\\
	&-&\alpha\left\{8\pi\rho_{\rm H}+\kappa_1\left(2+\kappa_2\right)\Theta\right\}\bigg]e^{-(r/r_0)^2},\label{eq:eth}
\end{eqnarray}
\begin{eqnarray}
	\left(\partial_t-\beta^k\partial_k\right){\tilde A}_{ij}&=&-W^2\left(D_iD_j\alpha-\frac{1}{3}\gamma_{ij}D^kD_k\alpha\right)\nonumber\\
	&+&W^2\alpha\left(R_{ij}-\frac{1}{3}\gamma_{ij}R\right)\nonumber\\
	&+&\alpha\left[\left({\hat K}+2\Theta\right){\tilde A}_{ij}-2{\tilde A}_{ik}{\tilde A}^k_j\right]\nonumber\\
	&-&8\pi W^2\alpha\left(S_{ij}-\frac{1}{3}\gamma_{ij}S\right)\nonumber\\
	&+&{\tilde A}_{ik}\partial_j\beta^k+{\tilde A}_{jk}\partial_i\beta^k\nonumber\\
	&-&\frac{2}{3}{\tilde A}_{ij}\partial_k\beta^k,
\end{eqnarray}
\begin{eqnarray}
	\left(\partial_t-\beta^j\partial_j\right){\tilde \Gamma}^i&=&-2{\tilde A}^{ij}\partial_j\alpha+2\alpha\left[{\tilde \Gamma}^i_{jk}{\tilde A}^{jk}-\frac{3}{W}{\tilde A}^{ij}\partial_jW\right.\nonumber\\
	&-&\left.\frac{1}{3}{\tilde \gamma}^{ij}\partial_j\left(2{\hat K}+\Theta\right)-8\pi{\tilde \gamma}^{ij}j_j\right]\nonumber\\
	&+&{\tilde \gamma}^{jk}\partial_j\partial_k\beta^i+\frac{1}{3}{\tilde \gamma}^{ij}\partial_j\partial_k\beta^k-{\tilde \Gamma}^k_{\rm d}\partial_k\beta^i\nonumber\\
	&+&\frac{2}{3}{\tilde \Gamma}^i_d\partial_k\beta^k-2\alpha\kappa_1\left({\tilde \Gamma}^i-{\tilde \Gamma}^i_{\rm d}\right),
\end{eqnarray}
where $D_i$ denotes a covariant derivative associated with $\gamma_{ij}$, ${\tilde \Gamma}^i_{\rm d}=-\partial_j {\tilde \gamma}^{ij}$, and $\kappa_1$ and $\kappa_2$ are coefficients associated with the constraint damping. An overall factor, $e^{-(r/r_0)^2}$,  is multiplied in the right-hand side of Eq.~(\ref{eq:eth}) so that $\Theta$ plays a role only in the inner region of the simulation box. In our simulation, we set $\kappa_1=\kappa_2=0$, and $r_0=L/2$, where $L$ is the size of the computational domain on one side (see Table.~\ref{tb:amr}). The spatial derivatives in the evolution equations are evaluated by fourth-order centered finite differencing except for the advection terms, which are evaluated by fourth-order upwind finite differencing. A fourth-order Runge-Kutta scheme is employed for the time evolution.

 Following~\cite{bib09}, we employ a moving-puncture gauge in the form
\begin{eqnarray}
	\left(\partial_t-\beta^i\partial_i\right)\alpha&=&-2\alpha K,\\
	\left(\partial_t-\beta^j\partial_j\right)\beta^i&=&\frac{3}{4}B^i,\\
	\left(\partial_t-\beta^j\partial_j\right)B^i&=&\left(\partial_t-\beta^j\partial_j\right){\tilde \Gamma}^i-\eta_{\rm B} B^i,
\end{eqnarray}
where $B^i$ is an auxiliary variable, and $\eta_{\rm B}$ is a coefficient introduced to suppress a strong oscillation of the shift vector. In this work, we set $\eta_{\rm B}=0.16/M_\odot$.

 The EOS is basically the same as those described in~\cite{bib:bhns9, bib:bhns10}: We decompose the pressure and the specific internal energy into a zero-temperature part and a thermal part as 
\begin{equation}
	P=P_{\rm cold}+P_{\rm th}\ ,\ \varepsilon=\varepsilon_{\rm cold}+\varepsilon_{\rm th}.
\end{equation}
Here, $P_{\rm cold}$ and $\varepsilon_{\rm cold}$ are functions determined by the piecewise polytropic EOS. Then, the thermal part of the specific internal energy is calculated by $\varepsilon_{\rm th}=\varepsilon-\varepsilon_{\rm cold}$, where $\varepsilon$ is given from the hydrodynamics. Finally, we determine the thermal part of the pressure using a simple $\Gamma$-law, ideal-gas EOS as
\begin{equation}
	P_{\rm th}=\left(\Gamma_{\rm th}-1\right)\rho\varepsilon_{\rm th},
\end{equation}
where $\Gamma_{\rm th}$ is an adiabatic index for the thermal part, which we choose $\Gamma_{\rm th}=1.8$.  As is discussed in Appendix A. of \cite{bib:bhns13}, the difference of the disk mass and the ejecta mass among the different values of $\Gamma_{\rm th}$ is expected to be small compared to the numerical errors due to finite gridding, thus, we do not study the dependence on $\Gamma_{\rm th}$ in this paper.

 Since the vacuum is not allowed in any conservative hydrodynamics scheme (see~\cite{bib15} for details), we put an artificial atmosphere of small density  outside the NS. The atmosphere density is set  to be $\rho_{\rm atm}=10^{-12}\rho_{\rm max}\sim 10^{3}~{\rm g/cm^3}$ for the inner region, $r<R_{\rm crit}$, and $\rho_{\rm atm}=10^{-12}\rho_{\rm max}\left(R_{\rm crit}/r\right)^3$ for the outer region, $r\ge R_{\rm crit}$, with $R_{\rm crit}\approx L/16$. The total rest mass of the atmosphere is always less than $10^{-6}M_\odot$, and hence we can safely neglect the effect of the artificial atmosphere as far as appreciable tidal disruption occurs.

\subsection{Diagnostics}\label{ssec:diag}
 We estimate the mass and the spin angular momentum of BH with misaligned spin assuming that the deviation from Kerr spacetime is negligible in the vicinity of the BH, at least for the case that the separation of the binary is large or the system is approximately regarded as a steady state. In a steady state with stationary slicing, the event horizon (EH) agrees with apparent horizon (AH). Thus, the irreducible mass of the BH is determined by the area of the AH, $A_{\rm AH}$, as 
\begin{equation}
	M_{\rm irr}=:\sqrt{\frac{A_{\rm AH}}{16\pi}}.
\end{equation}
In Kerr spacetime, a relation
\begin{equation}
	M_{\rm BH}^2=\frac{2M_{\rm irr}^2\left( 1-\sqrt{1-\chi^2}\right)}{\chi^2}
\end{equation}
 holds between the gravitational and irreducible masses of the BH.
 If the irreducible mass and the BH spin are known, we can calculate the BH mass using this relation. To determine the spin angular momentum, we use the relation between the spin and the intrinsic scalar curvature of the horizon, $R^{(2)}_{\rm EH}$, following~\cite{bib:chi}. In Kerr spacetime, $R^{(2)}_{\rm EH}$ is written as
\begin{equation}
	R^{(2)}_{\rm EH}\left(\theta\right)=\frac{2\left({\hat r}_+^2+\chi^2\right)\left({\hat r}_+^2-3\chi^2{\rm cos}^2\theta\right)}{M_{\rm BH}^2 \left({\hat r}_+^2+\chi^2{\rm cos}^2\theta\right)^3}.
\end{equation}
Here, ${\hat r}_+=1+\sqrt{1-\chi^2}$ is a normalized radius of the EH, and $\theta$ is the latitude in Boyer-Lindquist coordinates. The minimum and maximum values of $R^{(2)}_{\rm EH}$ at $\theta=0$ and $\pi/2$, i.e., at the pole and the equator of the BH are, respectively,
\begin{eqnarray}
	R^{(2)}_{\rm min}&=&\frac{-1+2\sqrt{1-\chi^2}}{2M_{\rm irr}^2},\\
	R^{(2)}_{\rm max}&=&-\frac{2\left(-2+\chi^2+2\sqrt{1-\chi^2}\right)}{M_{\rm irr}^2\chi^4}.
\end{eqnarray}
Solving these equations with respect to $\chi$, we have
\begin{eqnarray}
	\chi_{\rm min}^2&=&1-\left(\frac{1}{2}+M_{\rm irr}^2R^{(2)}_{\rm min}\right)^2,\\
	\chi_{\rm max}^2&=&\frac{-2+2\sqrt{2M_{\rm irr}^2R^{(2)}_{\rm max}}}{M^2_{\rm irr}R^{(2)}_{\rm max}}.
\end{eqnarray}
Using these relations, we can estimate the BH spin approximately from the intrinsic scalar curvature of the AH. The direction of the BH spin can be determined from the location of $R^{(2)}_{\rm min}$. We note that the direction of the BH spin is not  gauge invariant in this definition. However, we expect that it gives a reasonable measure of the spin direction if the tidal forces are negligible~\cite{bib:chi,bib:chi2}. For all the models, the values of $\chi_{\rm min}$ and $\chi_{\rm max}$ agree with each other up to significant ($\sim3$) digits . Thus, we use the value of $\chi_{\rm max}$ for calculating $M_{\rm BH}$ and ${\bf S}_{\rm BH}$ in this paper.

 We define the total angular momentum of the system, ${\bf J}=\left(J^x, J^y, J^z\right)$,  from a rotational invariance of the gravitational Hamiltonian at spatial infinity as~\cite{bib:tam},
 \begin{equation}
 	J^i:=\frac{1}{8\pi}\epsilon^{ijk}\oint_{r\rightarrow \infty}x_j\left(K_k^l-K\gamma_k^l\right)dS_l,
 \end{equation}
 where $\epsilon^{ijk}$ is the spatial Levi-Civita tensor.

 Finally, we define the time at which the binary merges. For this purpose, we define the rest mass inside the AH as
\begin{equation}
	M_{\leq {\rm AH}}:=\int_{r\leq r_{\rm AH}}\rho_*d^3x	,
\end{equation}
  where, $r_{\rm AH}=r_{\rm AH}\left(\theta,\varphi\right)$ is the radius of the AH as a function of the angular coordinates, and $\rho_*=\rho\alpha u^t\sqrt{\gamma}$ is the conserved rest-mass density. Then, we define the merger time, $t_{\rm merge}$, as the time at which $M_{\leq {\rm AH}} \geq 10^{-2}M_\odot$ is achieved.

\subsection{Setups for AMR grids}
\begin{table}\begin{center}
\caption{Setups of the grid structure for the simulation with our AMR algorithm. $\Delta x$ is the grid spacing at the finest-resolution domain. $R_\mathrm{diam}$ is the semimajor diameter of the NS in the direction perpendicular to the axis of helical symmetry. $L$ is a half of the edge length of the largest domain. $\lambda_0 =\pi /\Omega_0$ is the gravitational wavelength of the initial configuration.}
\label{tb:amr}
\begin{tabular}{l|cccc}\hline
	Model & $\Delta x/ M_0$ & $R_{\rm diam}/\Delta x$ & $L/\lambda_0$ &	 $L~({\rm km})$\\\hline\hline
	APR4i30	&	0.0134	&	101	&	2.357	&	2444	\\
	APR4i60	&	0.0133	&	102	&	2.343	&	2429	\\
	APR4i90	&	0.0136	&	100	&	2.398	&	2486	\\\hline
	ALF2i30	&	0.0153	&	102	&	2.687	&	2786	\\
	ALF2i60	&	0.0153	&	102	&	2.687	&	2786	\\
	ALF2i90	&	0.0156	&	101	&	2.729	&	2829	\\\hline
	H4i30	&	0.0172	&	102	&	3.032	&	3144	\\
	H4i60	&	0.0172	&	102	&	3.032	&	3144	\\
	H4i90	&	0.0173	&	101	&	3.046	&	3158	\\\hline
	MS1i30	&	0.0186	&	102	&	3.273	&	3394	\\
	MS1i60	&	0.0188	&	101	&	3.308	&	3430	\\
	MS1i90	&	0.0188	&	101	&	3.308	&	3430	\\		\hline
\end{tabular}
\end{center}
\end{table}
	In {\tt SACRA}, the Einstein and hydrodynamical equations are solved in an AMR algorithm described in~\cite{bib15}. Here, we briefly describe the settings for AMR grids, and the details are found in~\cite{bib15}. In this work, we prepare nine refinement levels with different grid resolutions and domain sizes. Each domain is composed of the uniform vertex-centered cubic grid with the grid number $(2N+1, 2N+1, 2N+1)$ for $(x,y,z)$. We always chose $N=60$ for the best resolved runs in this work. We also performed simulations with $N=40$ and $48$ to check the convergence of the result. As described in~\cite{bib15}, the AMR domains are classified into two categories: One is the coarser domains which cover wider regions with their origin fixed approximately at the center of the mass of the system. The other is the finer domains which cover the regions around the BH or the NS and comove with it. We set four coarser domains and five pairs of finer domains for all the simulations which we performed in this paper. The grid spacing for each domain is $h_l=L/(2^lN)$, where $2L$ is the edge length of the largest cubic domain and $l$ is the depth of the domain. 
	
	Table~\ref{tb:amr} summarizes the parameters of the grid structure for the simulations. In all the simulations, the semimajor diameter of the NS in the direction perpendicular to the axis of helical symmetry is covered with $\approx 100$ grid points. For $N=60$, the total memory required for the simulation with 14 domains is about 35 GBytes. We perform all the simulations using personal computers of 64--128 GBytes memory and 6--24 processors with OpenMP library. The typical computational time required to perform one simulation is 9 weeks for the 24 processors case.

\section{Numerical Results}\label{sec:sec4}
We present numerical results of our simulations in this section.
\subsection{Orbital Evolution}\label{ssec:orbevo}
\begin{figure*}
\begin{tabular}{ccc}
 	\includegraphics[width=60mm]{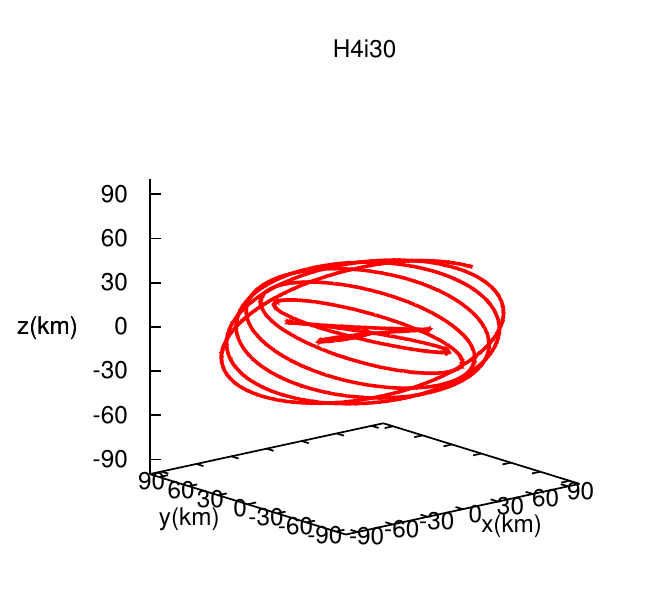} &
 	\includegraphics[width=60mm]{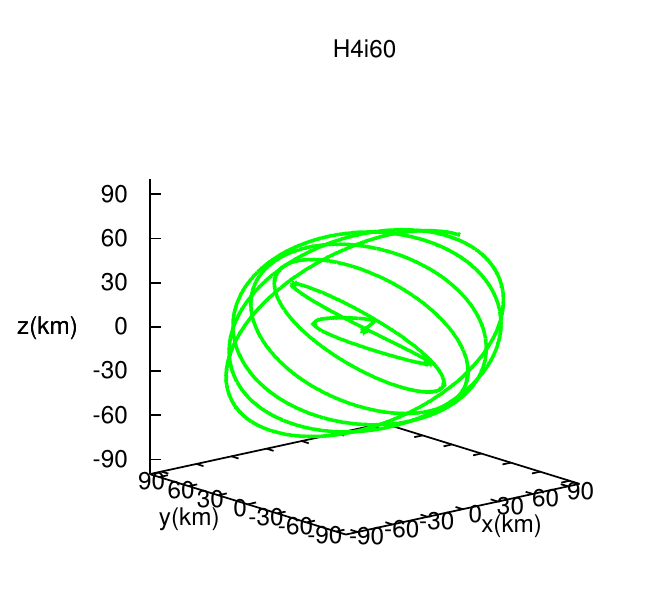} &
 	\includegraphics[width=60mm]{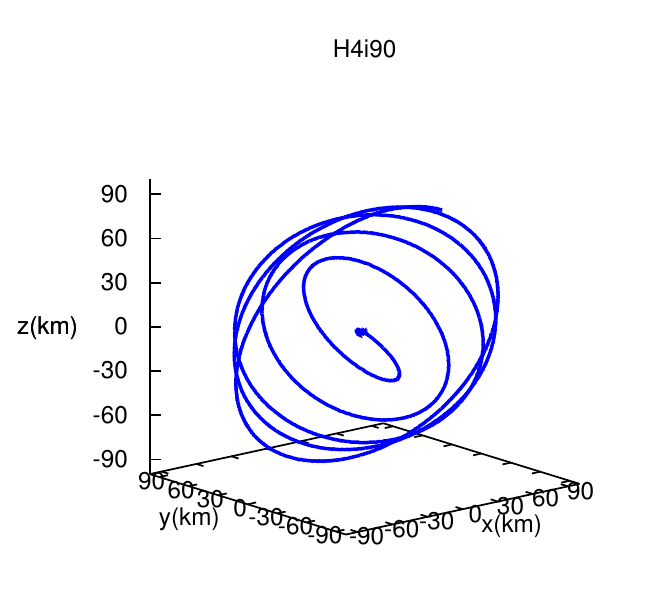} \\
\end{tabular}
\caption{Evolution of the orbital separation $x_{\rm sep}^i:=x_{\rm NS}^i-x_{\rm BH}^i$ of binaries with $\left(Q,M_{\rm NS},\chi\right)=\left(5,1.35M_\odot,0.75\right)$. The left, middle, and right panels show the results with H4i30, H4i60, and H4i90, respectively.}
\label{fig:orbH4}
\end{figure*}
Figure~\ref{fig:orbH4} plots bird's-eye views for the evolution of the coordinate separation $x^i_{\rm sep}:=x^i_{\rm NS}-x^i_{\rm BH}$ for models H4i30, H4i60, and H4i90. Here $x^i_{\rm NS}$ denotes the location of the maximum rest-mass density, and $x^i_{\rm BH}$ is the location of the puncture. This figure shows that the number of the orbits before the merger decreases with the increase of $i_{\rm tilt,0}$ as $\approx 8.5, 7.5$, and $5.5$, respectively for models H4i30, H4i60, and H4i90. This dependence stems primarily from the general relativistic spin-orbit interaction, which is well known for inducing an ``orbital hung up'' effect~\cite{bib:ohu1,bib:pn,bib:ohu2,bib09}. The additional energy of the spin-orbit interaction is written, in the leading order, as 
\begin{equation}
	E_{\rm SO}=\frac{1}{r^3}\frac{M_{\rm NS}}{M_{\rm BH}}{\bf L}\cdot{\bf S}_{\rm BH},
\end{equation} 
and hence the spin-orbit interaction essentially weakens the attractive force of gravity if ${\bf L}\cdot{\bf S}_{\rm BH}>0$. For this situation, the angular velocity of the binary, $\Omega$, decreases, and so does the luminosity of gravitational waves, which is proportional to $\Omega^{10/3}$. The reduction of the gravitational-wave luminosity makes the approaching velocity smaller, and thus the time to merger becomes longer. Since ${\bf L}\cdot{\bf S}_{\rm BH}\propto {\rm cos}\,i_{\rm tilt}$, this effect can be significant when the BH spin is aligned with the orbital angular momentum, and thus, the binary with a small value of $i_\mathrm{tilt,0}$ merges later.

 The figure also illustrates that the orbits of the binaries are precessing. This is also primarily due to the spin-orbit interaction. For models H4i30, H4i60, and H4i90, the elevation angles of the orbits measured from the $xy$-plane are always $\approx 15^\circ, 30^\circ$, and $45^\circ$, respectively. We note that gravitational waves are radiated primarily to the direction of the orbital angular momentum, and the direction of the total angular momentum ${\bf J}$ changes during the inspiral phase due to the gravitational radiation reaction. However, the angle between ${\bf J}$ and the $z$-axis is always smaller than $5^\circ$.
 
 \begin{figure}
	\begin{center}
		\includegraphics[width=80mm]{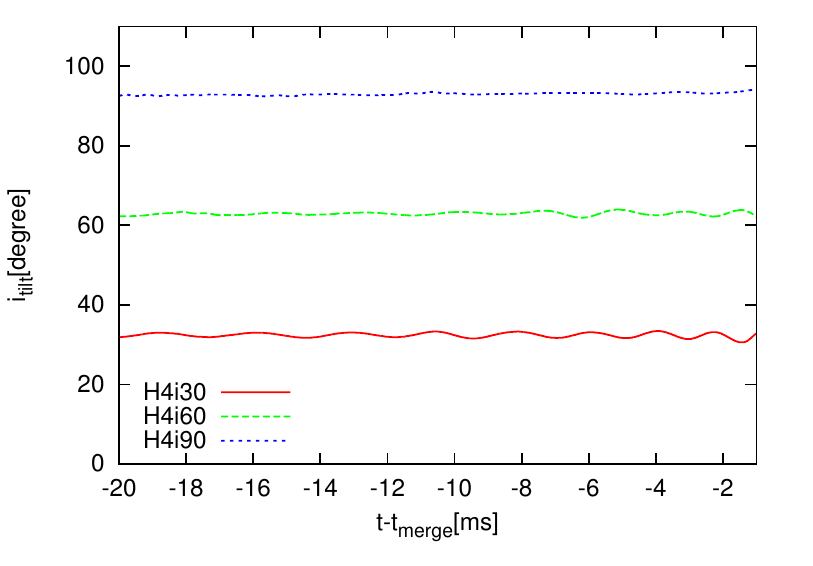}
	\end{center}
	\caption{Evolution of $i_{\rm tilt}$  for the models with H4.}
	\label{fig:itilt}
\end{figure}
 
It is known that, at least approximately, $i_{\rm tilt}$ is a constant of motion~\cite{bib:pn} for the case that the spin of the NS is absent. This feature is also seen in our simulation: see Fig.~\ref{fig:itilt}, in which we plot the time evolution of $i_{\rm tilt}$. Irrespective of $i_{\rm tilt,0}$ and EOS, we indeed find that $i_{\rm tilt}$ approximately keeps their initial values and their fluctuation is smaller than $\approx 3^\circ$ irrespective of models. Therefore $i_{\rm tilt}$ would be regarded approximately as the value which is determined when the BH-NS binary was born, even just before the merger. Also this property ensures that setting of the simulation models is well-defined; the models with different values of $i_{\rm tilt,0}$ describes entirely different physical systems. 

  \begin{figure}
	\begin{center}
		\includegraphics[width=80mm]{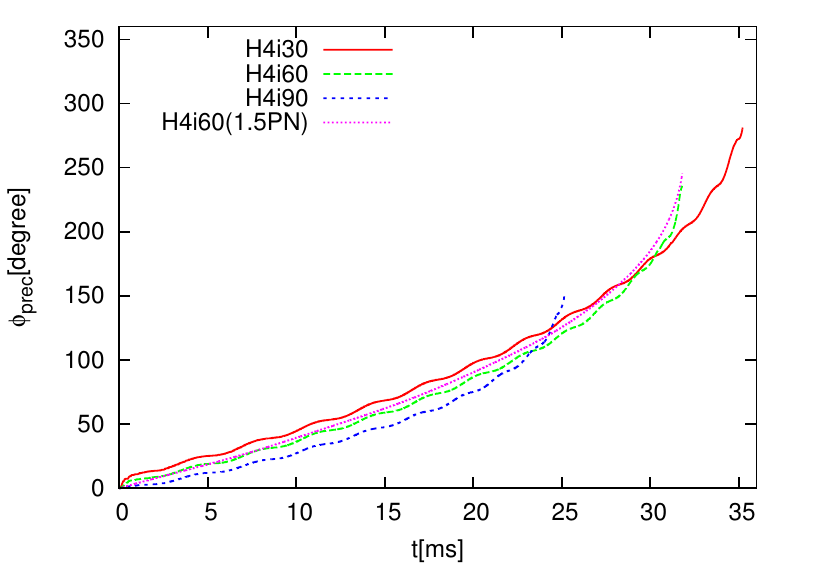}
	\end{center}
	\caption{Evolution of the precession angle, $\phi_{\rm prec}$, for the models with H4.}
	\label{fig:prec}
\end{figure}

Next, we analyze the evolution of the precession angle. We define the precession angle of the orbit as
 \begin{equation}
 	\phi_{\rm prec}:={\rm cos}^{-1}\left[\frac{\left({\bf L}\times{\bf S}_{\rm BH}\right)\cdot {\bar {\bf y}}}{\left|{\bf L}\times{\bf S}_{\rm BH}\right|}\right].
 \end{equation}
Here, ${\bar {\bf y}}$ is the coordinate basis of the $y$-axis. The evolution of $\phi_{\rm prec}$ is shown in Fig.~\ref{fig:prec} up to the time of merger. We also plot the value obtained by integrating the leading post-Newtonian formula of the precession angular velocity~\cite{bib:pn},
 \begin{equation}
 	\omega_{\rm prec}^{\rm PN}=\frac{\left|J\right|}{2 r^3}\left(3Q^{-1}+4\right),
	\label{eq:ome_prec}
 \end{equation}
 using instantaneous values of $J$ and $r$ for the simulation of the model H4i60.

 For models H4i30, H4i60 and H4i90, the final values of $\phi_{\rm prec}$ are $\approx 290^\circ , 235^\circ$, and $145^\circ$, respectively, while the number of the orbits are $\sim8.5$--$5.5$. We find that the precession angular velocity $\omega_{\rm prec}$ computed by the time derivative of $\phi_{\rm prec}$ is always smaller than orbital angular velocity by an order of magnitude. The evolution of $\phi_{\rm prec}$ agrees quantitatively with the one calculated with the leading-order post-Newtonian formula, despite the difference in the gauge condition.  Figure~\ref{fig:prec} shows that the final value of $\phi_{\rm prec}$ is smaller for the case that $i_{\rm tilt,0}$ is larger. It is simply because a longer inspiral phase is realized for binaries of a smaller value of $i_{\rm tilt,0}$.

\subsection{Tidal disruption}

  \begin{figure*}
	\begin{center}
		\includegraphics[width=88mm]{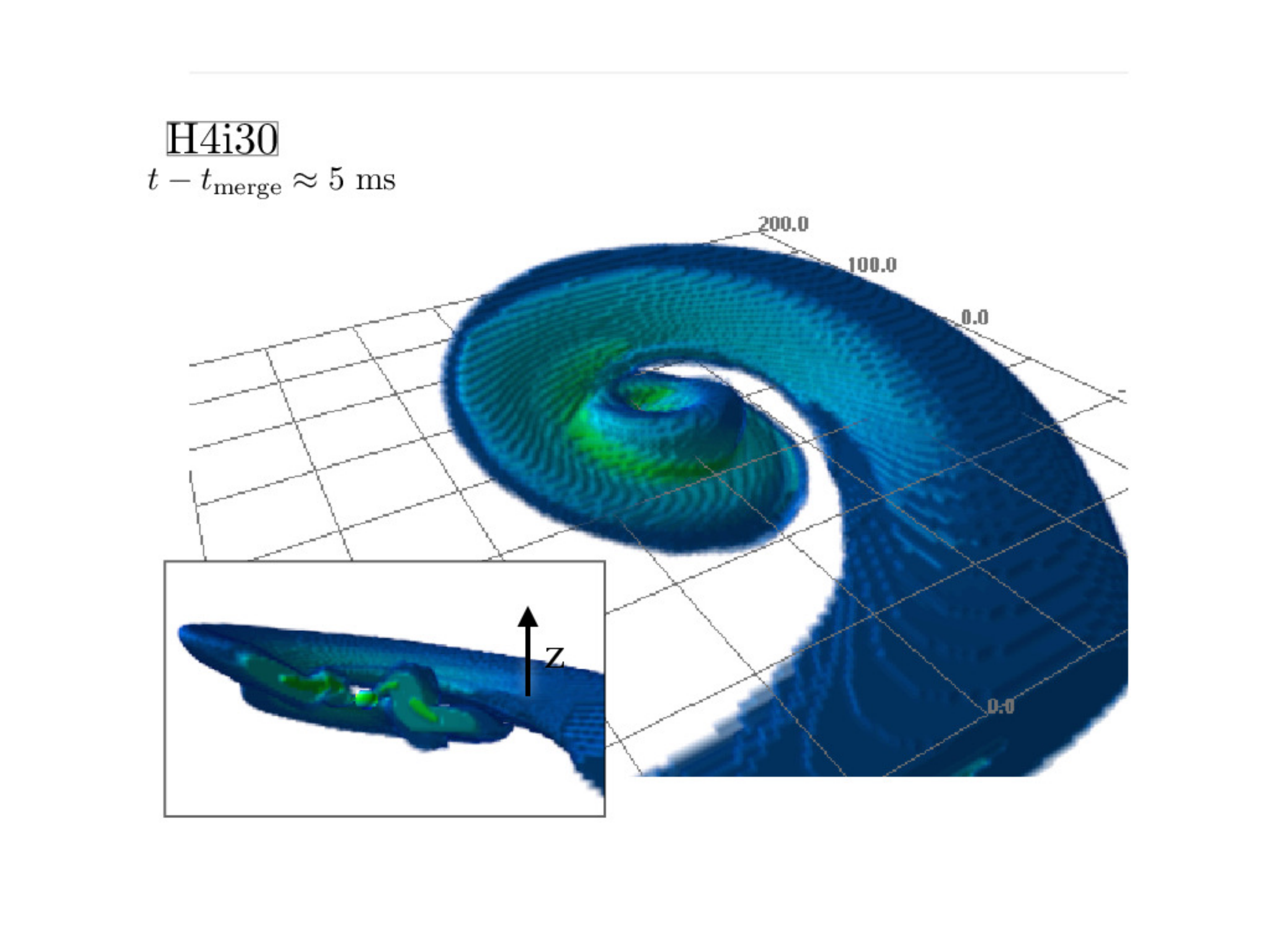}
		\includegraphics[width=88mm]{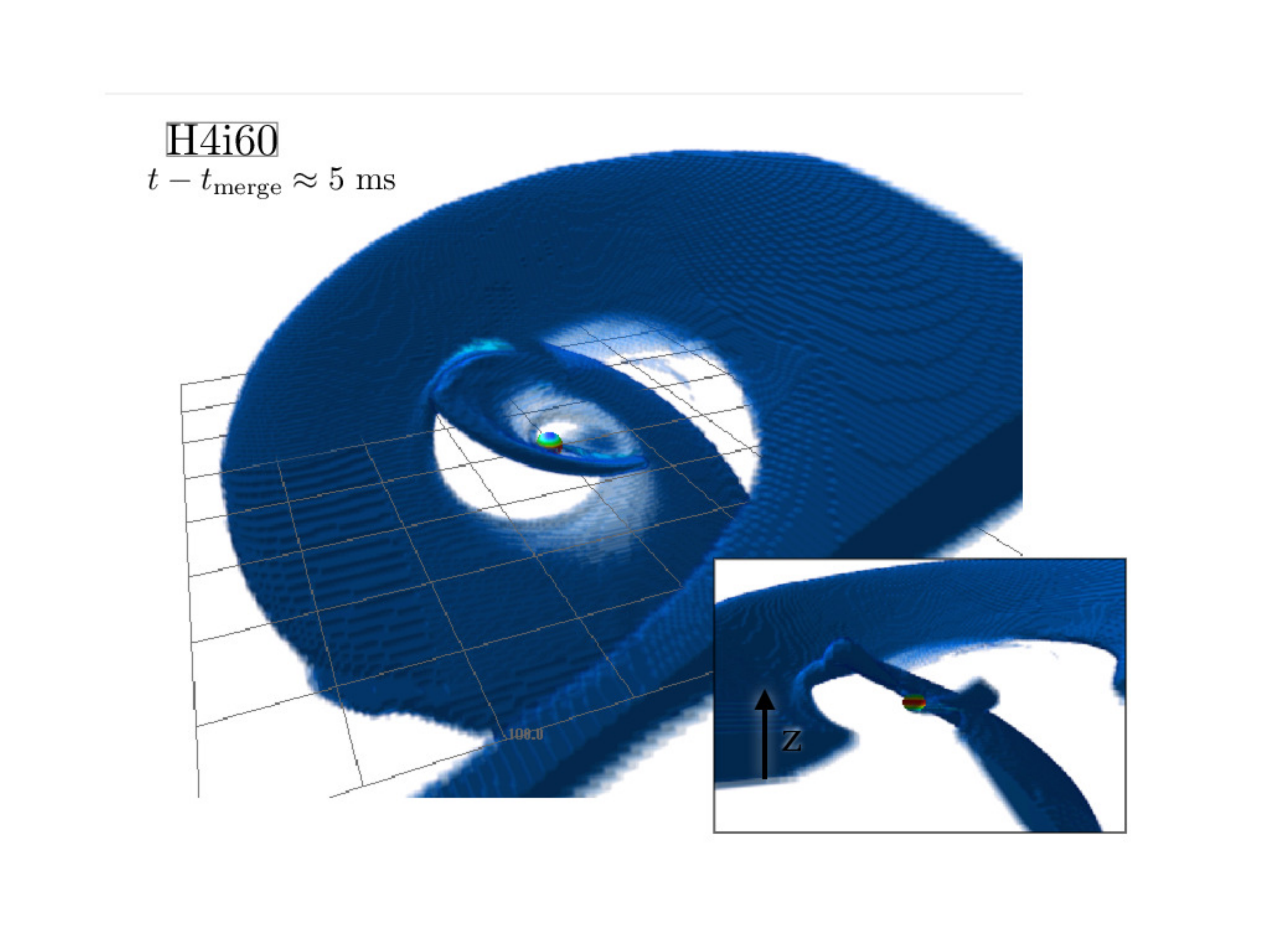}
	\end{center}
	\caption{Snapshots of the volume rendered density map as well as the location of AHs at $\approx5~{\rm ms}$ after the onset of merger for the models H4i30 (left panel) and H4i60 (right panel). The inner regions of the simulation with side lengths of $\approx 300~{\rm km}$ are shown.}
	\label{fig:3dfig}
\end{figure*}

	Figure~\ref{fig:3dfig} shows the rest-mass density together with the location of apparent horizons for models H4i30 and H4i60. The images are generated using a volume rendering method truncating the density below $10^{10}~{\rm g/cm^3}$, and the color on the AH surfaces describes the value of 2-dimensional Ricci scalar on it (cf. Sec.~\ref{ssec:diag}).
	
	For the model H4i30, the NS is tidally disrupted forming a one-armed tidal tail around the BH. An efficient angular momentum transport process works, and a fraction of the NS material becomes gravitationally unbound during this phase. At a few milliseconds after the onset of the tidal disruption, a large fraction of the NS material is swallowed by the BH, and $\approx20\%$ of the NS material remains outside the BH. The inner part of the tidal tail is subsequently wound around the BH, and a disk with its radius $\approx150~{\rm km}$ is formed. Also, some material, which was not able to get enough kinetic energy to escape from the system, falls back to the disk continuously.
	
	 Initially, the tidal tail around the BH is tilted with the angle less than $\approx15\rm ^\circ$, which reflects the elevation angle of the orbit. Also, the tidal tail is slightly warping due to the spin-orbit interaction. However, the disk formed finally is nearly aligned with the BH spin, and the morphology of the disk for this model resembles the disk formed with aligned-spin BH-NS mergers. The remnant BH spin axis agrees approximately with the direction of the initial total angular momentum, i.e., the $z$-axis.

	For the model H4i60, general features of the tidal disruption of the NS is similar to the model H4i30, while the precession of the tidal tail is more appreciable for this model. More than $\approx 90\%$ of the NS material falls into the BH in a few milliseconds after the tidal disruption for this model. This is a result of the fact that the tidal disruption occurs in the vicinity of the ISCO of the BH. The elevation angle of the tidal tail measured from the $xy$-plane is different for each part, and it is pointed out in~\cite{bib:bhns_t2} that this may prevent the collision of the tidal tail. However, the tidal tail still collides with itself to form a disk or torus for this model. This might be due to the difference of the initial parameters of the binary, that a binary with larger mass ratio and dimensionless spin parameter are employed in~\cite{bib:bhns_t2}, for which larger elevation angle is expected to be achieved. Moreover, we performed the simulation for a longer time after the merger than in~\cite{bib:bhns_t2}, and this also makes more chances for the tidal tail to collide with itself. The disk appears to be misaligned with the BH spin by $\approx 20{\rm ^\circ}$ at least at $\approx 10~{\rm ms}$ after the tidal disruption, while the BH spin is aligned approximately with the $z$-axis.
	
	For the model H4i90, the NS is tidally disrupted very weakly by the BH and a tiny tidal tail is formed. Since the tidal disruption occurs at a close orbit to the ISCO or perhaps inside the ISCO of the  BH, most of the  NS material is swallowed by the BH. Thus, only a tiny accretion disk is formed and the total amount of the ejecta is not appreciable for this model.

\begin{figure*}
\begin{center}
 \begin{tabular}{ll}
 	 \includegraphics[width=80mm]{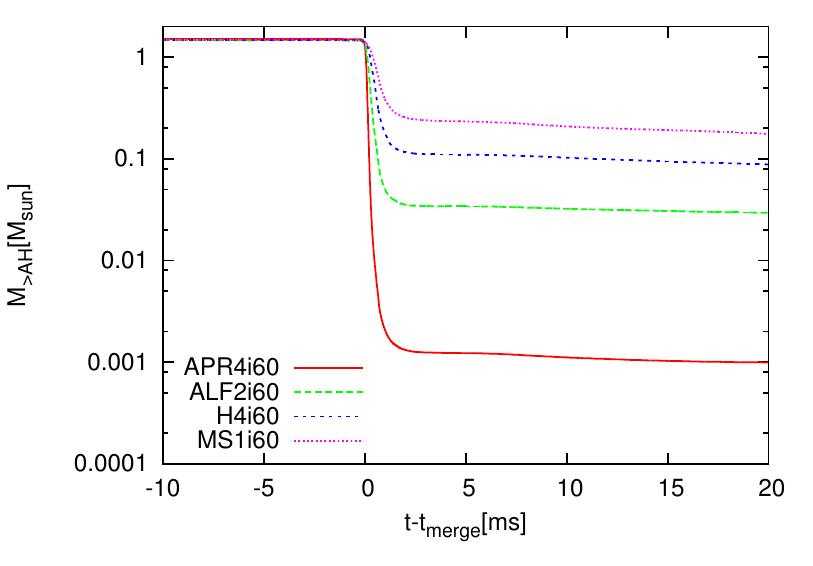}&
	 \includegraphics[width=80mm]{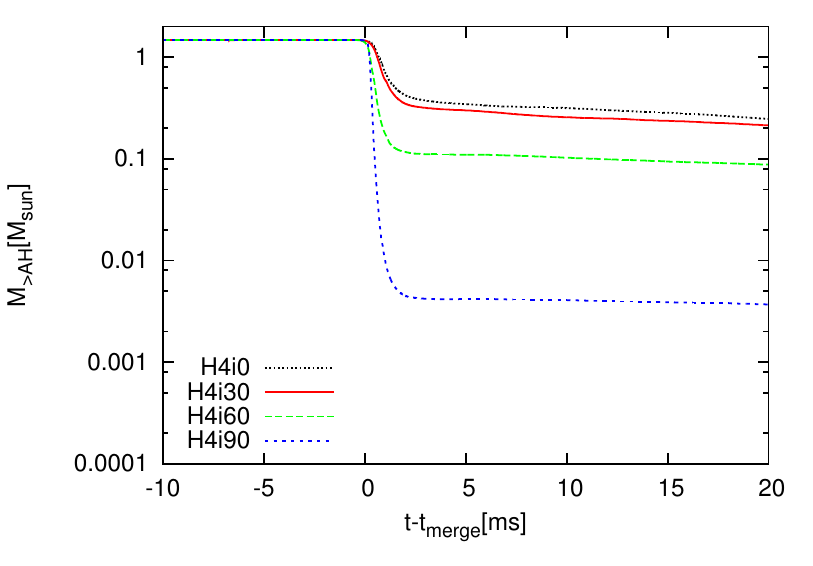}
   \end{tabular}
\end{center}
\caption{Evolution of the rest mass outside the apparent horizon $M_{> {\rm AH}}$ for the models with $i_{\rm tilt,0}\approx60^\circ$ (left figure) and with H4 (right figure). The result for the aligned-spin case with $Q=5$ are picked up from~\cite{bib:bhns13}.}
	\label{fig:mahEOS}
\end{figure*}

 Figure~\ref{fig:mahEOS} plots the time evolution of the total rest mass outside the BH defined by
 \begin{equation}
	M_{> {\rm AH}}:=\int_{r>r_{\rm AH}}\rho_*d^3x,
\end{equation}
   for a variety of EOSs and $i_{\rm tilt,0}$. From the comparison among the models with the same value of $i_{\rm tilt,0}$, we find that the values of $M_{> {\rm AH}}$ increases in the order of APR4, ALF2, H4, and MS1, i.e., in the order of the compactness. This result shows that the tidal disruption occurs at a more distant orbit for the case that the compactness of the NS is small. This dependence of the tidal disruption on the compactness is the same as the dependence which was found in the study on BH-NS mergers with aligned spins~\cite{bib:bhns9,bib:bhns10}.
 
 For a fixed EOS, the values of $M_{> {\rm AH}}$ after the merger decrease as the values of $i_{\rm tilt,0}$ increase, and thus the increase of $i_{\rm tilt,0}$ prevents the tidal disruption of the NS. This result agrees qualitatively with the result of previous study on the misaligned-spin BH-NS binary merger~\cite{bib:bhns_t1,bib:bhns_t2}, and is explained primarily by the reduction of the spin-orbit coupling. According to the study on the aligned-spin BH-NS merger~\cite{bib:bhns10}, a larger mass remains outside the BH after the merger for the case that the BH spin is larger and parallel with the orbital angular momentum. This is because the spin-orbit interaction works as a repulsive force and the ISCO radius of the BH becomes small for this situation. Since the spin-orbit interaction energy is proportional to ${\bf L}\cdot{\bf S}_{\rm BH}=L S_{\rm BH} {\rm cos}\,i_{\rm tilt,0}$ at the leading PN order, the spin-orbit interaction is weakened as $i_{\rm tilt,0}$ increases. Thus, the increase of $i_{\rm tilt,0}$ enlarges the ISCO radius of the BH effectively and reduces the remnant mass after the merger.

 We summarize the value of $M_{> {\rm AH}}$ at $\approx 10~{\rm ms}$ after the onset of merger in Table~\ref{tb:rm}. We compare these results with the fitting formula for the aligned-spin case obtained in~\cite{bib:rmfit}. Since the spin-orbit interaction is proportional primarily to ${\bf L}\cdot{\bf S}_{\rm BH}$, we compare our numerical results with the values derived by the fitting formula using the effective spin parameter defined by  \footnote{In~\cite{bib:disk_t, bib:grb_tilt, bib:bhns_t2}, the effective spin parameter is defined by a different form.} 
  \begin{equation}
 	\chi_{\rm eff}=\chi {\rm cos}\,i_{\rm tilt,0}.\label{eq:chieff}
 \end {equation}
 The deviations of the value calculated by fitting formula $\Delta_{\rm fit}=|M_{\rm fit}-M_{> {\rm AH}}|/M_{\rm fit}$ are within $50\%$ for $M_{\rm >AH} \agt0.1 M_\odot$ and within $30\%$ for $M_{\rm >AH} \agt0.2 M_\odot$. As far as the value of $M_{>{\rm AH}}$ is larger than $0.1 M_\odot$, $M_{\rm fit}$ gives a reasonable estimate for $M_{>{\rm AH}}$.

\subsection{Disk formation and mass ejection}
\subsubsection{Mass of ejecta and disk}
\begin{table*}
\caption{The list of $M_{> {\rm AH}}$, $M_{\rm disk}$, $M_{\rm eje}$, $v_{\rm ave}$, and $v_{\rm eje}$. The subscripts $10~{\rm ms}$ and $20~{\rm ms}$ denote the values evaluated at $\approx10~{\rm ms}$ and $\approx20~{\rm ms}$ after the onset of merger, respectively. The results for the aligned-spin case are obtained from~\cite{bib:bhns13}. -- implies that we were not able to take the data for them. For the model APR4i0, the simulation was stopped before $t-t_{\rm merge}\approx 20~{\rm ms}$. For the model APR4i90, the mass of the disk and ejecta are so small that accurate values cannot be derived for them. For the i90 models, the data for $P_{{\rm eje}, i}$ was not output.}
\begin{center}
 \begin{tabular}{l|ccccccc} \hline
 Model 	& $M_{>{\rm AH},10{\rm ms}}[M_\odot]$& $M_{{\rm disk},10{\rm ms}}[M_\odot]$& $M_{{\rm disk},20{\rm ms}}[M_\odot]$& $M_{{\rm eje},10{\rm ms}}[M_\odot]$ & $v_{{\rm ave},10{\rm ms}}[c]$ & $v_{{\rm eje},10{\rm ms}}[c]$\\ \hline\hline
 APR4i0   	& 0.068	& 0.059 			& --				& $8\times 10^{-3}$	&0.26	&0.099\\
 APR4i30 	& 0.022	& 0.017			& 0.014 			& $5\times 10^{-3}$ 	& $0.30$	&0.057\\
 APR4i60 	& $4\times 10^{-3}$	& $2\times 10^{-3}$	& $2\times 10^{-4}$	& $1\times 10^{-4}$ 	& $0.27$	&0.078\\
 APR4i90 	& --	& -- 				& --				& $< 10^{-4}$	    & $0.24$	&--\\\hline
 ALF2i0    	& 0.24	& 0.20 			& 0.16 			& 0.046  	&0.21	&0.15\\
 ALF2i30  	& 0.16	& 0.13 			& 0.10 			& 0.033 	& $0.27$	&0.17\\
 ALF2i60  	& 0.026	& 0.016 			& 0.013 			& 0.010 	& $0.28$	&0.048\\
 ALF2i90  	& $2\times 10^{-4}$			& $1\times 10^{-4}$	& $< 10^{-4}$	& $< 10^{-4}$ 		& $0.26$	&--\\\hline
 H4i0        	& 0.32	& 0.27 			& 0.21 			& 0.050  	&0.22	&0.18\\
 H4i30      	& 0.25	& 0.21 			& 0.19 			& 0.042 	& $0.27$	&0.21\\
 H4i60      	& 0.084	& 0.072 			& 0.061 			& 0.012 	& $0.25$	&0.14\\
 H4i90      	& $3\times 10^{-3}$	& $2\times 10^{-3}$	& $2\times 10^{-3}$	& $1\times 10^{-3}$ 	& $0.28$	&-\\\hline
 MS1i0     	& 0.36	& 0.28			& 0.22			& 0.079  	&0.24	&0.19\\
 MS1i30   	& 0.30	& 0.23 			& 0.19 			& 0.070 	& $0.28$	&0.23\\
 MS1i60   	& 0.18	& 0.14			& 0.11 			& 0.041 	& $0.27$	&0.21\\
 MS1i90   	& 0.022	& 0.012			& 0.011 			& 0.010 	& $0.27$	&--\\\hline
 \end{tabular}
\end{center}
\label{tb:rm}
\end{table*}

  \begin{figure*}
	\begin{center}
	\begin{tabular}{ll}
		\includegraphics[height=80mm]{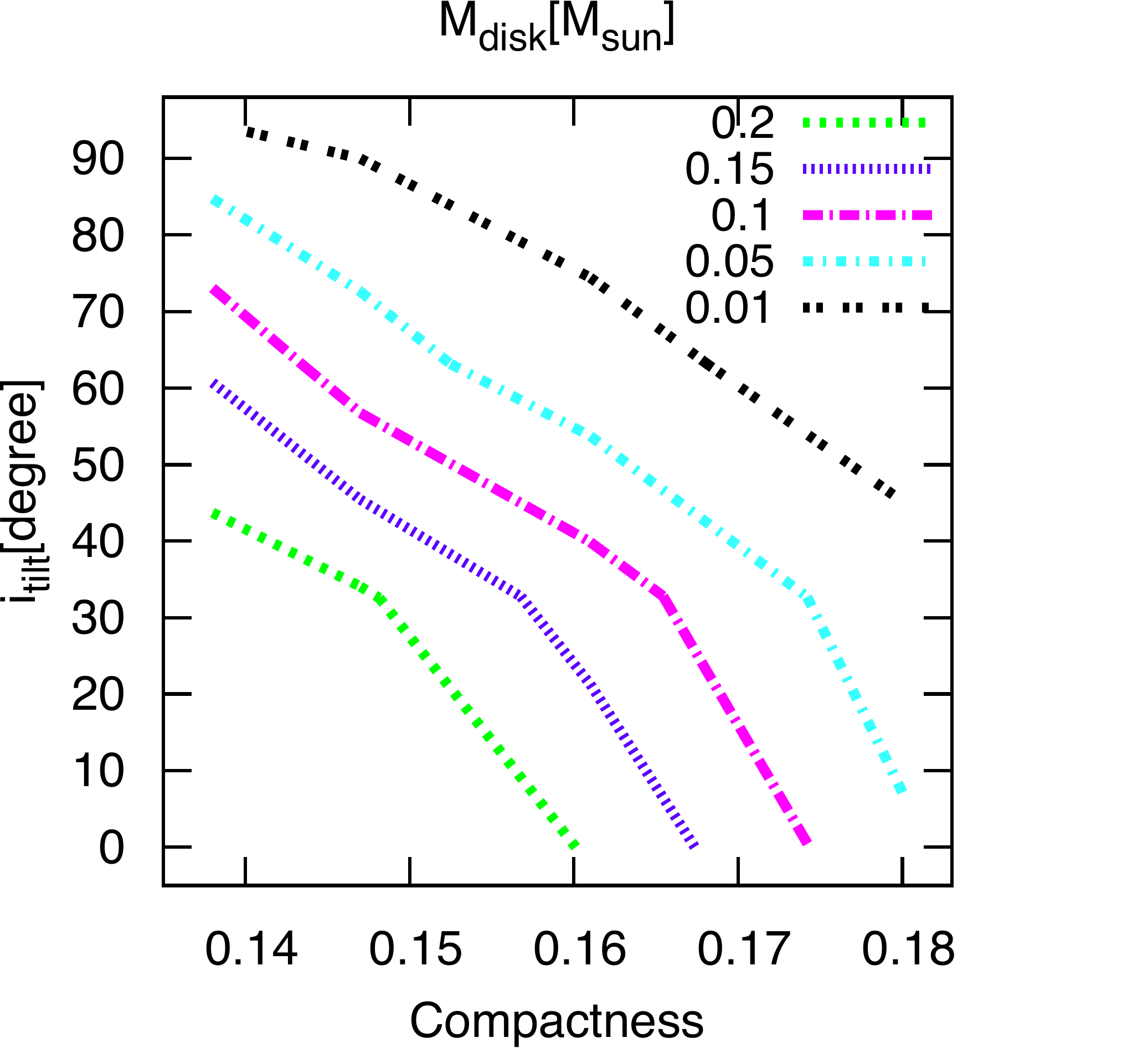} &
		\includegraphics[height=80mm]{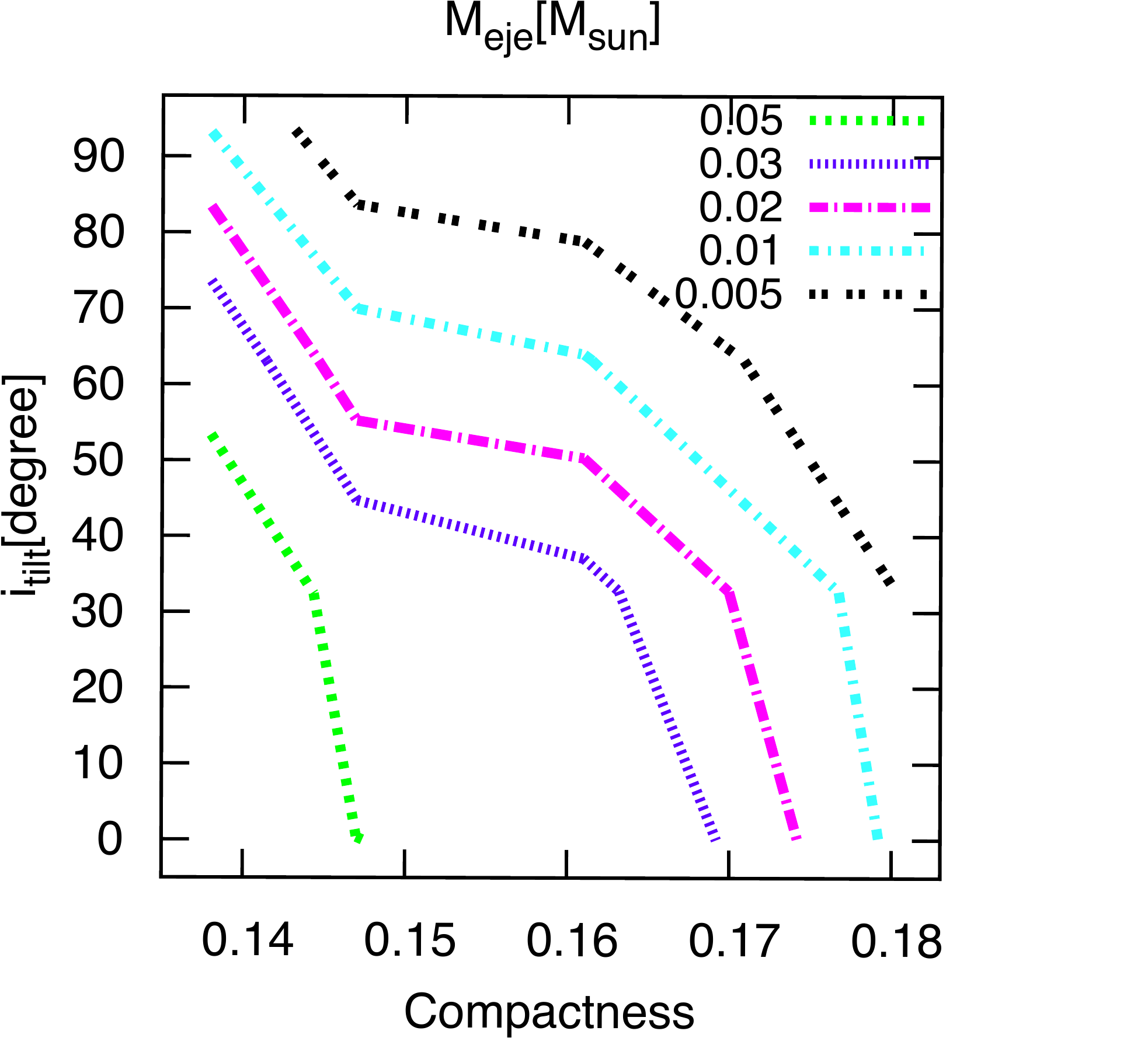}
	\end{tabular}
	\end{center}
	\caption{The contour for $M_{\rm disk}$ (left panel) and $M_{\rm eje}$ (right panel) evaluated at $\approx 10~{\rm ms}$ after the onset of merger in the plane of the NS compactness ${\cal C}$ and initial value of $i_{\rm tilt}$.}
	\label{fig:rmcont10ms}
\end{figure*}

  \begin{figure*}
	\begin{center}
	\begin{tabular}{ll}
		\includegraphics[height=60mm]{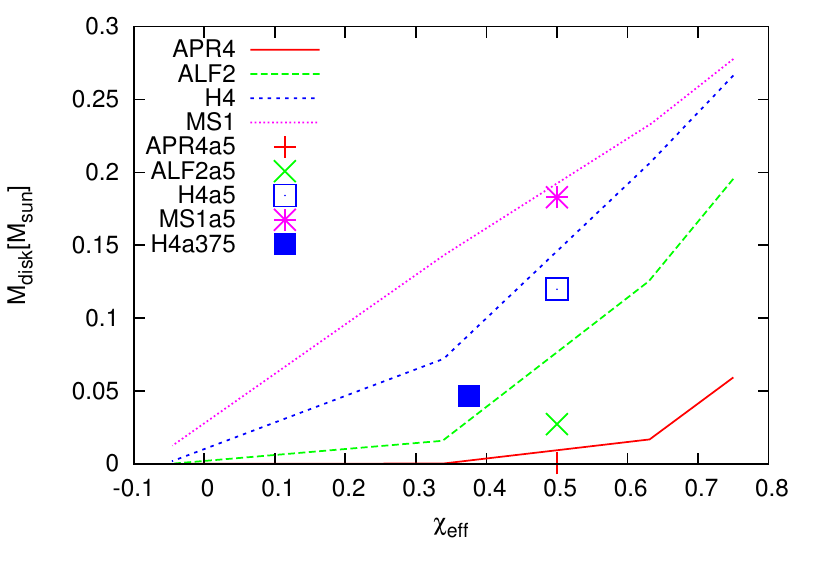} &
		\includegraphics[height=60mm]{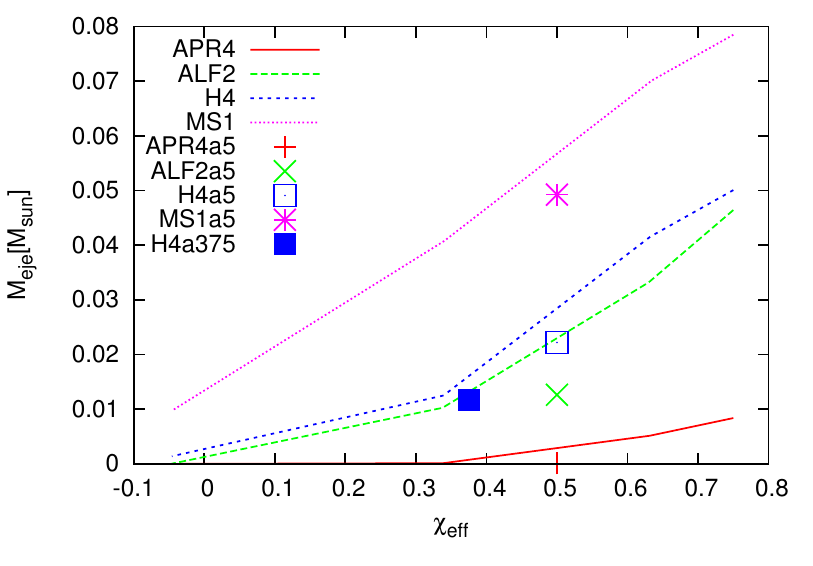}
	\end{tabular}
	\end{center}
	\caption{$M_{\rm disk}$ (left panel) and $M_{\rm eje}$ (right panel) evaluated at $\approx 10~{\rm ms}$ after the onset of merger as a function of an effective spin parameter $\chi_{\rm eff}=\chi{\rm cos}\,i_{\rm tilt,0}$. The plotted lines are the linear interpolation for the results of misaligned BH-NS mergers obtained in this paper, and the plotted points are the results for aligned-spin BH-NS mergers with $Q=5$ obtained in~\cite{bib:bhns13}. ``a5'' and ``a375'' denote the aligned models with $\chi=0.5$ and $\chi=0.375$, respectively. }
	\label{fig:rmcomp}
\end{figure*}

  Next we evaluate the mass of disk and ejecta, which are the key quantities for the electromagnetic emission from the remnant of BH-NS mergers. We calculate the ejecta mass by
\begin{equation}
	M_{\rm eje}:=\int_{r>r_{\rm AH},u_t<-1}\rho_*d^3x.
\end{equation}
 Here, we assume that the contribution of the internal energy of the ejecta is negligible for estimating the unbound material. Then, we define the rest mass of remnant disks by
\begin{equation}
	M_{\rm disk}:=M_{> {\rm AH}}-M_{\rm eje}.
\end{equation}
	We note that $M_{\rm disk}$ is described as $M_{\rm bd}$ in~\cite{bib:bhns13}.

  We list the values of $M_{\rm disk}$ and $M_{\rm eje}$ at $\approx10~{\rm ms}$ and $\approx20~{\rm ms}$ after the onset of merger in Table~\ref{tb:rm}. This shows that $M_{\rm disk}$ and $M_{\rm eje}$ monotonically decrease with the increase of $i_{\rm tilt,0}$. This reflects the fact that the effective ISCO radius of the BH increases with the increase of $i_{\rm tilt,0}$. We find that $M_{\rm disk}$ and $M_{\rm eje}$ with $i_{\rm tilt,0}\agt 30^\circ$ are appreciably smaller than those for $i_{\rm tilt,0}=0^\circ$. In particular, little amount of disk and ejecta are produced for all the EOS with $i_{\rm tilt,0}\approx 90^\circ$. For the moderate misalignment angle, $i_{\rm tilt,0}\approx 60^\circ$, the values of $M_{\rm disk}$ and $M_{\rm eje}$ are sensitive to the EOS: For the MS1 EOS, the disk with $M_{\rm disk}>0.1M_\odot$ and ejecta  with $M_{\rm eje}>10^{-2}M_\odot$ are produced. On the other hand, $M_{\rm disk}<10^{-2}M_\odot$ and $M_{\rm eje}<10^{-3}M_\odot$ for the APR4 EOS.

  To clarify the dependence of $M_{\rm disk}$ and $M_{\rm eje}$ on $i_{\rm tilt,0}$ and EOS, we plot contours of $M_{\rm disk}$ and $M_{\rm eje}$ at  $\approx10~{\rm ms}$ after the onset of merger as functions of  $i_{\rm tilt,0}$ and the compactness parameter ${\cal C}$ of the NS in Fig.~\ref{fig:rmcont10ms}. The dependence of $M_{\rm disk}$ and $M_{\rm eje}$ is clear: Both of them decrease monotonically with the increases of ${\cal C}$ and $i_{\rm tilt,0}$. For a moderate value of compactness ${\cal C}=0.160$, $i_{\rm tilt,0}$ should be smaller than $50^\circ$ for $M_{\rm disk}$ to be larger than $0.1M_\odot$. On the other hand, $M_{\rm disk}\agt 0.1M_\odot$ even if $i_{\rm tilt,0}\approx70^\circ$ for a stiff EOS that realizes ${\cal C}=0.140$. For a soft EOS with which ${\cal C}\agt0.175$,  $M_{\rm disk}$ is smaller than $0.1M_\odot$ for any value of $i_{\rm tilt,0}$. $M_{\rm eje}>0.01M_\odot$ is possible for $i_{\rm tilt,0}<85^\circ$ with ${\cal C}=0.140$, for $i_{\rm tilt,0}<65^\circ$ with ${\cal C}=0.160$, and for $i_{\rm tilt,0}<30^\circ$ with ${\cal C}=0.175$.

 Figure~\ref{fig:rmcomp} compares $M_{\rm disk}$ and $M_{\rm eje}$ obtained by numerical simulations for aligned-spin BH-NS mergers ~\cite{bib:bhns13} with those for the misaligned-spin cases. Each line describes the results of $M_{\rm disk}$ and $M_{\rm eje}$ for the misaligned-spin BH-NS mergers interpolated linearly for $\chi_{\rm eff}$. Each point in Fig.~\ref{fig:rmcomp} shows the results of the aligned-spin BH-NS mergers with the same mass ratio $(Q=5)$ and the same EOS as we employed in this paper, but with smaller BH spin $\chi=0.5$. We also plot a new result for the model with $Q=5$, H4 EOS, and $\chi=0.375$. For both $M_{\rm disk}$ and $M_{\rm eje}$, the results of the aligned-spin case agree approximately with the interpolated line in the error margin due to the finite grid resolution (see Appendix~\ref{app:err}),  while slightly larger mass is realized for the misaligned-spin case. The slope between H4a375 and H4a5 also agrees with the slope of the interpolated line for H4. Exceptionally for models with ALF2, results obtained by the aligned-spin BH-NS mergers deviate from the interpolated plot lines by $\sim 100\%$ for $M_{\rm disk}$ and $\approx 30\%$ for $M_{\rm eje}$, which might be a little bit larger than the error margin due to the finite grid resolution. We note that the deviation depends on the interpolation method and a systematic error associated with employing $\chi_{\rm eff}$.
 
\subsubsection{Disk morphology and accretion}
\begin{figure*}
	\begin{center}
	\begin{description}
	\item[(a)]\begin{tabular}{lll}
		\includegraphics[width=59mm]{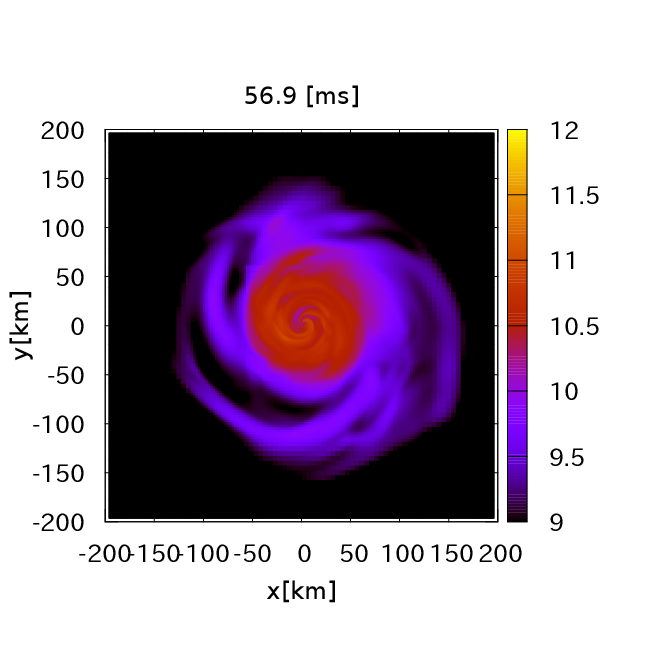}&
		\includegraphics[width=59mm]{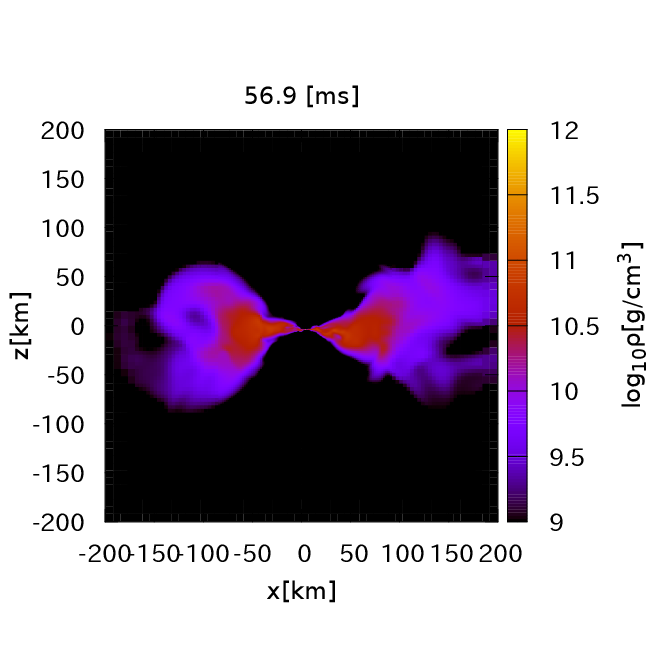}&
		\includegraphics[width=59mm]{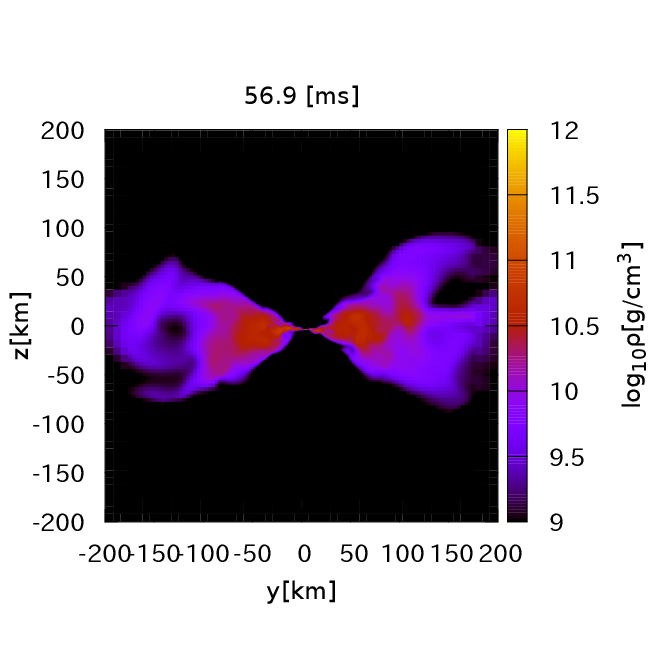}
	\end{tabular}
	\item[(b)]\begin{tabular}{lll}
		\includegraphics[width=59mm]{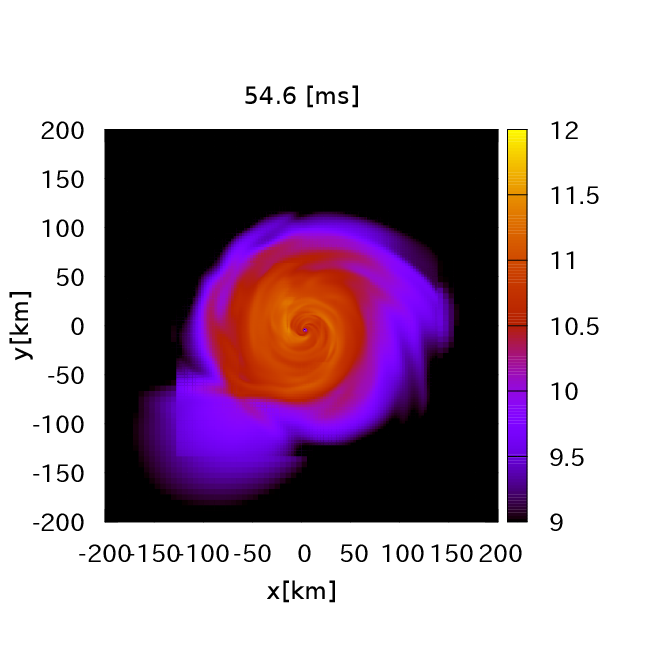}&
		\includegraphics[width=59mm]{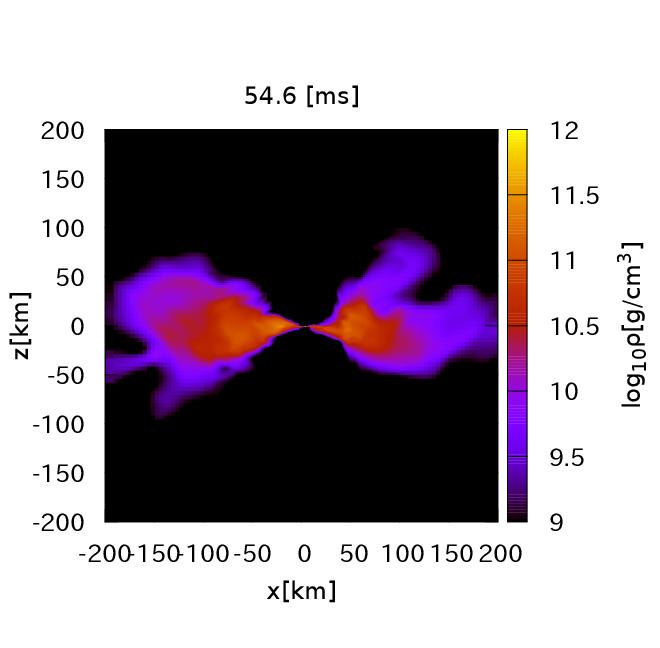}&
		\includegraphics[width=59mm]{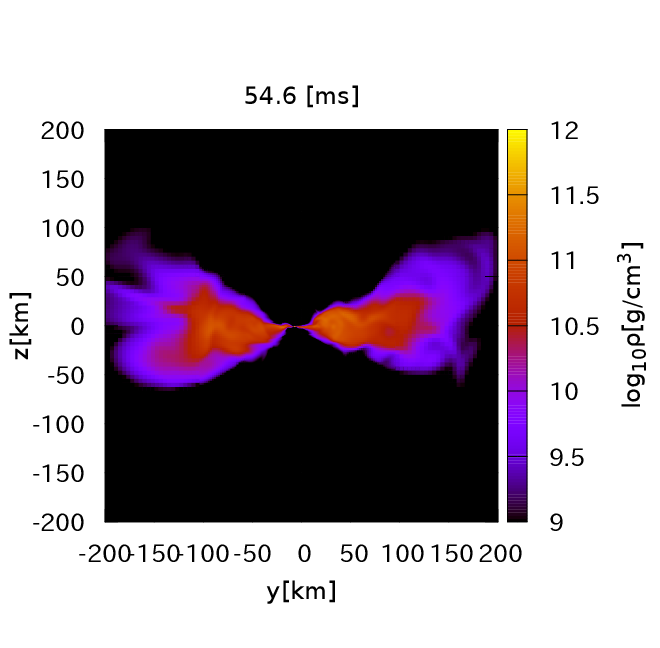}
		\end{tabular}
	\item[(c)]\begin{tabular}{lll}
		\includegraphics[width=59mm]{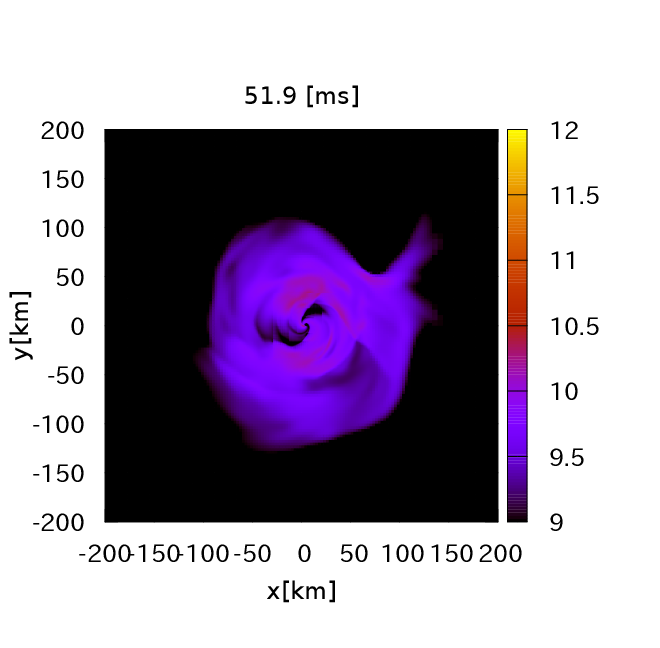}&
		\includegraphics[width=59mm]{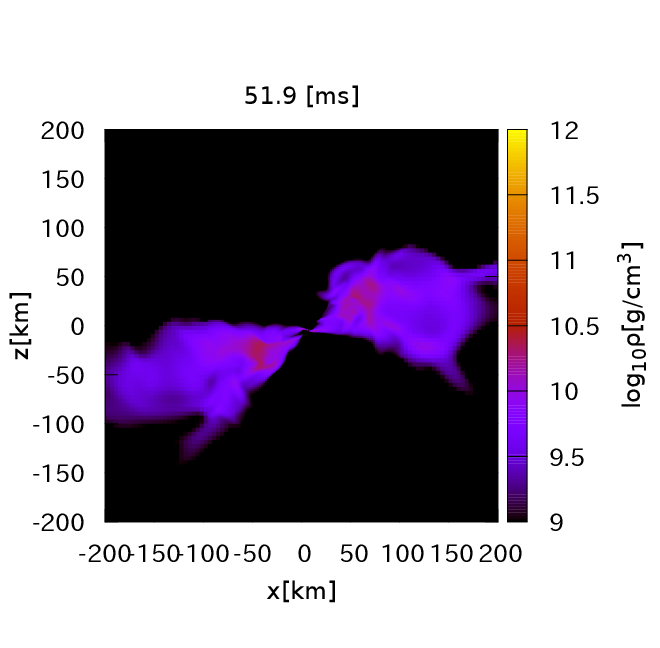}&
		\includegraphics[width=59mm]{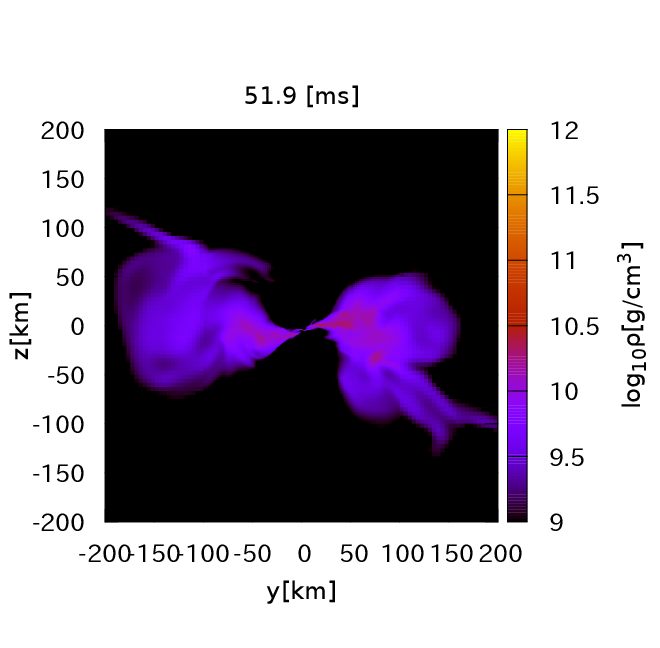}
		\end{tabular}
	\item[(d)]\begin{tabular}{lll}
		\includegraphics[width=59mm]{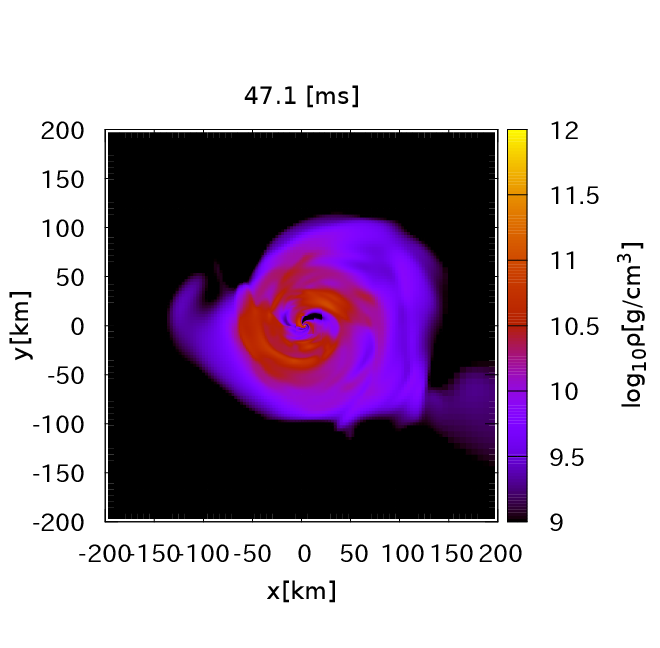}&
		\includegraphics[width=59mm]{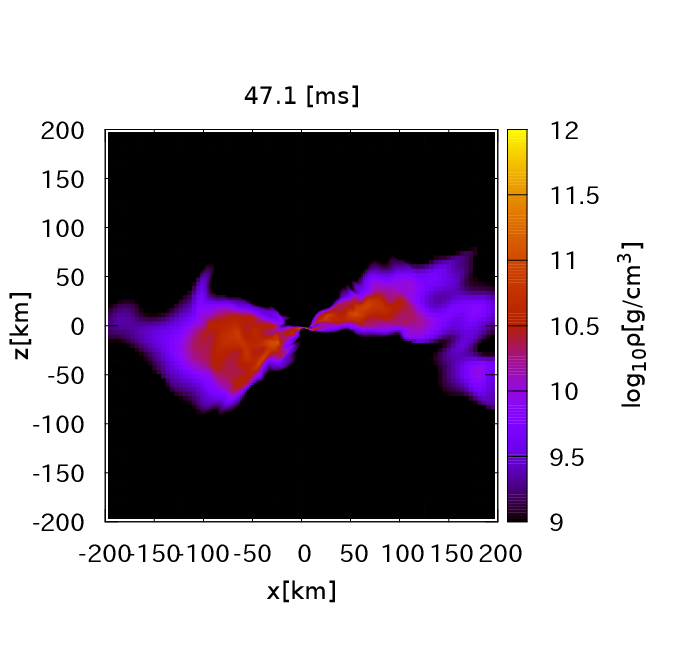}&
		\includegraphics[width=59mm]{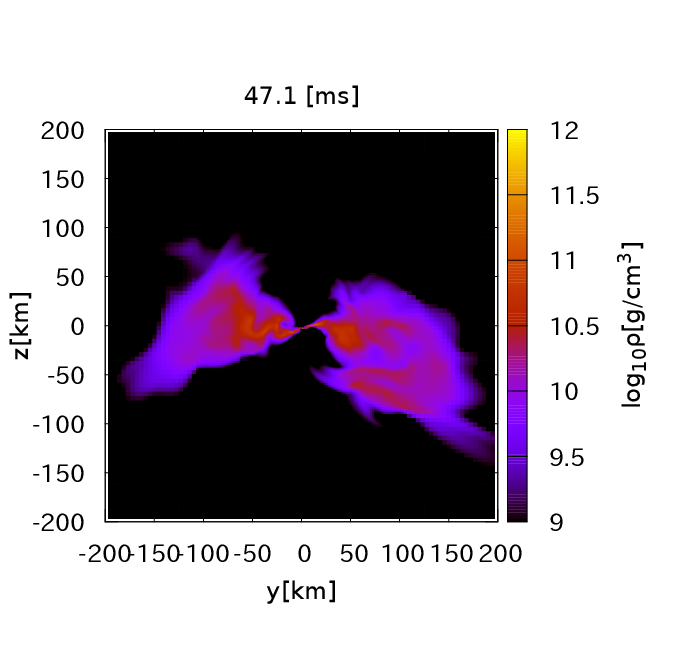}
	\end{tabular}
	\end{description}
	\end{center}
	\caption{The density profiles of the accretion disk at $\approx 20~{\rm ms}$ after the onset of merger for selected models. The left, middle, and right columns show the plots for the $xy$, $xz$, and $yz$-planes, respectively. The top, second top, third top, and bottom raws show the plots for models ALF2i30, H4i30, H4i60, and MS1i60, respectively.}
	\label{fig:disk_1}
\end{figure*}

\begin{figure}
 	\includegraphics[width=80mm]{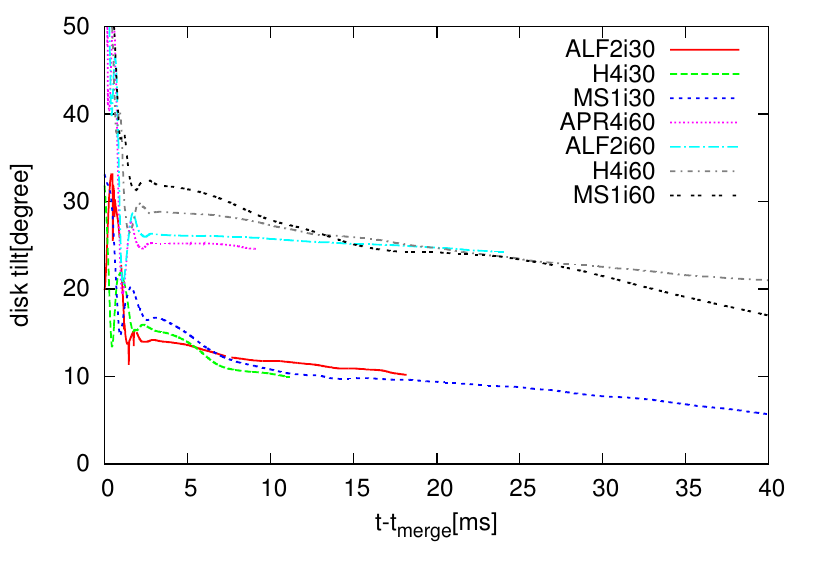} 
	\caption{Evolution of the tilt angle between $J_{\rm disk}^i$ and the BH spin for the models with $i_{\rm tilt,0}\approx30^\circ$ and $60^\circ$.}
	\label{fig:rmj_tilt}
\end{figure}

\begin{figure}
 	\includegraphics[width=70mm]{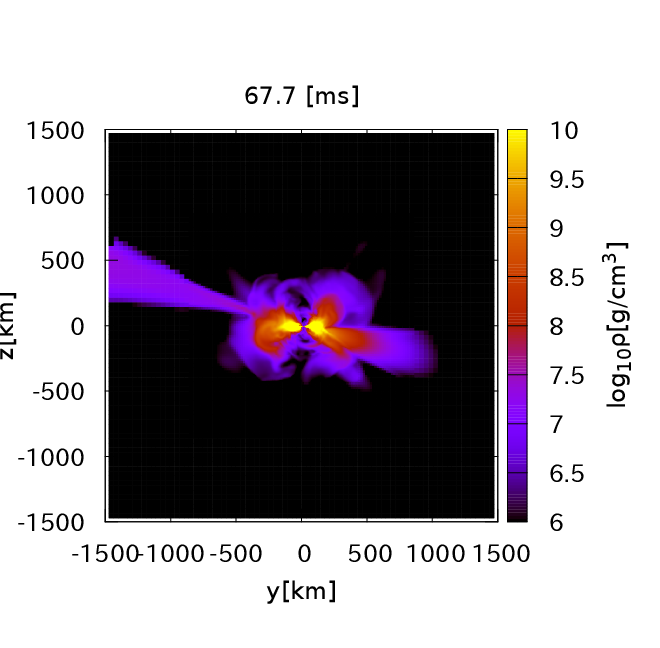} 
	\caption{The density profiles on the $yz$ cross section for the model MS1i30 at $\approx~40{\rm ms}$ after the onset of merger.}
	\label{fig:disk_3}
\end{figure}

\begin{figure*}
	\begin{center}
	\begin{tabular}{ccc}
		\includegraphics[width=60mm]{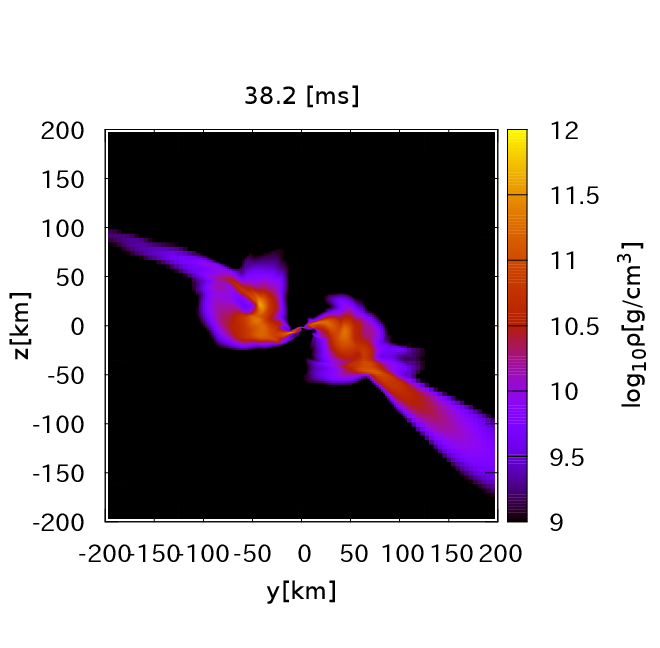}&
		\includegraphics[width=60mm]{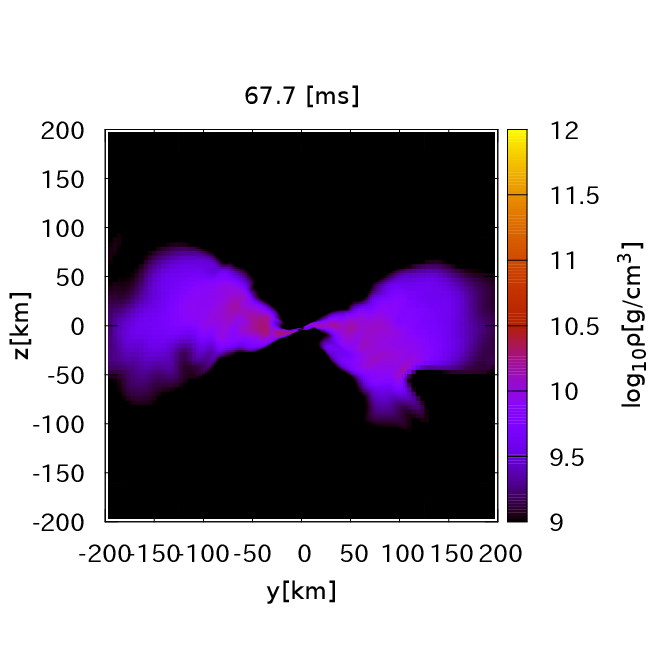}&
		\includegraphics[width=60mm]{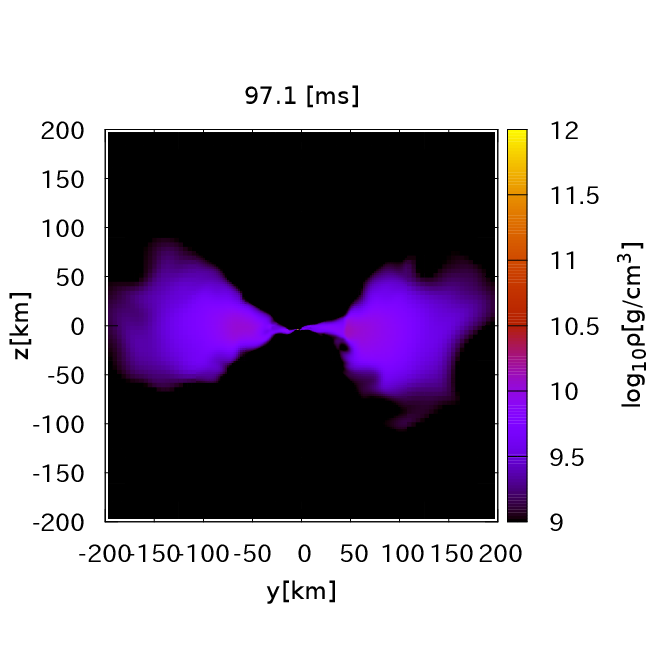}
	\end{tabular}
	\end{center}
	\caption{The density profiles on the $yz$ cross section for the model MS1i60 at $\approx 10, 40$, and $70~{\rm ms}$ after the onset of merger.}
	\label{fig:disk_2}
\end{figure*}

   \begin{figure*}
	\begin{center}
	\begin{tabular}{ll}
		\includegraphics[width=80mm]{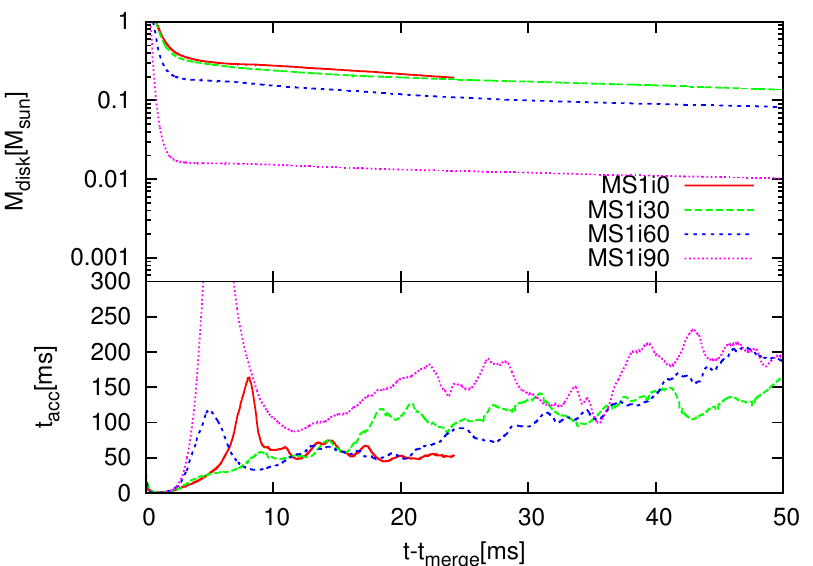} &
		\includegraphics[width=80mm]{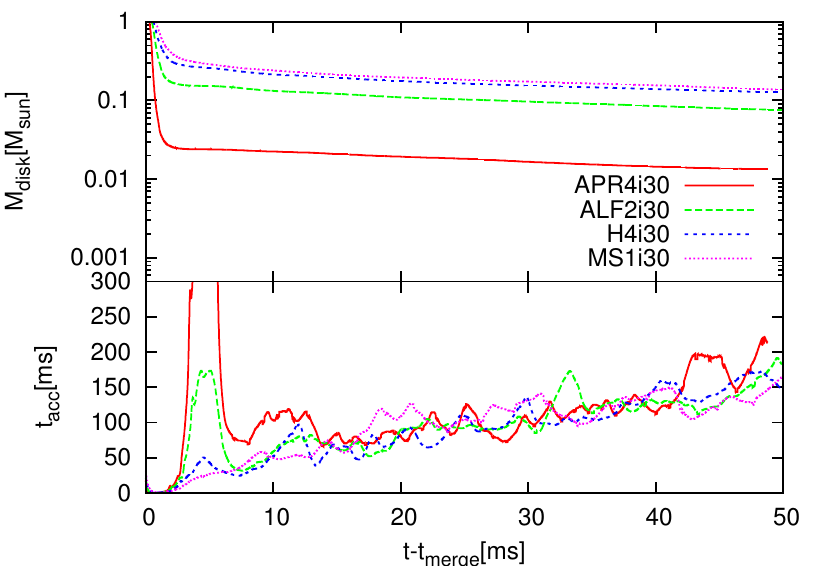}
	\end{tabular}
	\end{center}
	\caption{Evolution of $M_{\rm disk}$ and  the accretion time scale $t_{\rm acc}=M_{\rm disk}/{\dot M}_{\leq {\rm AH}}$ for the models with MS1 (left figure) and with $i_{\rm tilt,0}\approx30^\circ$ (right figure).}
	\label{fig:t_acc}
\end{figure*}
 
\begin{figure}
 	\includegraphics[width=80mm]{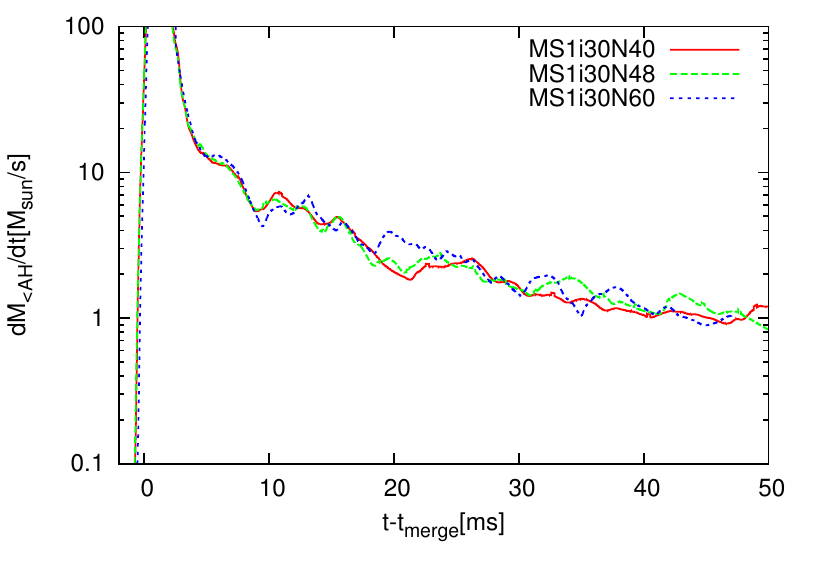} 
	\caption{Evolution of the accretion rate, ${\dot M}_{\leq {\rm AH}}$, to the AH for MS1i30 but with different grid resolutions.}
	\label{fig:mah_dot}
\end{figure}

	Figure~\ref{fig:disk_1} shows snapshots of the density profiles at $\approx 20~{\rm ms }$ after the onset of  merger for selected models (see figure caption). We find that the size of the region with $\rho>10^9~{\rm g/cm^3}$ is always $\sim 150~{\rm km}$, while that with $\rho>10^{10}~{\rm g/cm^3}$ depends on the mass of the disk: For the larger disk mass, the size of the region with $\rho>10^{10}~{\rm g/cm^3}$ becomes larger. The maximum density of the disk also becomes larger for larger disk mass: For instance, the maximum density exceeds $10^{11}~{\rm g/cm^3}$ for the model H4i30, while there is no region with $\rho >10^{11}~{\rm g/cm^3}$ for the model H4i60. The ratio of the disk height to disk radius depends only weakly on the binary parameters, and it is always $\sim 0.3$--$0.5$.

	For the models with $i_{\rm tilt,0}\approx 30^\circ$, the rotational axes of the disks are approximately aligned with the $z$-axis. However, the density distribution on the $xy$ cross section is not axisymmetric and an approximately stationary spiral-shape shock wave is seen in the disks. On the other hand, for $i_{\rm tilt,0}\approx 60^\circ$, the disks are misaligned with the $z$-axis. Also, in a similar manner to i30 models, the disks have a non-axisymmetric structure for $i_{\rm tilt,0}\approx 60^\circ$.

	  To quantify the misaligned structure of the disk, we define the total angular momentum of the disk $J_{\rm disk}^i$ as
\begin{equation}
	J_{\rm disk}^i=\int_{r>r_{\rm AH}, u_t>-1}\rho_*\epsilon^{ijk}x_j{\hat u}_k d^3x,
\end{equation}
 where ${\hat u}_i=hu_i$, and we plot the evolution of the tilt angle of $J_{\rm disk}^i$ measured from the direction of the BH spin after the merger in Fig~\ref{fig:rmj_tilt}. This shows that the tilt angle of $J_{\rm disk}^i$ at $\approx5~{\rm ms}$ after the onset of merger is $\approx 15^\circ$ and $\approx 30^\circ$ for models with $i_{\rm tilt,0}\approx 30^\circ$ and $i_{\rm tilt,0}\approx 60^\circ$, respectively. These tilt angles of $J_{\rm disk}^i$ reflect the elevation angle of the orbits just before the merger and gradually decrease as the system evolves. Figure~\ref{fig:rmj_tilt} shows that these tilt angles appear to be slightly larger than those expected from Fig.~\ref{fig:disk_1}: We should note that the tilt angle shown in Fig.~\ref{fig:rmj_tilt} does not really describe representative values for the tilt angle of the dense part of the disk but indicates the tilt angle of the remnant matter including the tidal tail with larger orbital radii (see below).

	Figure~\ref{fig:disk_3} shows a larger-scale density profile ($\approx 1500~{\rm km}$) on the $yz$-plane for the model MS1i60 at $\approx 40~{\rm ms}$ after the onset of merger. While there is a relatively dense torus in the central region ($\approx 200~{\rm km}$), the tidal tail is widely spreading with an elevation angle $\approx 30^\circ$. This tidal tail is not unbound although it has a large orbital angular momentum, and thus it contributes to the tilt of $J_{\rm disk}^i$ which we defined above.

	Figure~\ref{fig:disk_2} shows the density profiles of the disk on the $yz$ cross section for the model MS1i60 at $\approx 10, 40$, and $70~{\rm ms}$ after the onset of merger. While the dense part with $\rho>10^9~{\rm g/cm^3}$ is tilted by $\approx 30^\circ$ from the $xy$-plane at $\approx 10~{\rm ms}$ after the onset of merger, the tilt angle of the disk decreases gradually as the system evolves, and its axis is approximately aligned with the $z$-axis at $\approx 70~{\rm ms}$ after the onset of merger. The same feature can also be seen on the plot of the $xz$-plane, and thus we conclude that the disk has a tendency to align with the BH spin during the evolution. The time scale for the disk to align is $\approx 50~{\rm ms}$  and is comparable to or even bit shorter than the time scale of the disk precession  $t_{\rm prec}\sim 100~{\rm ms}~(r_{\rm disk}/150~{\rm km})^3$. In the presence of fluid viscosity, the so-called Bardeen-Petterson effect~\cite{bib:bp} is known as a mechanism for the disk to be aligned with the BH spin. Since any effect of viscosity is not taken account, the Bardeen-Petterson effect cannot play a role in our simulation. However, we suspect that Bardeen-Petterson-like effect induced by a purely hydrodynamical mechanism, such as angular momentum redistribution due to a shock wave excited in a non-axisymmetric manner of the disk, should work in the disk for the alignment.
	To summarize, a disk with tilt angle of $\approx 20^\circ$--$30^\circ$ would be formed for models with $i_{\rm tilt,0}\approx 60^\circ$. However, the dense part of the disk is subsequently aligned with the BH spin in $\approx 50~{\rm ms}$ while the tidal tail with large orbital radii keeps its elevation angle.

	Figure~\ref{fig:t_acc} plots the time evolution of $M_{\rm disk}$ for the models with the MS1 EOS and $i_{\rm tilt,0}\approx 30^\circ$. This figure shows that $M_{\rm disk}$ gradually decreases after a steep initial decrease. This shows that the infall of the material into the BH continues for a long time scale. This is induced primarily by a hydrodynamical process associated with the non-axisymmetric torque in the disk, as we can infer from Fig.~\ref{fig:disk_1}. Also, the fallback of the matter could give an impact to the disk material. To check that numerical viscosity is not a main source of the angular momentum redistribution, we plot the mass accretion rate of the disk material into the BH for three different grid resolutions in Fig.~\ref{fig:mah_dot}. Here the accretion rate is defined as the time derivative of $M_{\leq {\rm AH}}$. We find that the value of ${\dot M}_{\leq {\rm AH}}$ depends only very weakly on the grid resolution. Since the numerical viscosity should depend on the grid resolution, this result shows that the contribution of the numerical viscosity to the mass accretion is negligible, and we can safely consider that the accretion is induced by a physical process. The non-axisymmetric and dynamical feature of the disk would be responsible for this.

 Lower panels of Fig.~\ref{fig:t_acc} plot the time scale of the accretion, $t_{\rm acc}:=M_{\rm disk}/{\dot M}_{\leq {\rm AH}}$, for the models with the MS1 EOS and for $i_{\rm tilt,0}\approx30~{\rm ^\circ}$. We find that it depends only weakly on $i_{\rm tilt,0}$ and EOS and $t_{\rm acc}\approx100{\rm ms}$ for all the models at $10$--$20~{\rm ms}$ after the onset of merger. The time scale increases as the system relaxes, but still, $t_{\rm acc}$ is as short as $\approx 200~{\rm ms}$ at $50~{\rm ms}$ after the onset of merger. In other words, the mass accretion rate is as large as $\sim 0.5$--$1M_\odot /{\rm s}$ for $M_{\rm disk}=0.1M_\odot$ even at $\approx50~{\rm ms}$ after the onset of merger. This shows the importance of the non-axisymmetric structure of the disk that is preserved for a long time scale and governs the angular momentum transport.

\subsubsection{Ejecta morphology and velocity}
   \begin{figure*}
	\begin{center}
		\begin{tabular}{lll}
		\includegraphics[height=38mm]{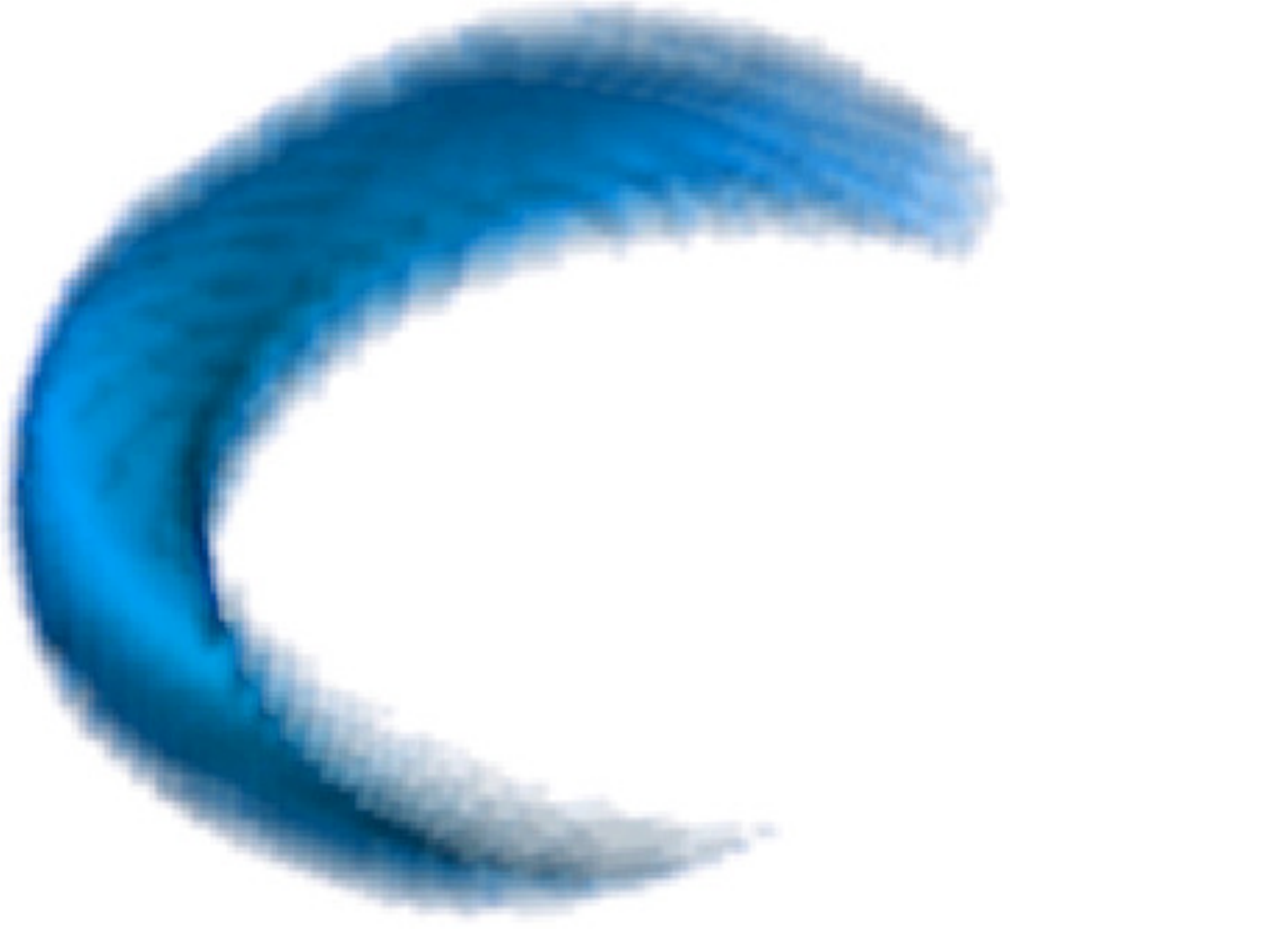}
		\includegraphics[height=38mm]{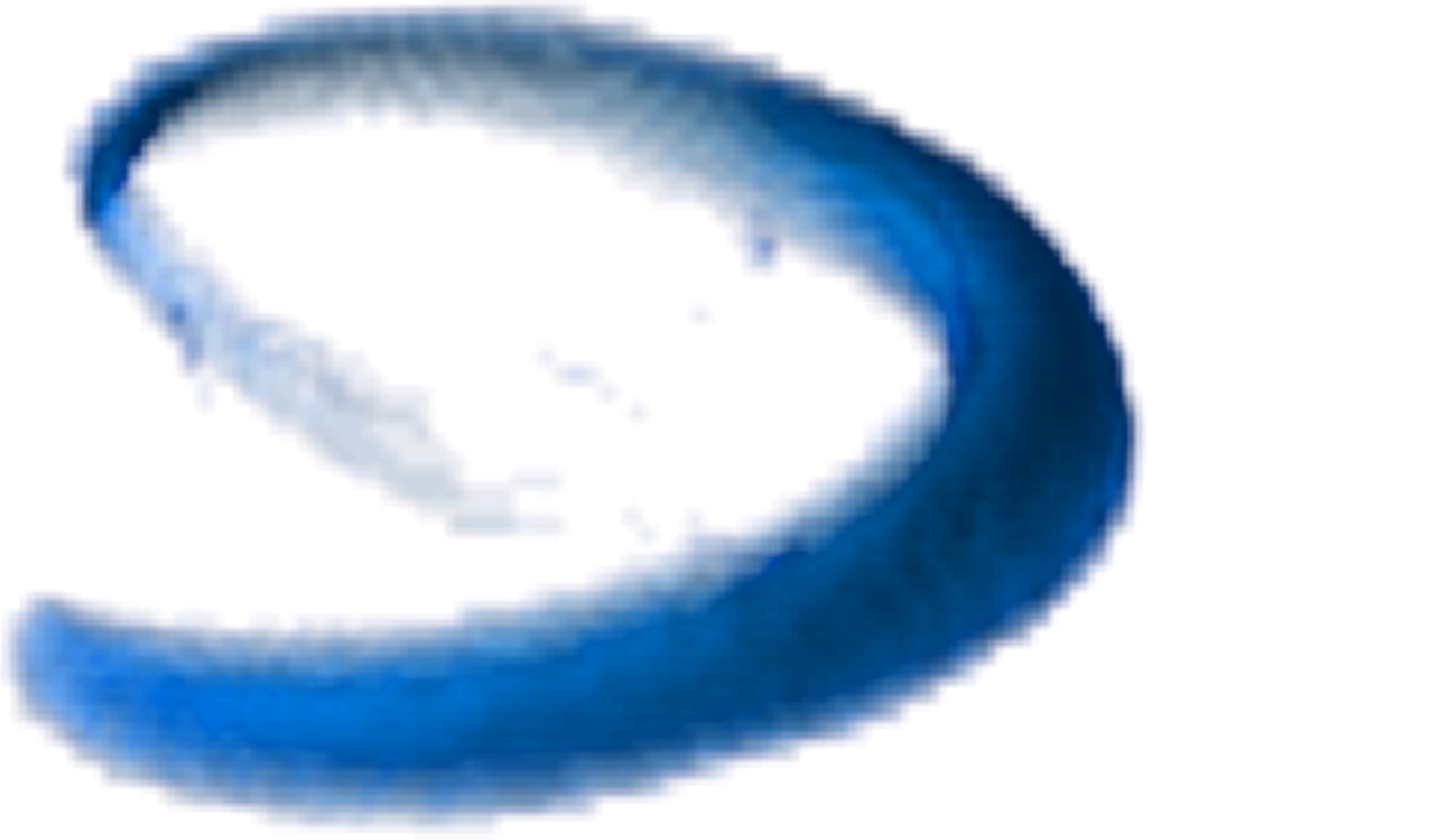}
		\includegraphics[height=38mm]{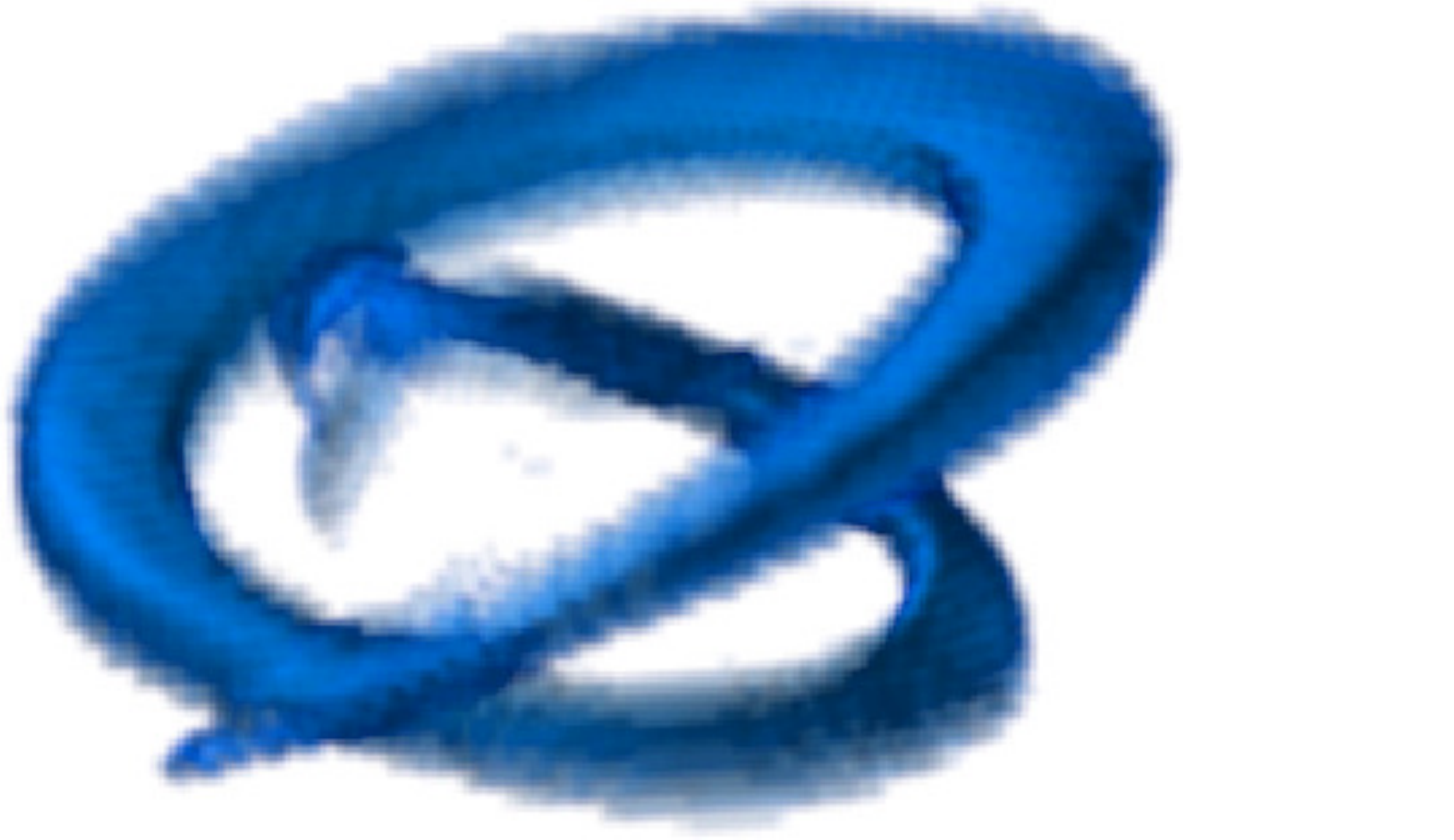}
		\end{tabular}
	\end{center}
	\caption{The volume-rendered density map of the ejecta at the time that the ejecta expands to $\approx1500~{\rm km}$ for models H4i30 (left), H4i60 (middle), and ALF2i60 (right), respectively.}
	\label{fig:3d_eje}
\end{figure*}
  
  In Fig.~\ref{fig:3d_eje}, we plot the image of the ejecta with volume rendering for models H4i30, H4i60, and ALF2i60. We find that for the case that the ejecta is massive (H4i30), it has a crescent-like shape with its opening angle $\approx180^\circ$. On the other hand, for the case that the ejecta  has small mass (ALF2i60), the opening angle becomes larger than 360$^\circ$ winding around the center of mass. We also find that the ejecta is warped due to the orbital precession for this case. In particular, the rear-end collision for the ejecta with its opening angle larger than $360^\circ$ is less pronounced in the misaligned-spin case than in the aligned-spin case \cite{bib:bhns13}.

	After the matter becomes gravitationally unbound, the ejecta expand in an approximately self-similar manner. To analyze the feature of the ejecta, we define an average velocity of the ejecta as,
\begin{equation}
	v_{\rm ave}=\sqrt{\frac{2T_{\rm eje}}{M_{\rm eje}}},
\end{equation}
where $T_{\rm eje}$ is kinetic energy of the ejecta, defined following~\cite{bib:hotoke}. We also compute the linear momentum of the ejecta defined by

\begin{equation}
	P_{{\rm eje},i}=\int\rho_* {\hat u}_id^3x,
\end{equation}
and calculate its magnitude by
\begin{equation}
	P_{\rm eje}=\sqrt{\sum_i \left(P_{{\rm eje},i}\right)^2}\hspace{1mm}.
\end{equation}
Using $P_{\rm eje}$, we define the bulk velocity of the ejecta $v_{\rm eje}$ by
\begin{equation}
	v_{\rm eje}=\frac{P_{\rm eje}}{M_{\rm eje}}.
\end{equation}
Here, note that $v_{\rm eje}$ may reflect the morphology of the ejecta: The linear momentum of the ejecta vanishes if its morphology is isotropic, while the value of $v_{\rm eje}$ becomes close to $v_{\rm ave}$ if the mass is ejected coherently to a particular direction.
    
 We summarize the values of $v_{\rm ave}$ and $v_{\rm eje}$ measured at $10~{\rm ms}$ after the onset of merger in Table~\ref{tb:rm}. This shows that irrespective of the models, $v_{\rm ave} \sim0.3c$ and their dependence on the misalignment angle and the EOS is weak. 
 Although the magnitude of $v_{\rm ave}$ is quite universal, $v_{\rm eje}$ varies from model to model. $v_{\rm eje}$ becomes large for the case that the ejecta mass is large ($\approx 0.03 M_\odot$). This is consistent with the result in Fig.~\ref{fig:3d_eje} that the mass ejection proceeds in an anisotropic manner for the case that $M_{\rm eje}$ is large. On the other hand, $v_{\rm eje}$ becomes small for the ejecta with small mass ($\alt 0.01 M_\odot$). This is also consistent with a quasi-axisymmetric ejection shown in Fig.~\ref{fig:3d_eje}. 

\subsection{The properties of remnant BH}

\begin{table*}
\caption{The quantities of the remnant BH evaluated at $\approx 10~{\rm ms}$ after the onset of merger. The irreducible mass of the BH $(M_{\rm irr,f})$, the mass of the BH $(M_{\rm BH,f})$, the dimensionless spin parameter $(\chi_{\rm f})$, the BH spin $(S_{\rm BH,f})$, the tilt angle of the BH spin $({\rm cos}^{-1}(S_{\rm BH}^z/S_{\rm BH}))$, and the dominant QNM frequency derived by the fitting formula Eq.~(\ref{eq:qnmfit}) using the result for $M_{\rm BH}$ and $\chi$ $(f_{\rm QNM})$, respectively. The results for the aligned-spin case are taken from~\cite{bib:bhns13}. For the model APR4i90, we failed to find the location of the AH because of inappropriate setting of the AMR domains.}
\begin{center}
 \begin{tabular}{l|cccccc} \hline
 Model & $M_{\rm irr,f}[M_\odot]$ & $M_{\rm BH,f}[M_\odot]$ & $\chi_{\rm f} $ &$S_{\rm BH,f}[GM_\odot^2/c]$ & ${\rm cos}^{-1}(S_{\rm BH}^z/S_{\rm BH})[^\circ]$ & $f_{\rm QNM}[{\rm kHz}]$\\ \hline\hline
 APR4i0 & 6.83 & 7.82  & 0.85 & 52	&$0^\circ$&2.56\\
APR4i30 	& 6.93	& 7.85	& 0.83	& 51	&$<1^\circ$	&2.49\\
APR4i60 	& 7.16	& 7.90	& 0.77	& 48	&$1^\circ$	&2.32\\
APR4i90 	& --	& --	& --	& --	& --& -- \\\hline
ALF2i0 & 6.78 & 7.68 & 0.83 & 49	&$0^\circ$&2.55\\
ALF2i30 	& 6.87	& 7.76	& 0.82	& 50&$<1^\circ$	&2.50\\
ALF2i60 	& 7.14	& 7.89	& 0.77	& 48	&$1^\circ$	&2.33\\
ALF2i90 	& 7.45	& 7.95	& 0.65	& 41	&$4^\circ$	&2.10\\\hline
H4i0   & 6.74 & 7.64  & 0.83 & 48	&$0^\circ$&2.56\\
H4i30   	& 6.83	& 7.71	& 0.82	& 49	&$1^\circ$	&2.51\\
H4i60   	& 7.11	& 7.86	& 0.77	& 48	&$1^\circ$	&2.34\\
H4i90   	& 7.45	& 7.95	& 0.65	& 41	&$4^\circ$	&2.10\\\hline
MS1i0  & 6.74 & 7.64 & 0.83 & 48	&$0^\circ$&2.56\\
MS1i30  	& 6.81	& 7.66	& 0.81	& 48	&$1^\circ$	&2.51\\
MS1i60  	& 7.06	& 7.78	& 0.76	& 46	&$2^\circ$	&2.35\\
MS14i90 	& 7.44	& 7.95	& 0.66	& 42	&$3^\circ$	&2.10\\\hline
\end{tabular}
\end{center}
\label{tb:bh1}
\end{table*}

 \begin{figure}
 	\includegraphics[width=80mm]{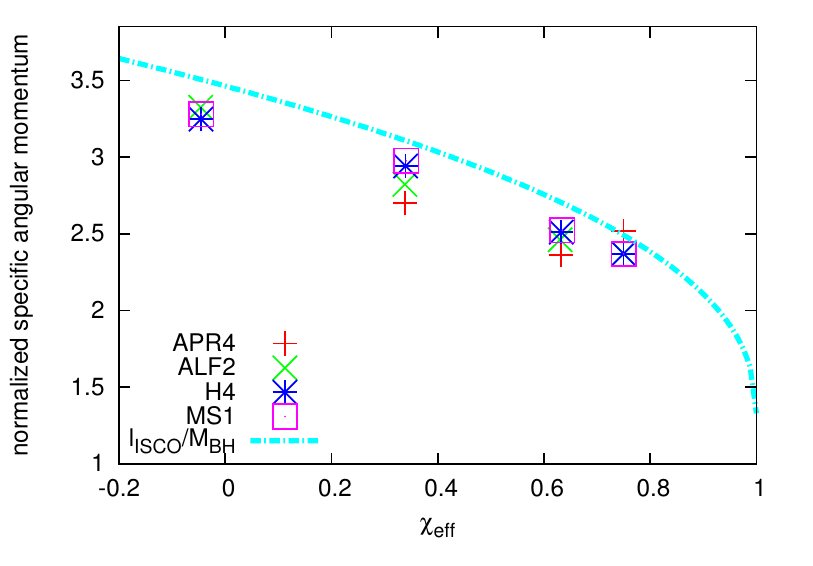} 

	\caption{The average specific angular momentum of material brought into BH normalized by $M_{\rm BH}$, $\Delta l/M_{\rm BH}$, as a function of an effective dimensionless spin parameter $\chi_{\rm eff}=\chi {\rm cos}\,i_{\rm tilt,0}$. The specific angular momentum at the ISCO of the BH is also plotted.}
		\label{fig:dlchi}
\end{figure}

 We list the values of $M_{\rm irr}$, $M_{\rm BH}$, $\chi$, $S_{\rm BH}$, and the tilt angle of the BH spin, ${\rm cos}^{-1}(S_{\rm BH}^z/S_{\rm BH})$, at $10~{\rm ms}$ after the onset of merger in Table~\ref{tb:bh1}. The mass of the remnant BH increases monotonically as the value of $i_{\rm tilt,0}$ increases and the EOS becomes stiff. The reason for this is that the tidal disruption is suppressed for larger values of $i_{\rm tilt,0}$ and stiffer EOSs, and more matter of the NS is swallowed by the BH.
 
 The dimensionless spin parameter decreases with the increase of $i_{\rm tilt,0}$, and it depends weakly on the EOS. In particular, for the models with $i_{\rm tilt,0}\approx90^\circ$, the final value of the dimensionless spin parameter becomes smaller than the intial value of 0.75. This is because the dimensionless spin parameter is defined by $S_{\rm BH}/M_{\rm BH}^2$, and the increase of $M_{\rm BH}^2$ is larger than $S_{\rm BH}$ for these models. In fact, we can see the increase of $S_{\rm BH}$. The value of ${\rm cos}^{-1}(S_{\rm BH}^z/S_{\rm BH})$ shows that the BH spin becomes approximately parallel with the $z$-axis after the merger. This is because the direction of the total angular momentum approximately preserves its initial direction, and the BH swallows nearly entire angular momentum of the system. The tilt angle of ${\bf S}_{\rm BH}$ in the final state becomes larger for larger values of $i_{\rm tilt,0}$. However, it is always smaller than $5^\circ$.

 The BH spin increases because the BH swallows the NS matter of positive angular momentum. From the increments of the BH mass $\Delta M_{\rm BH}:=M_{\rm BH,f}-M_{\rm BH,i}$ and the BH spin $\Delta S_{\rm BH}:=\left|{\bf S}_{\rm BH,f}-{\bf S}_{\rm BH,pm}\right|$, we calculate the mean value of the specific angular momentum $\Delta l:=\Delta S_{\rm BH}/\Delta  M_{\rm BH}$ that the BH gained due to the falling material. Here, $M_{\rm BH,i}$ is the initial BH mass and ${\bf S}_{\rm BH,pm}$ is the value of the BH spin evaluated just before the merger, $t\approx t_{\rm merge}-1~{\rm ms}$. For the aligned-spin model, the value of $\Delta l$ is expected to reflect the specific angular momentum at the ISCO. Thus, we compare $\Delta l$ with the specific angular momentum of the ISCO of the BH with an effective spin parameter which we introduced in the previous section [see Eq.~(\ref{eq:chieff})].

 In Fig.~\ref{fig:dlchi}, we plot $\Delta l/M_{\rm BH}$ as a function of $\chi_{\rm eff}$.  We also plot the specific orbital angular momentum at the ISCO of the aligned-spin BH as a function of $\chi_{\rm eff}$. The values of $\Delta l/M_{\rm BH}$ are approximately the same as the specific angular momentum at an effective ISCO. This result implies that the separation at which the orbital motion of the binary becomes unstable and the NS falls into the BH is given effectively by the ISCO in the equatorial motion around  the BH with $\chi_{\rm eff}$. Taking a closer look, $\Delta l/M_{\rm BH}$ tends to be smaller than the value for the effective ISCO. This is likely to stem from the gravitational-wave emission that dissipates the orbital angular momentum while the matter falls into the BH, or the redistribution of the specific angular momentum due to the tidal torque. 
  
\subsection{The gravitational waveform}
 The misalignment between the orbital angular momentum and the BH spin causes the precession of the orbit and induces the modulation in  gravitational waves. Also, the misalignment angle of the BH spin and the EOS of the NS affect the tidal-disruption process, and as a result, gravitational waveforms are modified by them.
  In {\tt SACRA}, we extract the out-going component of the complex Weyl scalar $\Psi_4$ at finite radii and project it onto the spin-weighted spherical harmonic functions. Here, we took the axis of the spherical harmonics to be $z$-axis: the initial direction of the total angular momentum. Then to obtain gravitational waveforms, we integrate  $\Psi_4$ twice in time as 
 \begin{equation}
 	h(t)=h_{+}(t)-ih_{\times}(t)=\int_{0}^{t} dt'\int_{0}^{t'} dt'' \Psi_4(t'').
 \end{equation}
  In the following, we plot the normalized amplitude $Dh/m_0$ or the amplitude observed at a hypothetical distance $D=100~{\rm Mpc}$ as a function of approximate retarded time defined by 
\begin{equation}
	t_{\rm ret}=t-D-2M_{0}{\rm ln}\frac{D}{M_0}.
\end{equation}
 
 The Fourier spectrum of the gravitational waveform could reflect more quantitative information. In this paper, we define the Fourier power spectrum of gravitational waves as the root mean square of two independent polarizations as
\begin{eqnarray}
	{\tilde h}(f)&=&\sqrt{\frac{|{\tilde h}_{+}(f)|^2+|{\tilde h}_{\times}(f)|^2}{2}},\\
	{\tilde h}_{A}(f)&=&\int h_{A}(t)e^{2\pi i ft}dt,\\
	&&\left(A\right.=\left. +,\times\right).\nonumber
\end{eqnarray}
  We will plot a dimensionless Fourier spectrum ${\tilde h}_{\rm eff}(f):=f{\tilde h}(f)$ observed at a hypothetical distance $D=100~{\rm Mpc}$ as a function of the frequency $f$, or a normalized spectrum  $D{\tilde h}_{\rm eff}(f)/m_0$  as a function of dimensionless frequency $fm_0$. 
  
  \begin{figure*}
	\begin{center}
	\begin{tabular}{ll}
		\includegraphics[width=90mm]{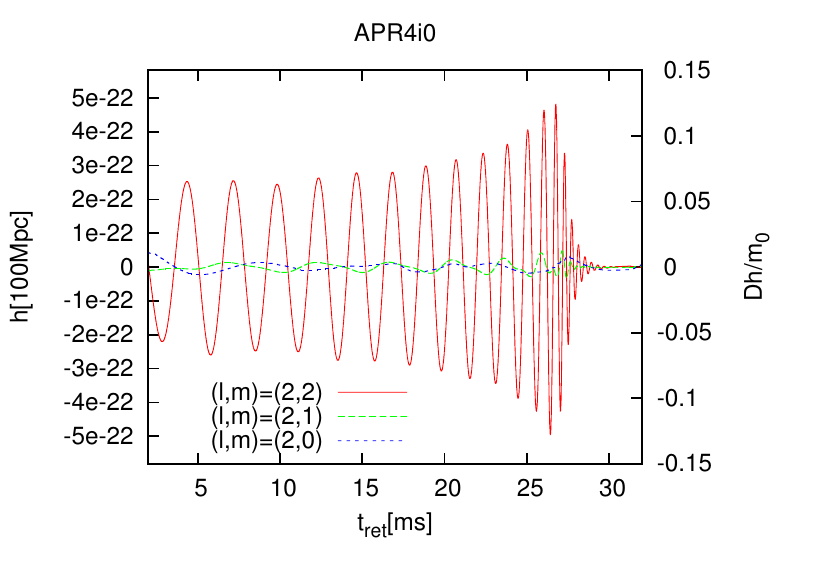} &
		\includegraphics[width=90mm]{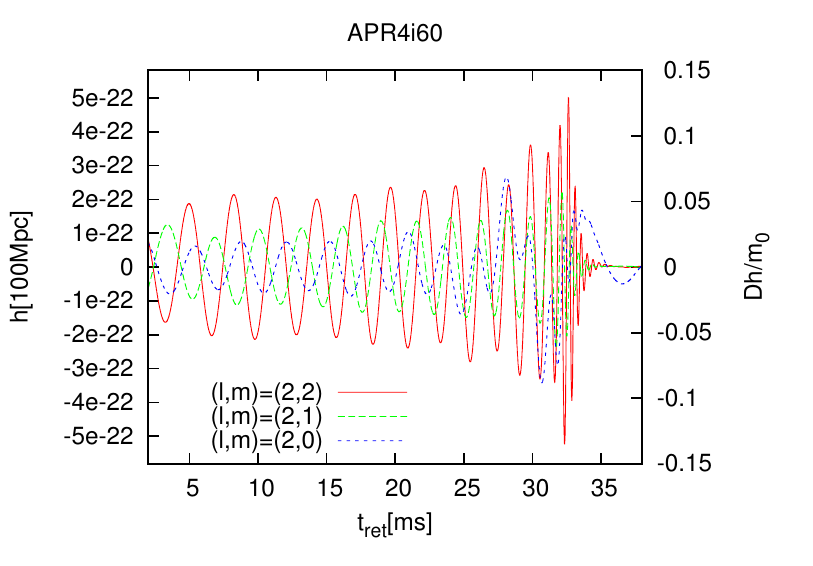} 
	\end{tabular}
	\end{center}
	\caption{ The plus-mode gravitational waveforms for $(l,m)=(2,2), (2,1)$, and $(2,0)$ for APR4i0 (the left panel) and APR4i60 (the right panel).}
	\label{fig:gwlmmode}
\end{figure*}

 Figure~\ref{fig:gwlmmode} shows plus-mode gravitational waveforms of $(l,m)=(2,2)$, $(2,1)$, and $(2,0)$ for models APR4i0 and APR4i60. While the amplitude of $(l,m)=(2,1)$ and $(2,0)$ is smaller than that of $(l,m)=(2,2)$ for the aligned-spin case, they could have a significant contribution to gravitational waveforms for the misaligned-spin case. This is because the direction of the orbital angular momentum does not always agree with the axis of the spin-weighted spherical harmonic function for the misaligned-spin case.
 
  As we show in Sec.~\ref{ssec:orbevo}, the angular velocity of the orbital precession is always smaller than the orbital angular velocity by an order of magnitude. Thus, gravitational waves for the misaligned-spin case have the feature similar to gravitational waves from the aligned-spin case observed from an inclined direction with respective to ${\bf L}$ for each instant. Indeed, it has already shown for precessing binary BH cases (see, e.g.,~\cite{bib:qa}) that the waveforms take a far simpler form in the quadrupole alignment  (QA) frame: the frame in which $z$-axis agrees with the instantaneous direction of ${\bf L}$. If we project gravitational waves onto the spherical-harmonic function in the QA frame, and describe these expansion coefficients by $(l',m')$, $(l',m')=(2,\pm 2)$ modes are the dominant modes. Under the rotational transformation, these components mixes not only into $(l,m)=(2,\pm 2)$ modes but also into different $m$ modes with $l=2$. This is consistent with the fact that the dominant frequency of $(l,m)=(2,1)$ and $(2,0)$ modes agrees with the frequency of $(2,2)$ mode rather than the half (this fact  is also pointed out in \cite{bib:bhns_t1}). We note that different $l$ modes do not mix under the rotation of the axis of spherical harmonics.

 \begin{figure}
 	\includegraphics[width=80mm]{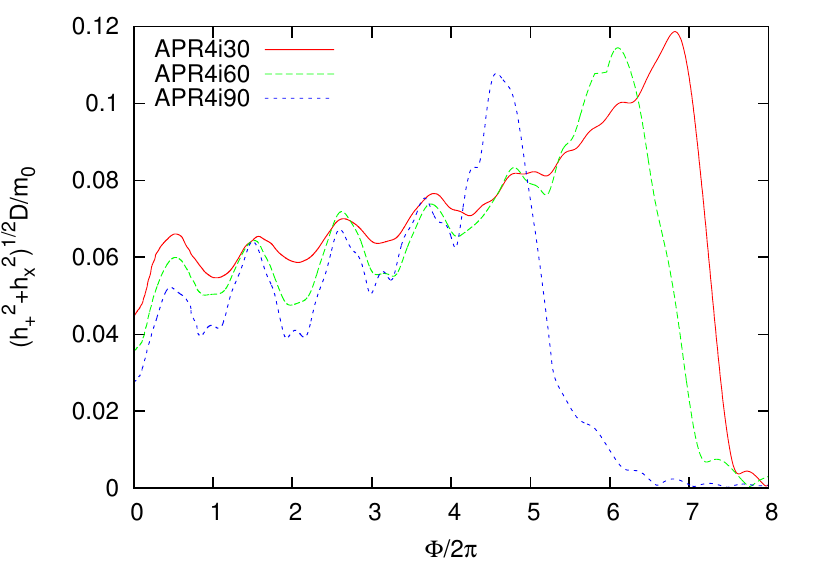} 

	\caption{The square of the gravitational-wave amplitude observed along the $z$-axis as a function of the orbital phase devided by $2\pi$, $\Phi/2\pi$, for models with the MS1 EOS.}
		\label{fig:phiamp}
\end{figure}

   Other $m'$ modes, such as $(l',m')=(2,\pm 1)$, $(3,\pm 1)$, and $(3,\pm 3)$ modes, also contribute to the gravitational waveform. Because the phase of the $m'$ mode is $m' \Phi$, where $\Phi$ is the phase of the orbit, the mixing among different $m'$ modes causes modulation in the amplitude of the waveforms. For example, when the $( l' , m' ) = ( 2 , 1 )$ mode is coupled with the $( 2 , \pm 2 )$ modes, the amplitude exhibits modulation with the periods of $2\pi$ and $6\pi$ in terms of the orbital phase, $\Phi$. Indeed, Fig.~\ref{fig:phiamp} shows that the amplitude observed along the $z$-axis modulates primarily with the period of $\approx 2\pi$ in terms of $\Phi$.
 
 Obviously, the mixing of several $m'$ modes in gravitational waves can occur for the aligned-spin case if we choose the axis of spherical harmonics which disagrees with the orbital angular momentum. One thing to be noted is that because the orbital angular momentum precesses for the misaligned-spin case, we cannot avoid the situation that the orbital angular momentum disagrees with the axis of spherical harmonics. This implies that the mixing among several $m'$ modes is unavoidable for the misaligned-spin case.

 \begin{figure*}
	\begin{center}
		\begin{tabular}{ll}
			\includegraphics[width=90mm]{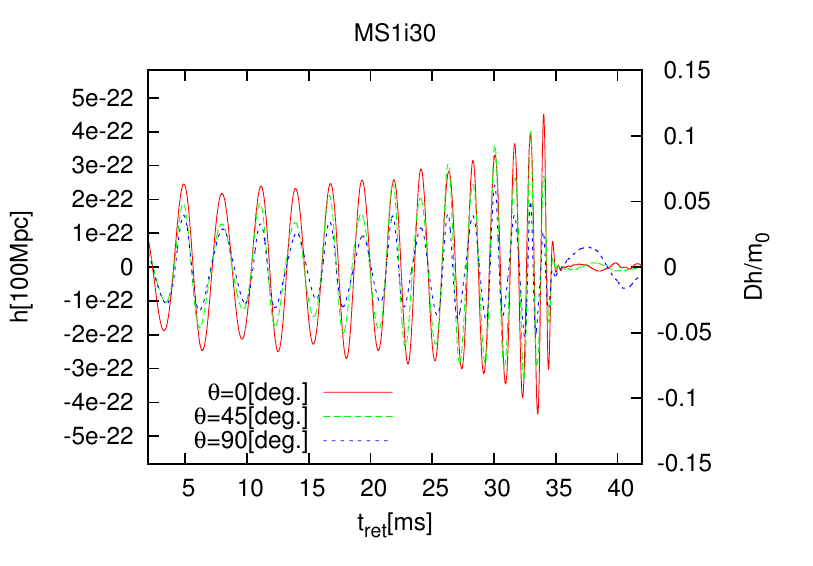}&
			\includegraphics[width=90mm]{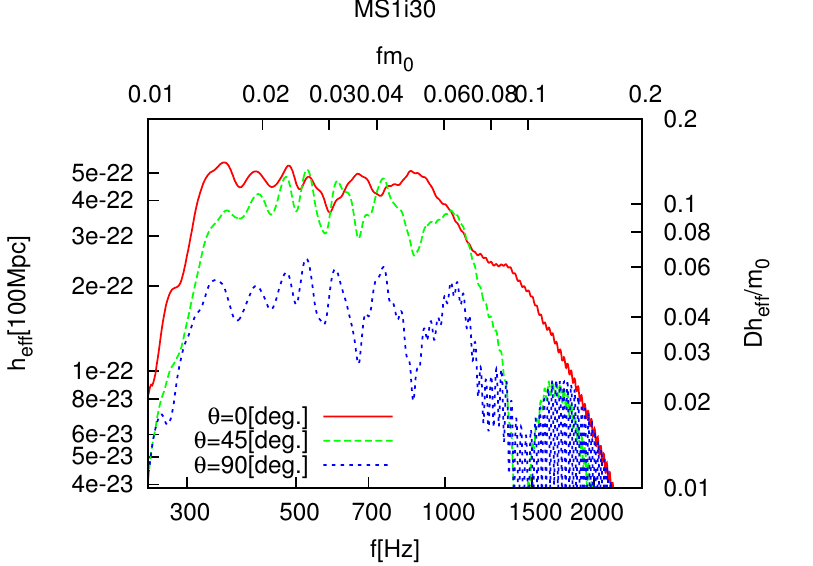}\\
			\includegraphics[width=90mm]{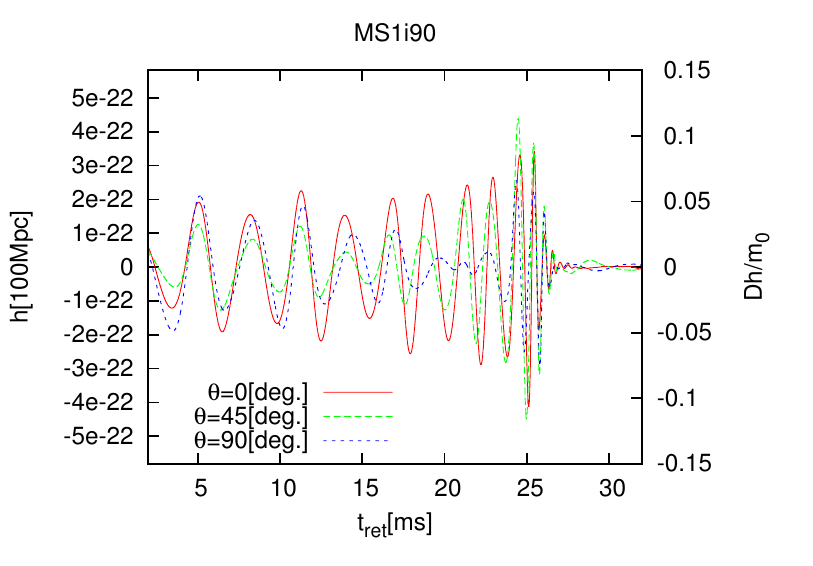}&
			\includegraphics[width=90mm]{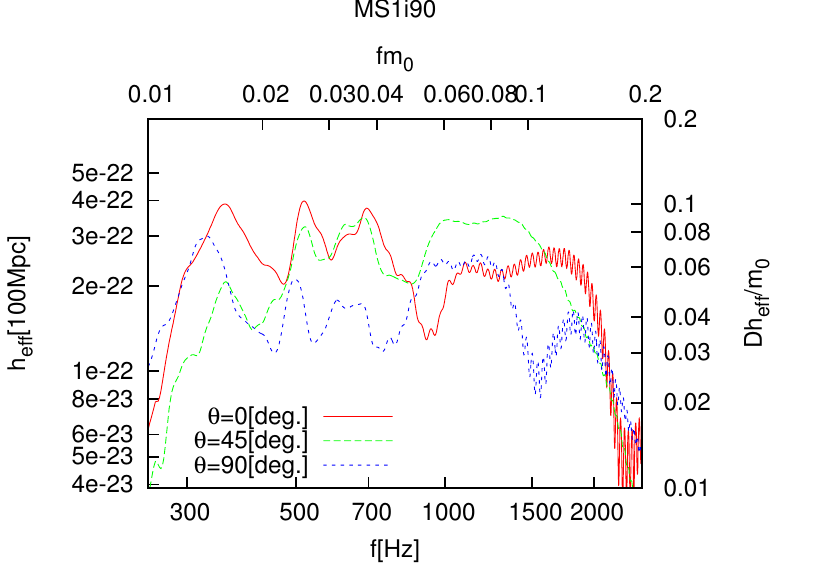}
		\end{tabular}
	\end{center}
	\caption{Plus-mode gravitational waveforms (left panels) and gravitational-wave spectra (right panels) observed from different inclination angles with respect to the $z$-axis. The top and bottom figures are the results for models MS1i30 and MS1i90, respectively. Gravitational waveforms are plotted as functions of retarded time, $t_{\rm ret}$. The left axes in the plots of the waveforms denote the amplitude observed at a hypothetical distance $D=100~{\rm Mpc}$, and the right does the normalized amplitude $Dh/m_0$. The upper axis in the plots of the spectra denotes the dimensionless frequency, $fm_0$, and the right axis denotes the normalized amplitude $D{\tilde h}_{\rm eff}(f)/m_0$. The bottom axis in the plots of spectra denotes the frequency of the gravitational waveform, in ${\rm Hz}$, and the left axis denotes the amplitude observed at a hypothetical distance $D=100~{\rm Mpc}$. }
	\label{fig:gwinc}
\end{figure*}

 \begin{figure*}
	\begin{center}
	\begin{tabular}{ll}
		\includegraphics[width=90mm]{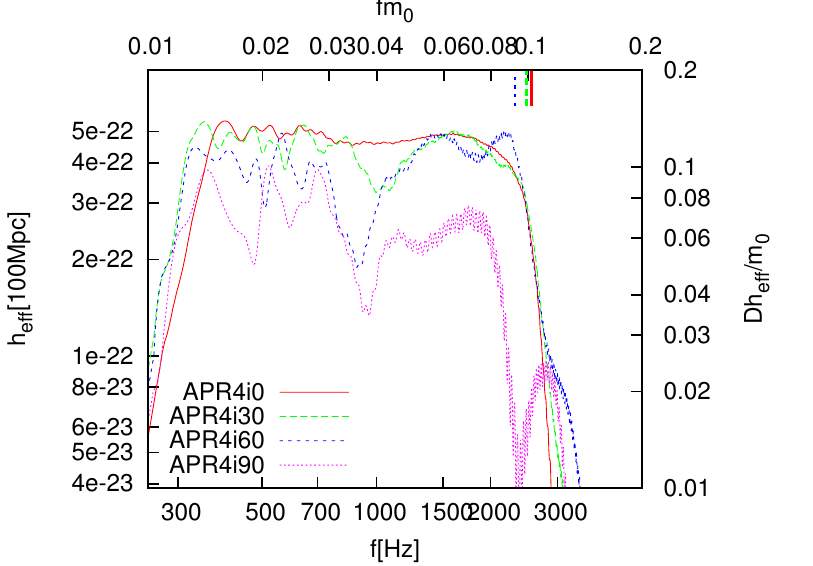}&
		\includegraphics[width=90mm]{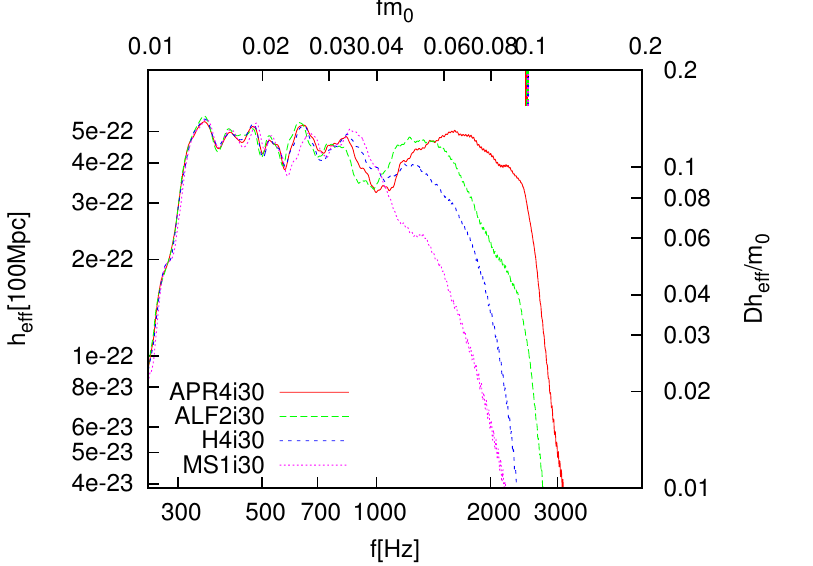}\\
		\includegraphics[width=90mm]{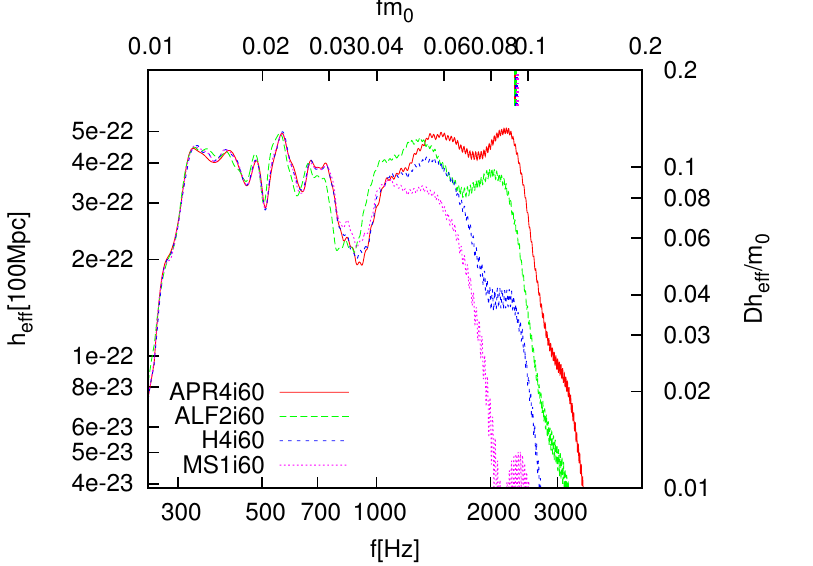}&
		\includegraphics[width=90mm]{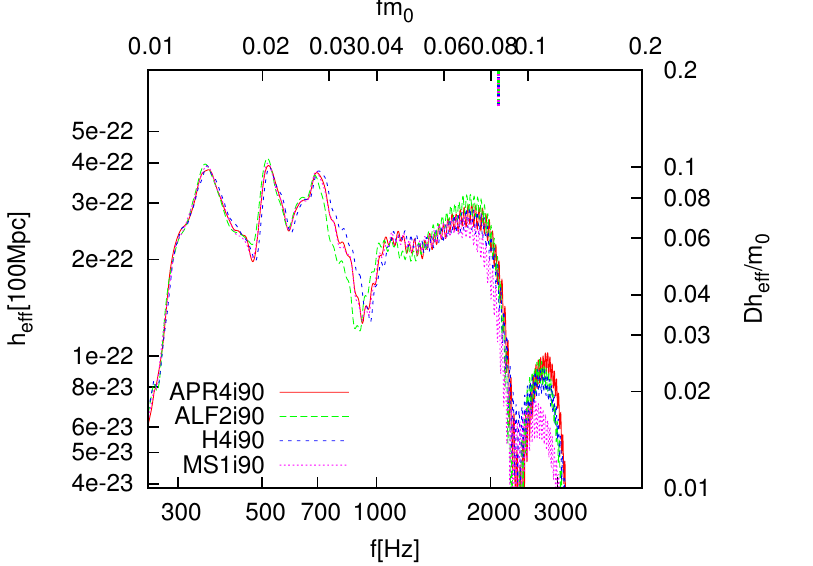}
	\end{tabular}
	\end{center}
	\caption{Gravitational wave spectra observed along the $z$-axis of the simulation. The top-left plot shows the spectra for the models with APR4 and with different initial values of $i_{\rm tilt,0}$. The top-right, bottom-left, and bottom-right plots compare the spectra among four different EOSs for $i_{\rm tilt,0}\approx 30^\circ$, $60^\circ$, and $90^\circ$, respectively. The vertical lines at the upper axis show the QNM frequency of the remnant BH for each models (see Table~\ref{tb:bh1}).}
	\label{fig:gwsdep}
\end{figure*}

	Figure~\ref{fig:gwinc} shows plus-mode gravitational waveforms (the left panels) and gravitational-wave spectra (the right panels) observed from different inclinations, $\theta =0^\circ$, $45^\circ$, and $90^\circ$ with respect to the $z$-axis. As we show in Sec.~\ref{ssec:orbevo}, ${\bf J}$ is always aligned approximately with the $z$-axis, and hence $\theta$ is regarded approximately as the angle between ${\bf J}$ and the direction of the observer. We take into account the contributions from $l=2-4$ modes in plotting the waveforms.
	
	For MS1i30, the waveforms observed from $\theta=0^\circ$ has a similar feature to those for the aligned-spin case; the amplitude and the frequency approximately monotonically increase as the system evolves. Only slight modulation in the amplitude and the frequency is found in the waveforms. There are much more appreciable modulation in the waveforms for $\theta=45^\circ$ and $90^\circ$. The amplitude of the waveforms becomes smaller for larger inclination angles. This dependence stems from the fact that for $(l',m')=(2,\pm2)$ modes the amplitude of the plus-mode waveform is proportional to $(1+{\rm cos}^2\,\theta)/2$. The phases of the waveforms agree among different values of $\theta$. 

	 For MS1i90, the modulation in the amplitude and the frequency are larger than that for MS1i30. There is no monotonic dependence of the amplitude of the waveforms on $\theta$. While the amplitude of the waveform for $\theta=0^\circ$ increases as the time evolves, that of $\theta=90^\circ$ approximately decreases until $t_{\rm ret}\approx 20~{\rm ms}$. The phase evolution of the waveforms is also different among different values of $\theta$. This is in particular significant for the last $\approx 10~{\rm ms}$  of the inspiral phase.  These features of the waveforms are due to the precession of the orbital plane. Although the orbital plane precesses, the angle between the $z$-axis and ${\bf L}$ is always approximately constant until the onset of merger for simulations in this study. Therefore, the angle between the direction of the observer and ${\bf L}$ is approximately unchanged for $\theta=0^\circ$, and the effect of the orbital precession to the waveform is small. On the other hand, for the observer at $\theta=90^\circ$, the orbital precession changes the angle between ${\bf L}$ and the direction of the observer, and affect the waveforms. In particular, the amplitude of the waveform is strongly suppressed when the orbital plane is edge-on to the direction of the observer.
	
	In the top-right and the bottom-right panels in Fig.~\ref{fig:gwinc}, we find that the amplitude of the spectra for MS1i30 decreases approximately monotonically for larger values of $\theta$, while there is no simple dependence of the amplitude on $\theta$ for MS1i90. While the spectrum for the aligned-spin model has a flat shape (see Fig.~\ref{fig:gwsdep} or ~\cite{bib:bhns9,bib:bhns10} for the spectra for the aligned-spin case), there are some bumps in the spectra for the misaligned-spin models. In particular, bumps at $\approx 1500~{\rm Hz}$ in the spectra for $\theta=45^\circ$ and $90^\circ$ change the feature of the cutoff. These bumps are the consequences of the mixing among the different modes of gravitational waves. We find that $l=4$ modes are negligible for the bumps in the spectra. Since $m'=2$ and $0$ do not contribute to the modulation, $(l',m')=(2,\pm 1)$, $(3,\pm 1)$, and $(3,\pm 3)$ are primarily responsible for the bumps.

	 The spectra of gravitational waves reflects the fate of the merger. In particular, as already clarified  by~\cite{bib:bhns6,bib:bhns9,bib:bhns10}, the location of the cutoff in the spectra has a correlation with the compactness of the NS. We expect that this also holds for the misaligned-spin case. However, the bumps of the spectra for the misaligned-spin models may make it difficult to determine the cutoff frequency, particular for the case that gravitational waves are observed from large inclination angles. Thus, in the following, we check the behavior of the bump in the spectra  and the dependence of the cutoff in the spectra on the models qualitatively as a first step to understand the gravitational-wave spectra from the misaligned-spin BHNS. Here, we only consider the waveforms for an observer along the $z$-axis, the direction of the initial total angular momentum, which would be the simplest case.

 We show the spectra of the waveforms for different values of $i_{\rm tilt,0}$ with APR4 in the top-left panel in Fig.~\ref{fig:gwsdep}. The bumps in the spectra become deeper as the value of $i_{\rm tilt,0}$ increases.  This is because the amplitude of $(l',m')=(2,\pm 1)$, $(3,\pm 1)$, and $(3,\pm 3)$ modes depend on the angle between ${\bf L}$ and the direction of the observer, and $\theta'$ becomes large for large values of $i_{\rm tilt,0}$ . For the models with APR4, the quasi-normal mode (QNM) is excited, and the cutoff frequency of the spectra reflects the frequency of the QNM. The frequency of the $(l,m)=(2,2)$ least-damped mode of the QNM is given approximately by the fitting fomula~\cite{bib:qnm} as
\begin{eqnarray}
	f_{\rm QNM}=\frac{1}{2\pi M_{{\rm BH},f}}\left[1.5251-1.1568(1-\chi_{{\rm f}})^{0.1292}\right].\label{eq:qnmfit}
\end{eqnarray}
 The values of the QNM frequency are summarized in Table~\ref{tb:bh1}. Figure~
\ref{fig:gwsdep} shows that the spectra have a cutoff around the QNM frequency, although the bumps of the spectra for the misaligned-spin models make it difficult to determine the cutoff frequency accurately.

 Next, we compare the spectra of the waveforms with different EOSs. The top right, bottom left, and bottom right panels of Fig.~\ref{fig:gwsdep} show the spectra for $i_{\rm tilt,0}\approx30^\circ$, $60^\circ$, and $90^\circ$, respectively with four different EOSs. We again find the bumps in the spectra, and they have approximately the same shape in $f\approx 400$--$1000~{\rm Hz}$ irrespective of the EOS. This reflects the fact that the waveform does not depend strongly on the EOS  for this late inspiral phase. The difference of the spectra due to the EOS becomes significant for $f>1000~{\rm Hz}$. The frequency of the spectral cutoff becomes lower in the order of APR4, ALF2, H4, and MS1. This is because for the stiffer EOS, tidal disruption occurs earlier and the waveform shuts down at lower frequency. Although the difference in the cutoff frequency is more appreciable for the spectra with $i_{\rm tilt,0}\approx30^\circ$ than with $i_{\rm tilt,0}\approx60^\circ$, we still find that the cutoff frequency should be, at least in principle, different among the models with $i_{\rm tilt,0}\approx60^\circ$. For example, if we say the cutoff frequency as the frequency at which $h_{\rm eff}\approx2\times 10^{-22}$ is achieved, the difference of the cutoff frequency between models with APR4 and H4 is $\approx40\%$ for both $i_{\rm tilt,0}\approx30^\circ$ and $i_{\rm tilt,0}\approx60^\circ$ cases, and the difference among the EOSs is always larger than $10\%$ for $i_{\rm tilt,0}\approx30^\circ$ and $6\%$ for $i_{\rm tilt,0}\approx60^\circ$. By contrast to the cases with $i_{\rm tilt,0}\approx30$ and $60^\circ$, the difference in the spectra is small among different EOSs for $i_{\rm tilt,0}\approx90^\circ$, and hence it might be difficult to distinguish the EOS from the spectra. The reason for this is that tidal disruption occurs so weakly that the difference in the EOS is not appreciable. The cutoff frequency at $f\approx 2100~{\rm Hz}$ for these models reflects the frequency of the QNM. Indeed, the value calculated by Eq.~(\ref{eq:qnmfit}) agrees with this value.

\subsection{Emitted energy, linear momentum, angular momentum by gravitational waves}\label{sec:gwemit}
\begin{table*}
\caption{The list of total radiated energy, $\Delta E$, its ratio to initial ADM mass $\Delta E/M_0$, total radiated angular momentum $\Delta J$, total radiated linear momentum normalized by the initial ADM mass, $P_{\rm GW}/M_0$, and the linear momentum of the ejecta normalized by the initial ADM mass, $P_{\rm eje}/M_0$, respectively. For the i90 models, the data for $P_{{\rm eje}, i}$ was not output.}
\begin{center}
	\begin{tabular}{l|cccc}\hline
	Model	&$\Delta E[M_\odot c^2]\left(\Delta E/M_0[\%]\right)$	&$\Delta J[GM_\odot^2/c]$ &$P_{\rm GW}/M_0[{\rm km/s}]$	&~~$P_{\rm eje}/M_0[{\rm km/s}]$~~		\\\hline\hline
	APR4i30	&0.15(1.9)	&11	&$8.1\times 10^{1}$	& $8.7$			\\
	APR4i60	&0.14(1.7)	&8.5	&$5.7\times 10^{2}$	& $5.7\times 10^{-1}$ 			\\
	APR4i90	&0.098(1.2)	&6.2&$5.4\times 10^{2}$	& $-$ 					\\\hline
	ALF2i30	&0.12(1.4)	&9.4	&$7.8\times 10^{1}$	& $2.0\times 10^{2}$			\\
	ALF2i60	&0.11(1.4)	&7.8	&$2.7\times 10^{2}$	& $2.0\times 10^{1}$			\\
	ALF2i90	&0.092(1.1)	&6.1&$3.3\times 10^{2}$	& $-$ 					\\\hline
	H4i30	&0.093(1.2)	&8.5&$7.8\times 10^{1}$	& $3.3\times 10^{2}$			\\
	H4i60	&0.093(1.2)	&7.2&$6.3\times 10^{1}$	& $6.6\times 10^{1}$			\\
	H4i90	&0.085(1.1)	&5.8&$3.6\times 10^{2}$	& $-$ 					\\\hline
	MS1i30	&0.074(0.93)	&7.6&$6.3\times 10^{1}$	& $5.7\times 10^{2}$			\\
	MS1i60	&0.075(0.93)	&6.6&$9.3\times 10^{1}$	& $2.8\times 10^{2}$			\\
	MS1i90	&0.073(0.90)	&5.5&$1.8\times 10^{2}$	& $-$ 					\\\hline
	\end{tabular}
\end{center}
\label{tb:gwemit}
\end{table*}

	A binary loses its orbital energy and orbital angular momentum by the gravitational radiation. The amount of energy and angular momentum emitted by gravitational waves depends on the binary parameter, and that is one of the interests in studying the binary merger. Gravitational waves could also carry linear momentum of the system. Non-zero linear momentum radiation causes recoil of the system. We evaluate these energy, linear momentum, and angular momentum emitted by gravitational waves, using the formula of~\cite{bib:gwe}. 

 In Table~\ref{tb:gwemit}, we list the total energy, $\Delta E$ (and its ratio to the initial ADM mass), the total linear momentum normalized by the initial ADM mass, $P_{\rm GW}/M_0$, and the angular momentum, $\Delta J$, emitted by gravitational waves. We also list the recoil velocity caused by the mass ejection,  $P_{\rm eje}/M_0$. We take into account the contributions from $l=2-4$ modes for the evaluation of the emitted quantities. The $l=2$ modes contribute to $\Delta E$ and $\Delta J$ by more than $83\%$, while $l=3$ by $\approx 10\%$, and $l=4$ by $\approx 3\%$. These fractions of the contribution depend only weakly on the EOS and $i_{\rm tilt,0}$. 
  
 Table~\ref{tb:gwemit} shows that $\Delta E$ and $\Delta J$ decrease monotonically with the decrease of the compactness of the NS. The same dependence of  $\Delta E$ and  $\Delta J$ on the compactness of the NS is found for the results obtained by the non-spinning and aligned-spin BH-NS mergers~\cite{bib:bhns9,bib:bhns10}, and it is due to the fact that a longer inspiral phase (i.e., longer gravitational-wave emission phase) is realized for a softer EOS. Table~\ref{tb:gwemit} also shows that, for a fixed EOS, $\Delta E$ and $\Delta J$ monotonically decrease with the increase of $i_{\rm tilt,0}$. The dependence of $\Delta E$ on $i_{\rm tilt,0}$ is weaker than that on the EOS as far as tidal disruption is appreciable, while $\Delta J$ depends appreciably on $i_{\rm tilt,0}$. 

	The recoil velocity induced by the gravitational-wave emission, $P_{\rm GW}/M_0$, decreases as the compactness of the NS becomes small. This is because the smaller compactness of the NS results in earlier tidal disruption during the insprial phase, resulting in an earlier shutdown of the gravitational-wave emission. The recoil velocity caused by the mass ejection is larger for the models with smaller compactness of the NS because the ejected mass becomes larger for these models. This opposite dependence of the recoil velocity on the compactness for $P_{\rm GW}$ and $P_{\rm eje}$ reverses the dominant component for the recoil~\cite{bib:bhns13}. While the recoil due to the gravitational-wave emission is dominant for models with a large compactness, the recoil induced by the mass ejection becomes dominant for models with a small compactness. These two components are comparable for the case that $M_{\rm eje}\sim 0.01M_\odot$.

\section{Summary and discussion}\label{sec:sec5}
	We performed numerical-relativity simulations for the merger of BH-NS binaries with various BH spin misalignment angles, employing four models of nuclear-theory-based EOSs described by a piecewise polytrope. We investigated the dependence of the orbital evolution in the late inspiral phase,  tidal-disruption process of the NS, properties and structures of the remnant disk and ejecta, properties of the remnant BH, gravitational waveforms and their spectra on the BH spin misalignment angle and the EOS of the NS.

 	We showed that a large BH spin misalignment angle suppresses the NS tidal disruption event by the reduction of the spin-orbit interaction. The remnant mass of the material outside the BH decreases as the misalignment angle of the BH increases. This dependence agrees with the previous results for misaligned-spin BH-NS mergers~\cite{bib:bhns_t1, bib:bhns_t2}. Also we reconfirm the findings in~\cite{bib:bhns9, bib:bhns10} that the remnant mass increases as the compactness of the NS decreases. In our study, this dependence on the compactness of the NS is shown irrespective of the BH misalignment angle. The deviation of the result from the prediction of the fitting formula~\cite{bib:rmfit}  is within $50\%$ for $M_{\rm >AH} \agt0.1 M_\odot$ and within $30\%$ for $M_{\rm >AH} \agt0.2 M_\odot$, even though we employ a simple definition of the effective spin parameter. This reconfirms the argument of~\cite{bib:bhns_t2} that the mass of the material outside the remnant BH can be modeled with a good accuracy by considering the result of the aligned-spin cases.
 	
 	Effects of the orbital precession are reflected in the tidal tail. The elevation angle of the tidal tail measured from the $xy$-plane is different for each part. Although it is pointed out in~\cite{bib:bhns_t2} that the elevation angle may prevent its elements to collide, the material of the tidal tail still collides with each other, and a weakly inclined torus is eventually formed at least for the model with $i_{\rm tilt,0}\alt 60^\circ$.

 	Monotonic dependence of the remnant disk mass and the ejecta mass on the orbital misalignment angle was shown. Both $M_{\rm disk}$ and $M_{\rm eje}$ decrease as the misalignment angle increases. However, we still found that if the compactness of the NS is moderate ${\cal C}=0.160$ or even small ${\cal C}=0.140$, BH-NS mergers with $i_{\rm tilt,0}\alt50^\circ$ and $i_{\rm tilt,0}\alt70^\circ$ can produce disks larger than $0.1M_\odot$. Such a system could be a candidate for the progenitors of sGRB. $M_{\rm eje}>0.01M_\odot$ is achieved for $i_{\rm tilt,0}<85^\circ$ with ${\cal C}=0.140$, $i_{\rm tilt,0}<65^\circ$ with ${\cal C}=0.160$, and $i_{\rm tilt,0}<30^\circ$ with ${\cal C}=0.175$. We note that if the magnitude of the BH spin becomes large, more massive disk and ejecta would be produced.
 	
	For the models with $i_{\rm tilt,0}\approx30^\circ$,  the structure of the disk is similar to the disk formed for the aligned-spin BH-NS mergers. In particular, the rotational axis of the dense part $(\agt 10^{9}~{\rm g/cm^3})$ of the disk is aligned approximately with the remnant BH spin. On the other hand, for the models with $i_{\rm tilt,0}\approx60^\circ$, we found that the axis of the disk is misaligned with the direction of the remnant BH spin initially with $\approx 30^\circ$,  although the misalignment angle of the dense part of the disk approaches zero in $\approx 50$--$60~{\rm ms}$. 
 While the dense part of the disk becomes aligned with the direction of the BH spin, an elevation angle of the tidal tail at large orbital radii is $\approx 15^\circ$ and $\approx 30^\circ$ for models with $i_{\rm tilt,0}\approx30^\circ$ and $\approx60^\circ$, respectively. This reflects the orbital elevation during the inspiral phase. It is pointed out in~\cite{bib:grb_tilt} that the misalignment of the disk may affect the light curve of the sGRB. However, since the high-density part of the disk with $\rho>10^{10}~{\rm g/cm^3}$ would play main roles, the effect of the BH spin misalignment may not be observable in the sGRBs,  because the dense part of the disk becomes aligned with the BH spin in a relatively short time scale.We suspect that Bardeen-Petterson-like effect induced by a purely hydrodynamical mechanism, such as angular momentum redistribution due to a shock wave excited in a non-axisymmetric manner of the disk, should work in the disk for the alignment.
 
	We found that the accretion time scale of the matter in the disk to the BH is typically $\approx 100~{\rm ms}$, and depends weakly on the binary parameters. The main mechanism of the accretion in the present context is the redistribution of the angular momentum due to the torque exerted by non-axisymmetric structure of the disk, which is seen in Fig.~\ref{fig:disk_1}. In reality,  the viscosity induced by the magnetorotational instability turbulence could play an important role for this phase~\cite{bib:mri}. Since we did not take those effects into account, the accretion rate for the late phase might not be very quantitative. However, the present result shows that the purely hydrodynamical effect is important for the accretion of the matter, and this effect should be considered whenever we study the evolution of the accretion disk formed by BH-NS. 
 	 
 	We found that the velocity of the ejecta is typically $0.2$--$0.3c$, and has only weak dependence on the misalignment angle of the BH spin and the EOS of the NS. We also found that the morphology of the ejecta changes depending on the ejecta mass: Crescent-like-shaped ejecta with its opening angle $\approx 180^\circ$ is formed for relatively massive ejecta $(\agt0.03M_\odot)$, while spiral-shaped ejecta with its opening angle larger than $360^\circ$ is formed for relatively less massive ejecta $(\alt0.01M_\odot)$. In particular, the spiral shape reflects the orbital precession. This dependence of the ejecta morphology was also found for the aligned-spin case~\cite{bib:bhns13}, and might be explained by the periastron advance in general relativity.

	 We found that the dimensionless spin parameter of the remnant BH depends only weakly on the EOS of the NS, but it depends strongly on the misalignment angle, $i_{\rm tilt,0}$. The final direction of the BH spin becomes aligned approximately with the initial direction of the total angular momentum. We also found an approximate relation between the misalignment angle and the increase of the BH spin, and that the  specific angular momentum that the BH gained during the merger approximately agrees with a specific angular momentum at the ISCO of the BH with $\chi_{\rm eff}=\chi {\rm cos}\,i_{\rm tilt,0}$ .

 	We showed that the mixing among the components of spherical harmonics occurs and causes the modulation in gravitational waveforms for the misaligned-spin case. In particular, we found that the period in the modulation is primarily $\approx 2\pi$ in terms of the orbital phase. We also studied the dependence of waveforms on the direction of the observer, and found that the modulation due to the orbital precession becomes significant for the case that the observer is located along the direction perpendicular to the total angular momentum. The bump-shape modulation in the power spectrum of gravitational waveforms is found, and the depth of the bump becomes large as $i_{\rm tilt,0}$ becomes large.
 	
 	In the presence of the bumps in the spectra, the location of the cutoff frequency becomes obscured. Nevertheless, for the case that gravitational waves are observed along the axis of total angular momentum, the differences of the location of the cutoff in the spectra among the EOSs are seen for $i_{\rm tilt,0}\alt 60^\circ$,  while they are hardly found for $i_{\rm tilt,0}\approx 90^\circ$. This result shows that, in principle, gravitational waves from BH-NS binaries with $Q=5$ and $\chi=0.75$ contain the information of the EOS of the NS even if the misalignment angle of the BH spin is large up to $\approx 60^\circ$. To discuss whether we can extract the information of the NS EOS from the waveform by the observation, we need to define a quantitative indicator which reflects the information of the EOS, such as a cutoff frequency in the spectra, in an appropriate manner even in the presence of the orbital precession, and discuss the detectability considering the noise in the signal. We leave these tasks for our future study.
 	
 	The dependence of the energy, linear momentum, and angular momentum radiated by gravitational waves on the misalignment angle and EOS was shown. We found that the recoil induced by the mass ejection dominates the total recoil velocity for the case that the ejecta mass is larger than $\approx 0.01M_\odot$, while the recoil induced by the gravitational radiation is dominant for the case that the ejecta mass is smaller. 
 	
	Finally, we list several issues to be explored in the future. In this paper we studied the models only with $Q=5$ and $\chi=0.75$ to focus on the dependence on the BH spin misalignment and the EOS of the NS. As it is known that the mass ratio and the BH spin magnitude influences on the merger process, we also need to clarify the dependence on these parameters with the spin misalignment systematically. In particular, the larger BH spin enhances the tidal disruption of the NS, and thus	 characteristic features of misaligned-spin BH-NS mergers could be revealed more clearly. Also, we plan to perform more detailed analysis of gravitational waveforms for the misaligned-spin cases, because the waveforms may contain rich information on the misaligned BH-NS system.
\begin{acknowledgements}
	Kyohei Kawaguchi is grateful to Hiroki Nagakura, Kenta Hotokezaka, Kunihito Ioka, Kenta Kiuchi, Sho Fujibayashi, Takashi Yoshida, and Yuichiro Sekiguchi for valuable discussions. Koutarou Kyutoku is grateful to Francois Foucart for valuable discussions. This work was supported by JSPS Grant-in Aid for Scientific Research (24244028) and RIKEN iTHES project. Kyohei Kawaguchi is supported by JSPS Research Fellowship for Young Scientists (DC1). Hiroyuki Nakano acknowledges support by JSPS Grant-in-Aid for Scientific Research (24103006). Keisuke Taniguchi acknowledges support by JSPS Grant-in-Aid for Scientific Research (26400267).
\end{acknowledgements}
\appendix
\section{Convergence with respect to the grid resolution}\label{app:err}
\begin{table}\begin{center}\begin{tabular}{l|c|cc}\hline
	Model	&	~~$N$~~	&	~~~$M_{>{\rm AH}}[M_\odot]$~~~	&	~~~$M_{\rm eje}[M_\odot]$~~~	\\\hline\hline
	ALF2i30	&	40	&	0.172	&	$3.43\times10^{-2}$	\\
			&	48	&	0.166	&	$3.30\times10^{-2}$	\\
			&	60	&	0.156	&	$3.32\times10^{-2}$	\\\hline
	H4i30	&	40	&	0.264	&	$4.10\times10^{-2}$	\\
			&	48	&	0.257	&	$4.23\times10^{-2}$	\\
			&	60	&	0.248	&	$4.16\times10^{-2}$	\\\hline
	H4i60	&	40	&	$0.103$	&	$1.42\times10^{-2}$	\\
			&	48	&	$9.19\times10^{-2}$	&	$1.24\times10^{-2}$	\\
			&	60	&	$8.45\times10^{-2}$	&	$1.25\times10^{-2}$	\\\hline
	MS1i90	&	40	&	$2.64\times10^{-2}$	&	$8.68\times10^{-3}$	\\
			&	48	&	$2.38\times10^{-2}$	&	$9.91\times10^{-3}$	\\
			&	60	&	$2.23\times10^{-2}$	&	$9.78\times10^{-3}$	\\\hline
\end{tabular}\caption{$M_{>{\rm AH}}$ and $M_{\rm eje}$ for runs with different grid resolutions for selected models.}\label{tb:conv}\end{center}\end{table} 
	Table~\ref{tb:conv} compares $M_{>{\rm AH}}$ and $M_{\rm eje}$ among different grid resolutions for selected models. If we assume the first-order convergence between $N=48$ and $N=60$, the errors with $N=60$ results are always smaller than $\approx 40\%$ and $\approx 32\%$ for  $M_{>{\rm AH}}$ and $M_{\rm eje}$, respectively. Errors become large for a smaller mass. In particular, the error for $M_{>{\rm AH}}$ is $\approx 16\%$ for model H4i30, while the error is $\approx 28\%$ for  model MS1i90.

\end{document}